\documentclass[aps,prl,twocolumn,groupeaddress]{revtex4}  

\usepackage{graphicx}
\usepackage{bm}        
\usepackage{amssymb}   

\usepackage{graphicx,amsmath,amssymb,amsfonts}
\usepackage{enumerate,setspace}

  \newcommand{\q}{\mathcal{Q}}

 \newcommand{\e}{{\rm e}}

\newcommand{\E}{{\cal E}}
\renewcommand{\L}{{\cal L}}

\newcommand{\para}[1]{\vspace{1ex} \noindent {\bf #1}}

\newcommand{\A}{{\bf A}}
\newcommand{\V}{{\bf V}}
\newcommand{\G}{{\cal G}}

\newcommand{\x}{{\bf{ x}}}
\newcommand{\y}{{\bf{ y}}}

\renewcommand{\k}{\mathcal{K}}
\newcommand{\T}{\mathcal{T}}

\newcommand{\C}{\mathcal{C}}
\newcommand{\EQ}{\begin{equation}}
\newcommand{\EE}{\end{equation}}

\newcommand{\EQA}{\begin{eqnarray}}
\newcommand{\EEA}{\end{eqnarray}}

\usepackage{lipsum}
\usepackage{capt-of}
\usepackage{wasysym}

\begin{document}
\title{Host-pathogen coevolution and the emergence of\\ broadly neutralizing antibodies in chronic infections}
\author{Armita Nourmohammad$^{1}{\footnote{Correspondence should be addressed to: Armita Nourmohammad (armitan@princeton.edu).}\footnote{Authors with equal contribution}}$,  Jakub Otwinowski$^{2\dagger}$,  Joshua B Plotkin$^{2}$}
\affiliation{$^1$ Joseph-Henri Laboratories of Physics and Lewis-Sigler Institute for Integrative Genomics, \\ Princeton University, Princeton, NJ 08544, USA\\ $^2$ Department of Biology, University of Pennsylvania, Philadelphia, PA 19104}

\begin{abstract}
 The vertebrate adaptive immune system provides a flexible and
diverse set of molecules to neutralize pathogens. Yet, viruses such as HIV can
cause chronic infections by evolving as quickly as the adaptive immune system,
forming an evolutionary arms race.  Here we introduce a mathematical framework to
study the coevolutionary dynamics of antibodies with antigens within a host.  We
focus on changes in the binding interactions between the antibody and antigen
populations, which result from the underlying stochastic evolution of genotype
frequencies driven by mutation, selection, and drift.  We identify the critical
viral and immune parameters that determine the distribution of antibody-antigen
binding affinities. We also identify definitive signatures of coevolution that
measure the reciprocal response between antibodies and viruses, and we introduce
experimentally measurable quantities that quantify the extent of adaptation during
continual coevolution of the two opposing populations. Using this
analytical framework, we infer rates of viral and immune adaptation based on
time-shifted neutralization assays in two HIV-infected patients. Finally, we
analyze competition between clonal lineages of antibodies and characterize the
fate of a given lineage in terms of the state of the antibody and viral
populations. In particular, we derive the conditions that favor the emergence of
broadly neutralizing antibodies, which may be useful in designing a vaccine
against HIV.
\end{abstract}

\maketitle
\begin{center}{\Large \bf Introduction} \end{center}

It takes decades for humans to reproduce, but our pathogens can reproduce in less than a day. How can we coexist with pathogens whose potential to evolve is $10^4$-times faster than our own? In vertebrates, the answer lies in their adaptive immune system, which uses recombination, mutation, and selection to evolve a response on the same time-scale at which pathogens themselves evolve.

\begin{figure}
\begin{center}
\includegraphics[width= \columnwidth]{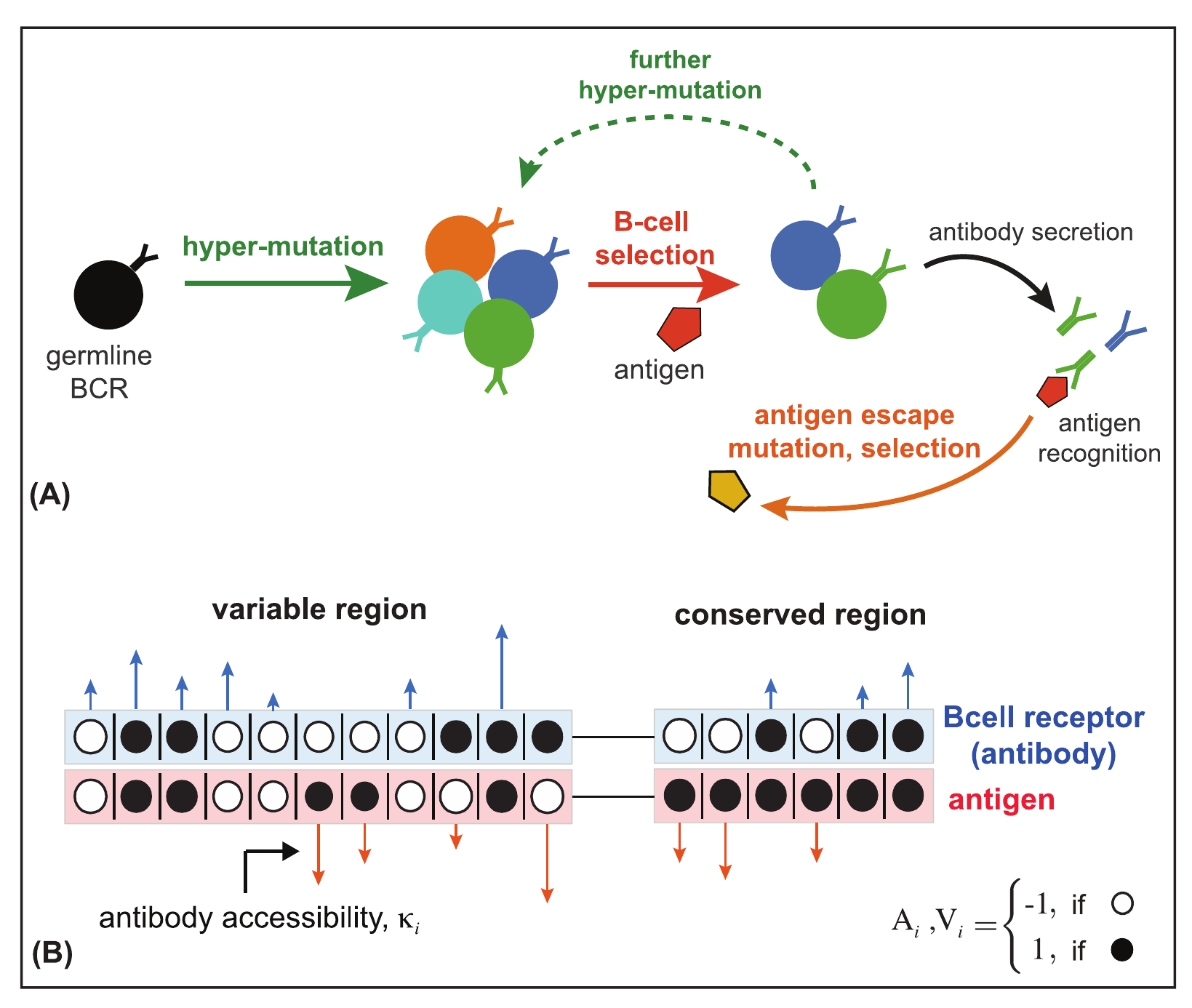}
\end{center}
\caption{{\bf coevolution of antibodies and viruses.} {\bf (A)} Schematic of affinity maturation in a germinal center. A  naive, germline B-cell receptor (black) with marginal binding affinity for the circulating antigen (red pentagon) enters the process of affinity maturation in a germinal center. Hypermutations produce a diverse set of B-cell receptors (colors), the majority of which do not increase the neutralization efficacy of B-cells, except for some beneficial mutations that increase binding affinity (dark blue and green) to the presented antigen. The selected B-cells may enter the blood and secrete antibodies, or enter further rounds of hypermutations to enhance their neutralization ability. Antigens mutate and are selected (yellow pentagon) based on their ability to escape the current immune challenge. {\bf(B)} We model the interaction between the genotype of a B-cell receptor and its secreted antibody (blue) with a viral genotype (red) in both variable and conserved regions of the viral genome. The black and white circles indicate the state of the interacting loci with values $\pm 1$. Loci in the conserved region of the virus are fixed at $+1$. The length of the arrows indicate the contribution  of each locus to the binding affinity, $\kappa_i$, which is a measure of the  accessibility of an antibody lineage to viral epitopes. The blue arrows indicate the interactions that increase binding affinity (i.e., loci with same signs in antibody and viral genotype), whereas red arrows indicate interaction that decrease the affinity (i.e., loci with opposite signs in antibody and viral genotype.) \label{fig:schematic}}
\end{figure}

One of the central actors in the adaptive immune system are B-cells, which recognize pathogens using highly diverse membrane-bound receptors. Naive B-cells are created by processes which generate extensive genetic diversity in their receptors via recombination, insertions and deletions, and hypermutations~\cite{Janeway:H7fnIHBf} which can potentially produce $\sim 10^{18}$ variants in a human repertoire~\cite{Elhanati:2015cp}. This estimate of potential lymphocyte diversity outnumbers the total population size of B-cells in humans, i.e., $\sim10^{10}$~\cite{Trepel:1974wx,Glanville:2009fp}.  During an infection, B-cells aggregate to form \emph{germinal centers}, where they hypermutate at a rate of about $\sim 10^{-3}$ per base pair per cell division over a region of 1-2 kilo base pairs~\cite{Odegard:2006hw}. The B-cell  hypermutation rate is approximately $4-5$ orders of magnitude larger than an average germline mutation rate per cell division in humans~\cite{Campbell:2013cp}.  Mutated B-cells compete for survival and proliferation signals from helper T-cells, based on the B-cell receptor's binding to antigens. This form of natural selection is known as \emph{affinity maturation}, and it can increase binding affinities up to 10-100 fold ~\cite{MeyerHermann:2012ja,Victora:2012gx, Cobey:2015bl}, see Fig.~\ref{fig:schematic}A. B-cells with high binding affinity may leave germinal centers to become antibody secreting plasma cells, or dormant memory cells that can be reactivated quickly upon future infections~\cite{Janeway:H7fnIHBf}. Secreted antibodies, which are the soluble form of B-cell receptors, can bind directly to pathogens to mark them for neutralization by other parts of the immune system. Plasma B-cells may recirculate to other germinal centers and undergo further hypermutation~\cite{Victora:2012gx}. 

Some viruses, such as seasonal influenza viruses, evolve quickly at the population level, but the adaptive immune system can nonetheless remove them from any given host within a week or two. By contrast, chronic infections can last for decades within an individual, either by pathogen dormancy or by pathogens avoiding neutralization by evolving as rapidly as B-cell populations. HIV mutation rates, for example, can be as high as $0.1-0.2$ per generation per genome~\cite{Duffy:2008is}. Neutralizing assays and phylogenetic analyses suggest an evolutionary arms race between B-cells and HIV populations during infection in a single patient~\cite{Richman:2003dc,Frost:2005,Moore:2009hv,Liao:2013gs,Luo:2015bi}. Viruses such as HIV have evolved to keep the sensitive regions of their structure inaccessible by the immune system e.g., through glycan restriction or immuno-dominant variable loops~\cite{Kwong:2002ec,Lyumkis:2013jv}. As a result, the majority of selected antibodies bind to the most easily accessible regions of the virus, where viruses can tolerate mutations and thereby escape immune challenge. Nonetheless, a remarkably large proportion of HIV patients ($\sim20\%$) eventually produce antibodies that neutralize a broad panel of virions~\cite{Simek:2009cn,DoriaRose:2010js} by attacking  structurally conserved regions, such as the CD4 binding site of HIV {\em env} protein~\cite{Walker:2009cd,Chen:2009ig,Zhou:2010gx,Walker:2011ew,Liao:2013gs}. These broadly neutralizing antibodies (BnAbs), can even neutralize HIV viruses from other clades, suggesting it may be possible to design an effective HIV vaccine if we can understand the conditions under which BnAbs arise~\cite{Walker:2009cd,Walker:2011ew,Mouquet:2013he,Kwong:2013ia,Klein:2013eb,Liao:2013gs,Wang:2015em}.

Recent studies have focused on mechanistic modeling of germinal centers in response to one or several antigens \cite{MeyerHermann:2012ja, Childs:2015fi}, and elicitation of BnAbs~\cite{Chaudhury:2014eu,Wang:2015em}. However, these studies did not model the coevolution of the virus and B-cell repertoire, which is important to understand how BnAbs arise \textit{in vivo}. Modeling of such coevolution is difficult because the mechanistic details of germinal center
activity are largely unknown~\cite{Luo:2015bi,Luo:2015gl}, and the multitude of parameters make it difficult to identify generalizable aspects of a model. While evidence of viral escape mutations and B-cell adaptation has been observed
experimentally~\cite{Richman:2003dc,Frost:2005,Moore:2009hv,Liao:2013gs} and modeled mechanistically~\cite{Chaudhury:2014eu,Wang:2015em}, it is not clear what are the generic features and relevant parameters in an evolutionary arms race that permit the development, or, especially, the early development of BnAbs. Phenomenological models ignore many details of affinity maturation and heterogeneity in the structure of germinal centers and yet produce useful qualitative predictions~\cite{Perelson:2002kw,Luo:2015bi,Luo:2015gl}. Past models typically described only a few viral types~\cite{Wang:2015em,Childs:2015fi}, and did not account for the vast genetic diversity and turnover seen in infecting populations. A recent study by Luo \& Perelson \cite{Luo:2015gl} described diverse viral and antibody populations, relying primarily on numerical simulations.

In this paper, we take a phenomenological approach to model the within-host coevolution of \emph{diverse} populations of B-cells and chronic viruses. We focus on the chronic infection phase, where the immune response is dominated by HIV-specific antibody-mediated mechanisms, which follow the strong response by the cytotoxic T-lymphocytes (i.e., CD8$+$ killers T-cells), around 50 days after infection~\cite{McMichael:2009aa}.  During the chronic phase, population sizes of viruses and lymphocytes are relatively constant but their genetic compositions undergo rapid turnover~\cite{Shankarappa:1999uk}. We characterize the interacting sites of B-cell receptors and viruses as mutable binary strings, with binding affinity, and therefore selection, defined by matching bits.  We keep track of both variable regions in the viral genome and conserved regions, asking specifically when B-cell receptors will evolve to bind to the conserved region, i.e., to develop broad neutralization capacity.  The main simplification that makes our analysis tractable is that we focus on the evolution of a shared interaction phenotype, namely the distribution of binding affinities between viral and receptor populations. Specifically, we model the effects of mutations, selection and reproductive stochasticity on the distribution of binding affinities between the two populations. Projecting from the high-dimensional space of genotypes to lower dimension of binding phenotypes allows for a predictive and analytical description of the coevolutionary process~\cite{NourMohammad:2013in}, whilst retaining the salient information about the quantities of greatest biological and therapeutic interest. 

Using this modeling approach we show that the evolution of the binding affinity does not depend on details of any single-locus contribution, but is an emerging property of all constitutive loci. Even though the coevolution of antibodies and
viruses is perpetually out of equilibrium, we develop a framework to quantify the amount of adaptation in each of the two populations by defining fitness and transfer flux, which partition changes in mean fitness. We discuss how to measure
the fitness and transfer flux from time-shifted experiments, where viruses are competed against past and future antibodies, and we show how such measurements provide a signature of coevolution. We use these analytical results to
interpret empirical measurements of time-shifted neutralization assays from two HIV-infected patients~\cite{Richman:2003dc}, and we infer two qualitatively different regimes of viral-antibody coevolution. We discuss the consequences of competition between clonal B-cell lineages within and between germinal centers. In particular, we derive analytic expressions for the fixation probability of a newly arisen, broadly neutralizing antibody lineage.  We find that BnAbs have an elevated chance of fixation in the presence of a diverse viral population, whereas specific neutralizing antibody lineages do not. We discuss the implications of these results for the design of preventive vaccines that elicit BnAbs against HIV.\\


\begin{center}{\Large \bf Model} \\
\vspace{1cm}
{{\bf Interaction  between antibodies and viruses}}
\end{center}

B-cell receptors undergo mutation and selection in germinal centers, whereas viruses are primarily affected by the receptors secreted into the blood, known as antibodies. Our model does not distinguish between antibodies and B-cells, so we will use the terms interchangeably. To represent genetically diverse populations we define genotypes for antibodies and viruses as binary sequences of  $\pm1$, where mutations change the sign of individual loci.  Mutations in some regions of a viral genome  are highly deleterious, e.g.~ at sites that allow the virus to bind target cell receptors, including CD4-binding sites for HIV. To capture this property we explicitly model a conserved region of the viral genome that does not tolerate mutations, so that its bits are always set to $+1$. We let viruses have variable bits at positions $i=1\,\dots\,\ell$, and conserved bits at positions $i=\ell+1,\dots,\ell+\hat \ell$; while antibodies have variable bits at positions $i=1\,\dots\,\ell+\hat \ell$; see Fig.~\ref{fig:schematic}B. 

Naive B-cells generate diversity by gene rearrangements (VDJ recombination), which differentiates their ability to bind to different epitopes of the virus; and then B-cells diversify further by somatic hypermutation and selection during affinity maturation. We call the set of B-cells that originate from a common germline sequence a clonal lineage.  A lineage with access to conserved regions of the virus can effectively neutralize more viral genotypes, since no escape mutation can counteract this kind of neutralization. 

The binding affinity between antibody and virus determines the likelihood of a given antigen neutralization by an antibody, and therefore it is the key molecular  phenotype that determines selection on both immune and viral populations. We model the binding affinity as a weighted dot product over all loci, which for antibody $A^\alpha$ chosen from the genotype space $\alpha \in 1 \dots 2^{\ell+\hat \ell}$ and virus $V^\gamma$ with $\gamma \in 1 \dots 2^{\ell}$ has binding affinity 

{ \begin{align}
\nonumber E_{{}_{\text{tot}}}^{\C} (A^\alpha,V^\gamma)&= \underbrace{ \sum_{i=1}^{\ell} \kappa_{i}^\C A^{\alpha}_i V^\gamma_i }_{\text{variable viral region}}+ \underbrace{\sum_{i=\ell+1}^{\ell+\hat\ell}\hat  \kappa^\C_{i}\,A^\alpha_i}_{\text{conserved viral region }} \\
&\equiv E_{\alpha\,\gamma}^\C +\hat E_\alpha^\C \label{binding}
\end{align}}

where, {\small $A^{\alpha}_i=\pm1$}  denotes the {\small $i^{th}$} locus of the $\alpha$ antibody genotype, and {\small $V^\gamma_i$} the {\small $i^{th}$} locus of the $\gamma$ viral genotype.  Matching bits at interacting positions enhance binding affinity between an antibody and a virus; see Fig.~\ref{fig:schematic}B. Similar models have been used to describe  B-cell maturation in germinal centers~\cite{Wang:2015em}, and T-cell selection based on the capability to bind external antigens and avoid self proteins~\cite{Detours:1999uq,Detours:2000wz}. The conserved region of the virus with $V_{i}=1$ is located at positions {\small $i=\ell+1,\dots,\ell+\hat \ell$} for all viral sequences. Consequently, the total binding affinity is decomposed into the interaction with the variable region of the virus, {\small $E_{\alpha\,\gamma}^\C $ } and with the conserved region of the virus, {\small $\hat E_\alpha^\C$}.  We call the lineage-specific binding constants {\small $\{\kappa_{i}^\C \geq 0 \}$} and {\small $\{\hat \kappa_{i}^\C\geq0\}$} the \emph{accessibilities}, because they characterize the intrinsic sensitivity of an antibody lineage to individual sites in viral epitopes. We begin by analyzing the evolution of a single  antibody lineage, and suppress the $\C$ notation for brevity. Coevolution with multiple antibody lineages is discussed in a later section.

Both antibody and viral populations are highly polymorphic, and therefore contain many unique genotypes.  While the binding affinity between a virus $V^\gamma$ and an antibody $A^\alpha$ is constant, given by eq.~(\ref{binding}), the frequencies of the antibody and viral genotypes, $x^\alpha$ and $y^\gamma$, and all quantities derived from them, change over time as the two populations coevolve.  To characterize the  distribution of binding affinities we define the genotype-specific binding affinities in each population, which are marginalized quantities over the opposing population: {\small $E_{\alpha\,\cdot } = \sum_\gamma E_{\alpha\,\gamma}  y^\gamma  $} for the antibody $A^\alpha$, and {\small $E_{.\,\gamma} = \sum_\alpha  E_{\alpha\,\gamma}x^\alpha$} for the virus $V^\gamma$. We will describe the time evolution of the joint distribution of {\small $E_{\alpha\,\cdot}$, $\hat E_{\alpha}$, and $E_{\cdot\,\gamma}$}, by considering three of its moments: (i) the mean binding affinity, which is the same for both populations {\small $\E=\sum_{\alpha} E_{\alpha\,\cdot} x^\alpha = \sum_{\gamma} E_{\cdot\,\gamma} y^\gamma$},  (ii) the diversity of  binding affinity in the antibodies, {\small $M_{A,2} =\sum_{\alpha} \left(E_{\alpha\,\cdot} - \E \right)^2 x^\alpha$} and (iii) the diversity of binding affinities in the viruses, {\small $ M_{V,2} =\sum_{\gamma}\left( E_{\cdot\,\gamma} - \E \right)^2 y^\gamma $}. Analogous statistics of binding affinities can be defined for the conserved region of the virus,  which we denote by $\hat \E$ for the mean interaction, and {\small $\hat M_{A,2}$} for the diversity across antibodies. The diversity of viral interactions in the conserved region must always equal zero, {\small $\hat M_{V,2}=0$}. \\

\vspace{1cm}
\begin{center}
{{\bf {Coevolution of an antibody lineage and viruses}} }\\
\end{center}

We first characterize the affinity maturation process of a single clonal antibody lineage coevolving with a viral population, which includes hypermutation, selection, and stochasticity due to population size in germinal centers, i.e., genetic drift. \\

\paragraph{Genetic drift and evolutionary time-scales.} Stochasticity in reproductive success, known as genetic drift, is an important factor that depends on population size, and therefore we model genetic drift by keeping populations at finite size $N_a$ for antibodies, and $N_v$ for viruses. Although the population of B-cells can reach large numbers within an individual host, significant bottlenecks occur in germinal centers, where there may be on the order of $\sim 10^3-10^4$ B-cells~\cite{MeyerHermann:2012ja}.  For HIV, estimates for intra-patient viral divergence suggests an effective population size of about $\sim 10^2-10^3$, which is much smaller than the number of infected cells within a patient $\sim 10^7-10^9$~\cite{Lemey:2006wb}.

Fluctuations by genetic drift define an important time-scale in the evolution of a polymorphic population: the neutral coalescence time is the characteristic time that two randomly chosen neutral alleles in the population coalesce to their most recent common ancestor, and is equal to $N$ generations. Neutral coalescence time is estimated by phylogenetic analysis, and is often interpreted as an effective population size, which may be different from the census population size. Coalescence time can be mapped onto real units of time (e.g., days) if sequences are collected with sufficient time resolution. Without loss of generality, we assume that generation times in antibodies and viruses are equal, but we distinguish between the neutral coalescence time of antibodies and viruses  by using distinct values for their population sizes, i.e., $N_a$ in antibodies and $N_v$ in viruses.\\

 \paragraph{Mutations.} In the bi-allelic model outlined in Fig.~\ref{fig:schematic}B, a mutation changes the sign of an antibody site, i.e.,~{\small $A_i^\alpha \rightarrow-A_i^\alpha$}, affecting binding affinity in proportion to the lineage's intrinsic accessibility at that site, $\kappa_i$. Therefore, a mutation in an antibody at position $i$ changes {\small $E_{\alpha\,.}$} by {\small $\delta_i E_{\alpha\,.} = - 2 \,\kappa_i  A_i^\alpha \sum_\gamma V_i^\gamma y^\gamma$}. Likewise, a mutation at position $j$ of a virus  {\small $V^\gamma_j \rightarrow - V^\gamma_j $} affects binding affinity in proportion to $\kappa_j$. We assume constant mutation rates in the variable regions of the viruses and antibodies: $\mu_v$ and $\mu_a$ per site per generation.  

Empirical estimates of per-generation mutation rates for viruses $\mu_v$ or hypermutation rates of BCR sequences $\mu_a$ are extremely imprecise, and so we rescale mutation rates by neutral coalescence times. To do this, we consider measurements of standing neutral sequence diversity, estimated from genetic variation in, e.g., four-fold synonymous sites of protein sequences at each position. Neutral sequence diversity for the antibody variable region, which spans a couple of hundred base pairs, is about $\theta_a= N_a \mu_a=0.05-0.1$~\cite{Elhanati:2015cp}.  Nucleotide diversity of HIV increases over time within  a patient, and ranges between $\theta_v= N_v\mu_v=10^{-3}-10^{-2}$ in the {\em env} protein of HIV-1 patients, with a length of about a thousand base pairs~\cite{Zanini:2015vb}. Interestingly, the total diversity of the variable region in BCRs is comparable to the diversity of its main target, the {\em env} protein, in HIV. Both populations have on the order of 1-10 mutations per genotype per generation, which we use as a guideline for parameterizing simulations of our model.\\

\paragraph{Selection.} Frequencies of genotypes change according to their relative growth rate, or fitness.  The change in the frequency of antibody $A^\alpha$ with fitness {\small $f_{A^\alpha} $ is $\Delta \, x^\alpha = (f_{A^\alpha}-F_A) x^\alpha$} per generation, where we define (malthusian) fitness as proportional to the growth rate, and {\small $F_A=\sum_\alpha f_{A^\alpha} x^\alpha$} denotes the mean fitness of the antibody population  (see Section~A of the Appendix). Likewise, the change in frequency of virus $V^\gamma$ due to selection per generation is, {\small $\Delta \,y^\gamma = (f_{V^\gamma}-F_V) y^\gamma$}, where $F_V$ denotes the mean fitness in the viral population. 

During affinity maturation in a germinal center, a B-cell's growth rate depends on its ability to bind to the limited amounts of antigen, and to solicit survival signals from helper T-cells~\cite{Victora:2012gx}. At the same time, viruses are neutralized by antibodies that have high binding affinity. The simplest functional form that approximates this process, and for which we can provide analytical insight, is  linear  with respect to the  binding affinity,

{ \begin{align} 
f_{A^\alpha} &= S_a (E_{\alpha\,\cdot} +\hat E_{\alpha}) \label{ABfitness} \\
 f_{V^\gamma} &= -S_v (E_{\cdot\,\gamma} +\hat E_{\cdot}) \label{Virfitness} 
 \end{align}}
for antibody $A^\alpha$ and virus $V^\gamma$.  The selection coefficient  {\small $S_a>0$} quantifies the strength of selection on the binding affinity of antibodies. The value of {\small $S_a$ } may decrease in late stages of a long-term HIV infection, as the host's T-cell count decays~\cite{Perelson:2002kw}, but we do not model this behavior. The viral selection coefficient {\small $S_v >0$} represents immune pressure impeding the growth of the virus. The contribution of the conserved region to the fitness of the virus is independent of the viral genotype in eq.~(\ref{Virfitness}), and it does not affect the relative growth rates of the viral strains.

The number of sites and the magnitude of their accessibilities affect the overall strength of selection on binding affinity. Therefore, it is useful to absorb the intrinsic effects of the phenotype magnitude into the selection strength, and use rescaled values that are comparable across lineages of antibodies, and across experiments. We therefore rescale quantities related to the binding affinity by the total scale of the phenotypes  {\small $E_0^2=\sum_i \kappa_i^2$} and  {\small $\hat E_0^2=\sum_i\hat \kappa_i^2$}, such that {\small $E_{\alpha\gamma}\to E_{\alpha\gamma}/E_0$} and {\small $\hat E_{\alpha\gamma}\to \hat E_{\alpha\gamma}/\hat E_0$}, resulting in rescaled mean binding affinities  $\varepsilon$ and {\small $\hat\varepsilon$}, and diversities {\small $m_{A,2}$}, {\small $\hat m_{A,2}$} and {\small $m_{V,2}$}  in variable and conserved regions of both populations. Accordingly, we define rescaled selection coefficients {\small $s_a = N_a S_a E_0$}, {\small $\hat s_a = N_a S_a \hat E_0$}, {\small $s_v= N_v S_v E_0$} and {\small $\hat s_v =N_v S_v \hat E_0$}, which describe the total strength of selection on binding affinity; see Section~B.1 of the Appendix for details. 

Many aspects of affinity maturation are not well known, and so it is worth considering other forms of selection. In Section~B.5 of the Appendix we describe fitness as a non-linear function of the binding affinity.  In particular, we consider fitness that depends on the antibody activation probability, which is a sigmoid function of the \emph{strongest} binding affinity among a finite number of interactions with antigens. The linear fitness function in eq.~(\ref{ABfitness}) is  a limiting case of this more general fitness model. While most of our analytical results are based on the assumption  of linear fitness function, we also discuss how to quantify adaptation for arbitrary fitness models, and we numerically study the effect of nonlinearity on the rate  of antibody adaptation during affinity maturation.\\

\begin{center}{\Large\bf Results} \\
\vspace{1cm}
{{\bf Evolution of the mean binding affinity}}\end{center}

We focus initially on understanding the (rescaled) mean binding affinity {\small $\varepsilon$, $\hat \varepsilon$} between a clonal antibody lineage and the viral population, since this is a proxy for the overall neutralization ability that is commonly monitored during an infection.  Combining genetic drift with mutation and selection, and assuming a continuous-time and continuous-frequency process, results in a stochastic dynamical equation for the evolution of rescaled mean binding affinity in the variable region,

{ { \begin{align}
\nonumber \frac{d}{d\tau}\varepsilon=& -2\left [\theta_a+\theta_v (N_A/N_v)\right ]\varepsilon+ s_a m_{A,2} -s_v m_{V,2}\\
&+ \sqrt{m_{A,2}+\frac{N_a}{N_v}m_{V,2}} \,\,\chi_\varepsilon \label{eq:dE}
\end{align}}}
and in the conserved region,
{ \EQ
\frac{d}{d\tau}\hat \varepsilon= -2\theta_a\hat \varepsilon+ s_a\hat  m_{A,2}+ \sqrt{{\hat m_{A,2}}}\chi_{\hat \varepsilon}\label{eq:dhatE}
\EE}

where {\small $\chi_\varepsilon$} and {\small $\chi_{\hat \varepsilon}$} are standard Gaussian noise terms, and time $\tau$ is  measured in units of the antibody coalescence time $N_a$. Our analysis neglects the correlation between the variable and the conserved regions of the virus, which is due to physical linkage of the segments. In Section~B.4 of the Appendix  we show that a difference in evolutionary time-scales between these regions reduces the magnitude of this correlation.

\begin{figure}
\centering\includegraphics[width=\columnwidth]{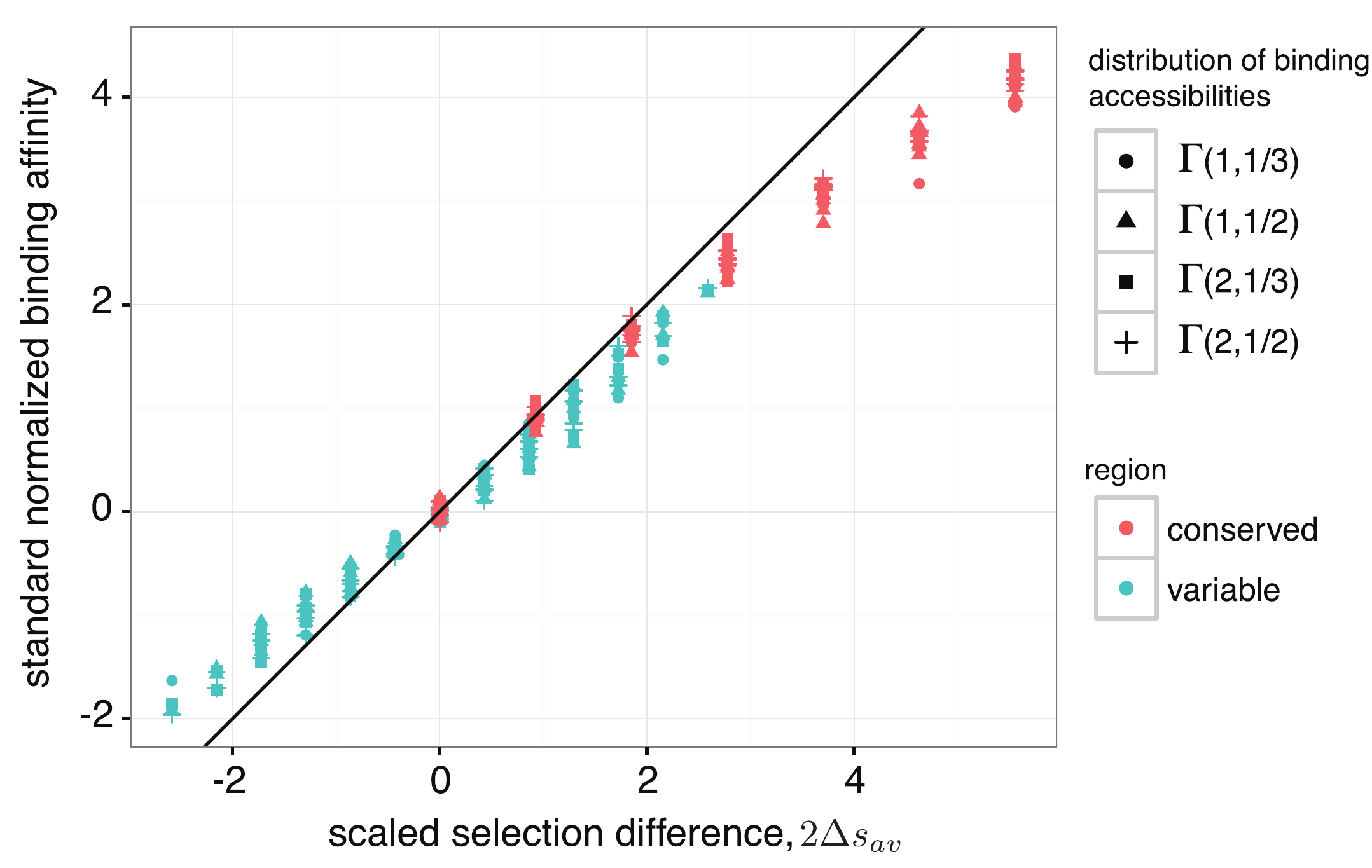}
\caption{{\bf Effect of selection on immune-virus binding affinity.} The stationary mean binding affinity, rescaled by antibody binding diversity ({\small $\E/\sqrt{M_{A,2}/4\theta_a}$}), on the y-axis, is well approximated by the scaled selection difference between antibody and viral populations, {\small $\Delta s_{av}$}, as predicted by our analysis (eq.~(\ref{Est})). Points show results of Wright-Fisher simulations, and the solid line has slope 1. Note that the mean binding affinity is insensitive to the details of heterogeneous binding accessibilities, $\kappa_i$, associated with an antibody lineage. Accessibilities $\kappa_i$ are drawn from several different $\Gamma$-distributions, shown in legend. Small deviations from the predicted mean binding are caused by higher moments of binding affinities, which can also be understood analytically (Fig.~S1). Simulation parameters are detailed in the Appendix. 
\label{fig:E}}
\end{figure}
As  eqs.~(\ref{eq:dE}, \ref{eq:dhatE}) reflect, mutations drive the mean affinity towards the neutral value, zero, whereas selection pushes it towards positive or negative values. The efficacy of selection on binding affinity is proportional to the binding diversity {\small$m_{A,2}$, $m_{V,2}$} in each of the populations. If a population harbors a large diversity of binding affinities then it has more potential for adaptation from the favorable tail of the distribution, which contains the most fit individuals in each generation~\cite{Fisher:1930wy,Price:1970vw}. It follows that selection on viruses does not affect the evolution of their conserved region, where the viral diversity of binding is always zero, {\small$\hat m_{V,2}=0$}. In Section~B.3 of the Appendix and Fig.~S2 we study the evolution of the higher central moments in detail.

The dynamics in eqs.~(\ref{eq:dE},\ref{eq:dhatE}) simplify in the regime where selection on individual loci is weak ({\small $N S \kappa <1$}), but the additive effects of selection on the total binding affinity are substantial ({\small $1\apprle s \ll \theta^{-1}  $}). This evolutionary regime is, in particular, relevant for  HIV escape from the humoral neutralizing antibody response~\cite{Zanini:2015vb}, that follows  the initial strong response to cytotoxic T-lymphocytes~\cite{Kessinger:2013aa}. In this parameter regime, the binding diversities are fast variables compared to the mean affinity, and can be approximated by their stationary ensemble-averaged values (Fig.~S3), which depend only weakly on the strength of selection even for substantial selection {\small  $s\sim 1$}: {\small $\langle m_{A,2} \rangle \simeq   4\theta_a $} and {\small $\langle m_{V,2}\rangle\simeq  4\theta_v$}. Higher-order corrections (Section~B.3 of the Appendix and Fig.~S2) show that strong selection reduces binding diversity.  The ensemble-averaged mean binding affinities relax exponentially towards their stationary values,

{ \begin{align}
\label{Est}\langle \varepsilon \rangle &\simeq \frac{2(s_a \theta_a - s_v\theta_v (N_a/N_v))}{\theta_a+\theta_v (N_a/N_v)} \equiv 2\, \Delta s_{av} \\
\label{Ehatst}\langle \hat\varepsilon \rangle &\simeq 2\,\hat s_a
\end{align}}
where {\small $\Delta s_{av}$} is an effective selection coefficient for binding affinity in the variable region, combining the effect of selection from both populations and accounting for their distinct genetic diversities. The stationary mean binding affinity quantifies the balance of mutation and selection acting on both populations. A strong selection difference between two populations {\small $\Delta s_{av}\gg 1$ } results in selective sweeps for genotypes with extreme values of binding affinity in each population, and hence, reduces the binding diversity. We validated our analytical solution for stationary mean binding, with corrections due to selection on binding diversity (Section~B.3 of the Appendix), by comparison with full, genotype-based Wright-Fisher simulations across a broad range of selection strengths (Figs.~S1 and~S2).

The weak dependence of binding diversity on selection allows for an experimental estimation  of the stationary rescaled mean binding affinity, using  measurements of the binding affinity distribution and neutral sequence diversities. The rescaled binding affinity can be approximated as: {\small $\varepsilon \approx \langle \E \rangle / \sqrt{\langle M_{A,2} \rangle /4\theta_a}$} and {\small $\hat \varepsilon \approx \langle \hat \E \rangle / \sqrt{\langle \hat M_{A,2} \rangle /4\theta_a}$}. Fig.~\ref{fig:E} demonstrates the utility of this approximation, and it shows that heterogeneous binding accessibilities, $\kappa_i$, drawn from several different distributions, do not affect stationary mean binding. Only the total magnitude of the accessibilities is relevant, as it determines the effect of selection on the whole phenotype. Although we have formulated a high-dimensional stochastic model of antibody-antigen coevolution in polymorphic populations, we can nonetheless understand the long-term binding affinities, which are commonly measured in patients, in terms of only a few key parameters.

In Section~B.5 of  the Appendix we numerically study the non-linear fitness landscapes described in the Model section, and their effect on the stationary mean binding and rate of adaptation (Fig.~S4). While the results differ quantitatively, we can qualitatively understand how the stationary mean binding affinity depends on the form of non-linearity.\\
\vspace{.5cm}
\begin{center}
{{\bf Fitness and transfer flux}}\\ 
\end{center} 

The antagonistic coevolution of antibodies and viruses is a non-equilibrium process, with each population  constantly adapting to a dynamic environment, namely, the state of the opposing population.  As a result, any time-independent quantity, such as the stationary mean binding affinity studied above, is itself not informative for the extent of coevolution that is occurring.  For example, a stationary mean binding affinity of zero (equivalently {\small $\Delta s_{av}=0$} in eq.~(\ref{Est})) can indicate either neutral evolution or rapid coevolution induced by equally strong selection in antibody and viral populations. 
 \begin{figure}
\begin{center}
\centering\includegraphics[width=\columnwidth]{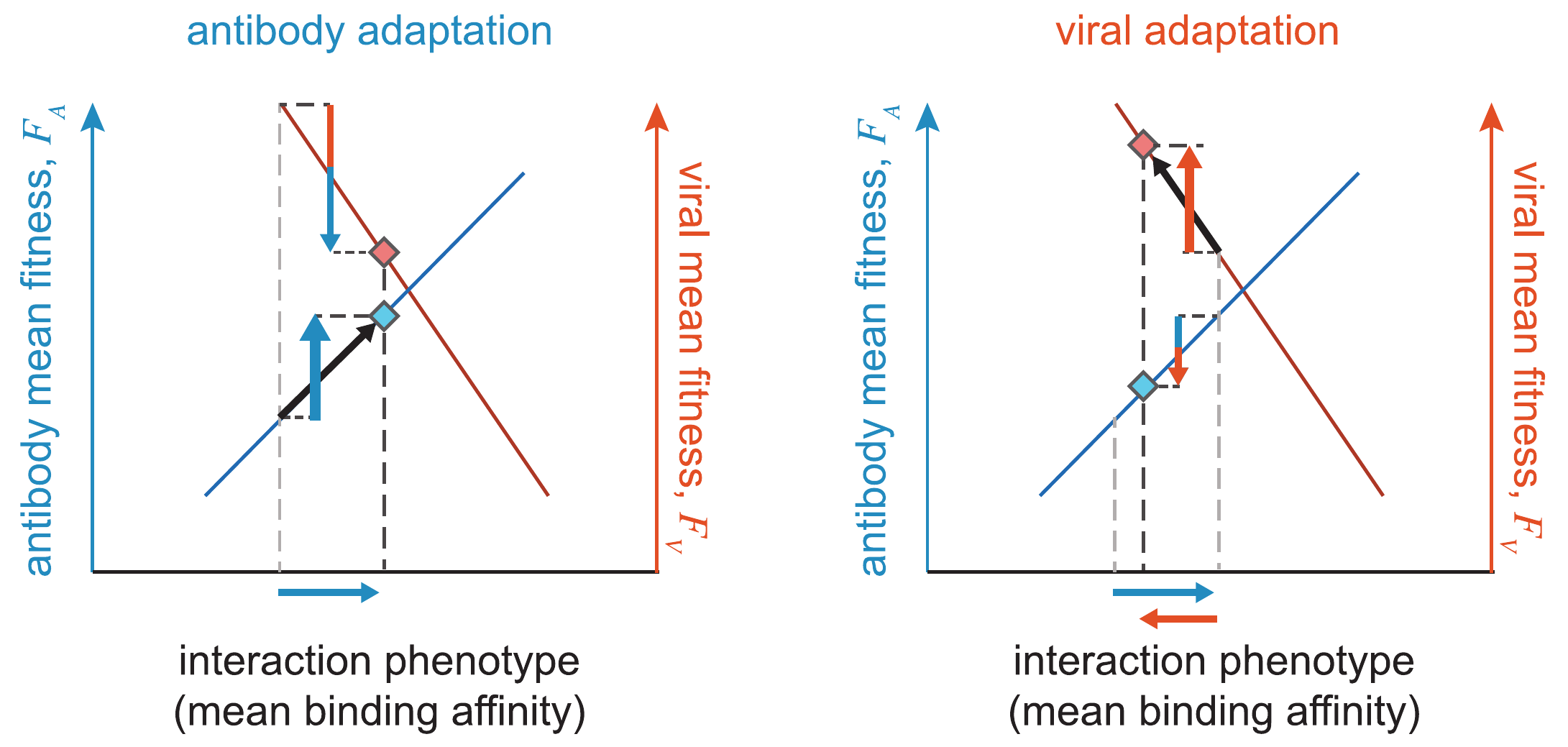}
\end{center}
\caption{\label{fig:flux} {\bf Fitness and transfer flux in antibody-viral coevolution.} The schematic diagram shows adaptation of antibody (blue diamond) and viral (red diamond) populations on their respective fitness landscapes, which depend on the common binding phenotype shown on the x-axis (i.e., the mean binding affinity).  During one step of antibody adaptation (left),  mean binding affinity increases (horizontal blue arrow) to enhance the fitness of the antibody population, with a  rate equal to  the antibody fitness flux $\phi_A$ (upward blue arrow). In the regime of strong selection, the fitness flux is proportional to the variance of fitness in the population; see eq.~(\ref{flux}). Adaptation of antibodies reduces the mean fitness in the viral population, with a rate proportional to the transfer flux from antibodies to viruses $\T_{A\to V}$ (downward red/blue arrow). On the other hand, viral adaptation (right) reduces the binding affinity and affects the fitness of both populations, with rates proportional to the viral fitness flux $\phi_V$ (upward red arrow) and the transfer flux from viruses to antibodies $\T_{V\to A}$ (downward blue/red  arrow); see eq.~(\ref{transfer}). Cumulative fitness flux (the sum of upward arrows) and cumulate transfer flux (the sum of downward arrows) over an evolutionary period quantify the amount of adaptation and interaction in the two antagonistic populations.
}
\end{figure}
To quantify the amount of adaptation and extent of interaction in two coevolving populations we will partition the change in mean fitness of each population into two components. We measure adaptation by the \emph{fitness flux}~\cite{Mustonen:2009vu,Mustonen:2010iga,Held:2014di}, which generically quantifies adaptation of a population in response to a changing environment (in this case the opposing population); see schematic Fig.~\ref{fig:flux}.  For our model, the fitness flux of the antibody population quantifies the effect of changing genotype frequencies on mean fitness, and is defined as {\small $\phi_A(t) =\sum_\alpha {\partial_{x^\alpha} F_A(t)} \,  d  x^\alpha(t)/dt $}, where {\small $F_A$ }denotes the mean fitness of antibodies, and the derivative {\small $dx^\alpha(t) /dt$} measures the change in frequency of the antibody  $A^\alpha$. The forces of mutation, drift, and selection all contribute to fitness flux, however the portion of fitness flux due to selection equals the population variance of fitness, in accordance with Fisher's theorem~\cite{Fisher:1930wy}. The second quantity we study, which we term the \emph{transfer flux}, measures the amount of  interaction between the two populations by quantifying the change in  mean  fitness  due to the response of the opposing population (schematic Fig.~\ref{fig:flux}). The transfer flux from viruses to antibodies is defined as {\small $\T_{V\to A}(t) = \sum_\gamma {\partial _{y^\gamma}F_A(t)}\,dy^\gamma(t)/dt$}. Analogous measures of adaptation and interaction can be  defined for the viral population (see Section~C of the Appendix).

The fitness flux and transfer flux represent rates of adaptation and interaction, and they are typically time dependent, except in the stationary state. The total amount of adaptation and interaction during non-stationary evolution, where the fluxes change over time, can be measured by the cumulative fluxes over a period of time: {\small $ \Phi_A(\tau_a)= N_a \int_{t'=0}^{t} \phi_A(t')\, dt'$ and  ${\bf T}_{V\to A}(\tau_a )= N_a \int_{t'=0}^{t} \T_{V\to A}(t') \,dt'$}, where time {\small $\tau_a=t/N_a$} is measured in units of neutral  coalescence time of antibodies $N_a$. In the stationary state, the ensemble-averaged cumulative fluxes grow linearly with time.   For coevolution  on the fitness  landscapes given  by equations~(\ref{ABfitness},\ref{Virfitness}),  the ensemble-averaged, stationary cumulative fitness flux and transfer flux in antibodies are

{  \begin{align}
 \label{flux} \langle \Phi_A(\tau_a) \rangle &= \big[ -2 \theta_a s_a \langle \varepsilon\rangle + s_a^2 \langle m_{A,2}\rangle \big]\tau_a\\
\label{transfer}  \langle {\bf T}_{V\to A} (\tau_a) \rangle&= \big[-2 \theta_v s_a   \langle \varepsilon\rangle -s_a s_v  \langle m_{V,2}\rangle  \big] (N_a/N_v )\tau_a
 \end{align}}
Note that the factor {\small $(N_a/N_v) \tau_a$ }in eq.~(\ref{transfer}), which is a rescaling of time in units of viral neutral coalescence time {\small $\tau_v=t/N_v$}, emphasizes the distinction between the evolutionary time scales of antibodies and viruses. The first terms on the right hand side of  eqs.~(\ref{flux}, \ref{transfer}) represent the fitness changes due to mutation, the second terms are due to selection, and the effects of genetic drift are zero in the ensemble average for our linear fitness landscape. Notably, the flux due to the conserved region of the virus is zero in stationarity, as is the case for evolution in a static fitness landscape (i.e., under equilibrium conditions).  In the stationary state, the cumulative fitness and transfer fluxes sum up to zero, {\small $\langle \Phi_A(\tau_a)\rangle+\langle {\bf T}_{V\to A}(\tau_a)\rangle=0$}.

Fitness flux and transfer flux are  generic quantities that are independent of the details of our model, and so they provide a natural way to compare the rate of adaptation in different evolutionary models or in different experiments.  In the regime of strong selection {\small $s_a,s_v\gtrsim1$}, non-linearity of the fitness function results in a more narrow distribution of fitness values in the antibody population, and hence, reduces the rate of adaptation and fitness flux; see Fig.~S4. In the following section we show how to use fitness and transfer flux to detect signatures of significant antibody-antigen coevolution.\\
\vspace{1cm}

\begin{center}{{\bf Signature of coevolution and inferences from time-shifted experiments}}\\ 
\end{center}
Measuring interactions between antibodies and viruses isolated from different times provides a powerful way to identify coevolution.  These ``time-shifted'' neutralization measurements in HIV patients have shown that viruses are more
resistant to past antibodies, from which they have been selected to escape, and more susceptible to antibodies from the future, due to selection and affinity maturation of B-cells~\cite{Richman:2003dc,Frost:2005,Moore:2009hv}.

\begin{figure*}[t!]
\begin{center}
\centering\includegraphics[width=2 \columnwidth]{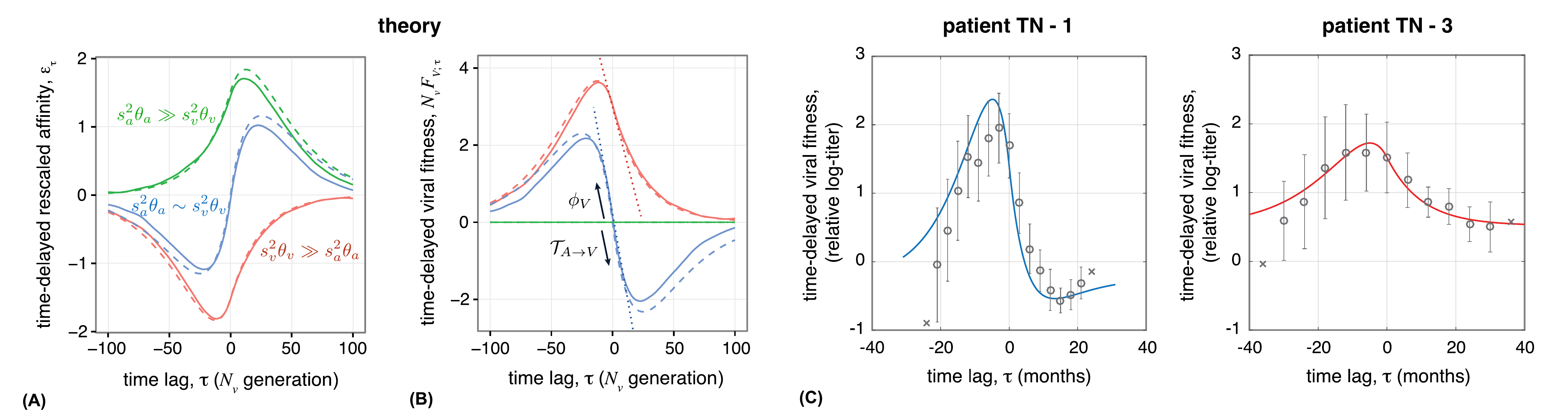}
\end{center}
\caption{\label{fig:delay} 
{\small {\bf Time-shifted binding assays between antigens and antibodies provide a definitive signature of viral-immune coevolution.} Viruses perform best against antibodies from the past and perform worst against antibodies from the future due to the adaptation of antibodies. {\bf(A)} Stationary rescaled binding affinity between viruses from time $t$ and antibodies from time $t+\tau$, averaged over all t: $\varepsilon_\tau= \langle \sum_{\alpha,\gamma} E_{\alpha\gamma} y^\gamma (t) x^\alpha(t+\tau) \rangle_t/E_0$, and {\bf (B)} time-shifted mean fitness of viruses $N_v F_{V;\tau}=-s_v \varepsilon_\tau$, are shown for three regimes of coevolutionary dynamics: strong  adaptation of both populations $s_a^2 \theta_a \sim s_v^2 \theta_v$, with $s_a=s_v=2$  (blue), stronger adaptation of viruses $s_v^2 \theta_v \gg s_a ^2\theta_a $ with $s_v=2, \,s_a=0$ (red), and stronger adaptation of antibodies $s_a^2\theta_a\gg s_v^2\theta_v $ with $s_v=0, \, s_a=2$ (green).  Wright-Fisher simulations (solid lines) are compared to the analytical predictions from eqs.~(S102, S103) in each regime (dashed lines).  The ``S"-shape curve in the blue regime is a signature of two antagonistically
coevolving populations $s_v\theta_v \sim s_a\theta_a$. For large time-shifts, binding relaxes to its neutral value, zero, as mutations randomize genotypes. In the absence of selection in one population, the time-shifted binding affinity reflects adaptation of one population against stochastic variation in the other due to mutation and genetic drift.  The slope of time-shifted fitness at lag $\tau=0$ is the viral population's fitness flux (slope towards the past) and the transfer flux from the opposing population (slope towards the future), which are equal to each other in the stationary state.  The slope of the dotted lines indicate the predicted fitness flux and transfer flux (eqs.~(\ref{flux}, \ref{transfer}).  Time-shifted fitness shown here does not include binding to the conserved region since that value is constant for all time-shifts in stationarity (see Fig.~S6 for non-stationary state). Simulation parameters are given in the Appendix. {\bf (C)} Empirical time-shifted fitness measurements of HIV based on a neutralization titer ($\text{IC}_{50}$)~\cite{Richman:2003dc}, averaged over all time points with equal time-shift $\tau$. Circles show averaged fitness $\pm$ 1 standard error, and crosses show fitness at time-shifts with only a single data point. Solid lines show analytical fits of our model to the data (see Section~F of Appendix).  In patient TN-1, viruses and antibodies experience a comparable adaptive pressure, with a similar time-shift pattern to the  blue ``S-curve" in panel (B).  In patient TN-3, however, adaptation in viruses is much stronger than in antibodies, resulting in an imbalanced shape of the time-shifted fitness curve, similar to the red curve in panel (B).  
}}
\end{figure*}

We can predict the form of time-shifted binding assays under our model; see Section~D of the Appendix for details. The rescaled time-shifted binding affinity between viruses at time $t$ and antibodies at time $t+\tau$ is given by {\small $\varepsilon_\tau(t)= \sum_{\alpha,\gamma}E_{\alpha\, \gamma} y^\gamma(t)\, x^\alpha(t+\tau) /E_0$} and {\small $\hat \varepsilon_\tau(t)= \sum_{\alpha}\hat E_{\alpha}  x^\alpha(t+\tau) /\hat E_0$} for the variable and the conserved region, respectively. The corresponding viral mean fitness at time $t$ against the antibody population at time  {\small $t+\tau$} is {\small $N_v F_{V;\tau}(t) = -s_v (\varepsilon_{\tau}(t)+\hat \varepsilon_\tau(t))$}. The slope of the time-shifted viral fitness at the time where the two populations co-occur (i.e., {\small $\tau=0$}), approaching from negative $\tau$, i.e., from the past, measures the amount of adaptation of the viral population in response to the state of the antibody population, and it is precisely equal to the fitness flux of viruses: {\small $\partial_\tau F_{V;\tau}(t-\tau) |_{\tau=0^-} = \phi_V(t)$}.  The slope approaching from positive time-shifts, i.e., from the future, measures the change in the mean fitness of the viral population due to adaptation of the antibody population, and it is precisely equal to the transfer flux from antibodies to viruses {\small $\partial_\tau F_{V;\tau}(t) |_{\tau=0^+} = \T_{A\rightarrow V}(t)$}. Similarly, we can define time-shifted fitness with antibodies as the focal population; see Section~D of the Appendix.  In stationarity, the sum of fitness flux and transfer flux is zero on average, and so the slopes from either side of $\tau=0$ are equal, as in Fig.~\ref{fig:delay} and Fig.~S5.  Note that these relationships between time-shifted fitness and the flux variables hold in general, beyond the specific case of a linear fitness landscape.  In a non-stationary state, the fitness flux and transfer flux are not balanced, and so {\small $\langle F_{V;\tau}(t)\rangle$ } has a discontinuous derivative at {\small $\tau=0$} (Fig.~S6). Therefore, observation of such a discontinuity provides a way to identify stationarity versus transient dynamics, given sufficient replicated experiments.
   
Whether in stationarity or not, the signature of out-of-equilibrium evolution is a positive fitness flux and negative transfer flux. For time-shifted fitness, this means that for short time shifts, where dynamics are dominated by selection, viruses have a higher fitness against antibodies from the past, and have lower fitness against  antibodies from the future. This is true even when one population is evolving neutrally and the other has substantial selection, as shown in Fig.~\ref{fig:delay}A. For long time shifts, the sequences are randomized by mutations and the fitness decays exponentially to the neutral value.  When selection and mutation are substantial on both sides the time-shifted fitness curve has a characteristic ``S" shape -- a signature of coevolution, whose inflected form can be understood in terms of the fitness and transfer fluxes. In Section~D of the Appendix we analytically  derive the functional form of the time-shifted binding affinity and fitness dependent on the evolutionary parameters. Fig.~\ref{fig:delay}A,B and Fig.~S5 show good agreement between Wright-Fisher simulations and our analytical predictions in eqs.~(S102, S103) for the stationary time-shifted binding affinity and viral fitness.

We can use our analytical results to interpret empirical measurements of time-shifted viral neutralization by a patient's circulating antibodies. We analyzed data from Richman {\em et~al.}~\cite{Richman:2003dc} on two HIV-infected
patients. We approximated the fitness of the virus against a sampled serum (antibodies)  as the logarithm of the neutralization titer {\small $F_V\simeq-\log\text{titer}$};  here titer is the reciprocal of antibody dilution where inhibition reaches 50\% ($\text{IC}_{50}$)~\cite{Blanquart:2013fq}. A signature of coevolution can sometimes be obscured when the fitnesses of antibodies and viruses also depend on time-dependent intrinsic and environmental factors, such as drug treatments~\cite{Blanquart:2013fq}. Therefore, we used fitness of a neutralization-sensitive virus (NL43) as a control measurement to account for the increasing antibody response during infection, shown in Fig.~S7. The relative time-shifted viral fitness in Fig.~\ref{fig:delay}C for the two HIV patients (TN-1 and TN-3), match well with the fits of our analytical equations (see Section~F of the Appendix).  The inferred parameter values indicate two distinct regimes of coevolutionary dynamics in the two patients. In patient TN-1, viruses and antibodies experience a comparable adaptive pressure, as indicated by the ``S-curve" in Fig.~\ref{fig:delay}C (blue line), whereas in patient TN-3, adaptation in viruses is much stronger than in antibodies, resulting in an imbalanced shape of the time-shifted fitness curve in Fig.~\ref{fig:delay}C (red line). We describe the inference procedure and report all inferred parameters in Section~F of the Appendix. The resolution of the data~\cite{Richman:2003dc} allows only for a qualitative interpretation of coevolutionary regimes. A more quantitative analysis can be achieved through longer monitoring of a patient, detailed information on the inhibition of viral replication at various levels of antibody dilution, and
directed neutralization assays against  HIV-specific antibody lineages.\\

\begin{center}{{\bf Competition between multiple antibody lineages}}\\  
\end{center}

B-cells in the adaptive immune system are associated with clonal lineages that originate from distinct ancestral naive cells, generated by germline rearrangements (VDJ recombination) and junctional diversification~\cite{Janeway:H7fnIHBf}. Multiple lineages may be stimulated within a germinal center, and also circulate to other germinal centers~\cite{Victora:2012gx}. Lineages compete for activation agents (e.g., helper T-cells) and interaction with a finite number of presented antigens~\cite{Victora:2012gx}. We extend our theoretical framework to study how multiple  lineages compete with each other and coevolve with viruses. This generalization allows us to  show that lineages with higher overall binding ability, higher fitness flux, and lower (absolute) transfer flux have a better chance of surviving. In particular, we show that an antibody repertoire fighting against a highly diversified viral population, e.g., during late stages of HIV infection, favors elicitation of broadly neutralizing antibodies compared to normal antibodies.

\begin{figure}
\begin{center}
\centering\includegraphics[width=\columnwidth]{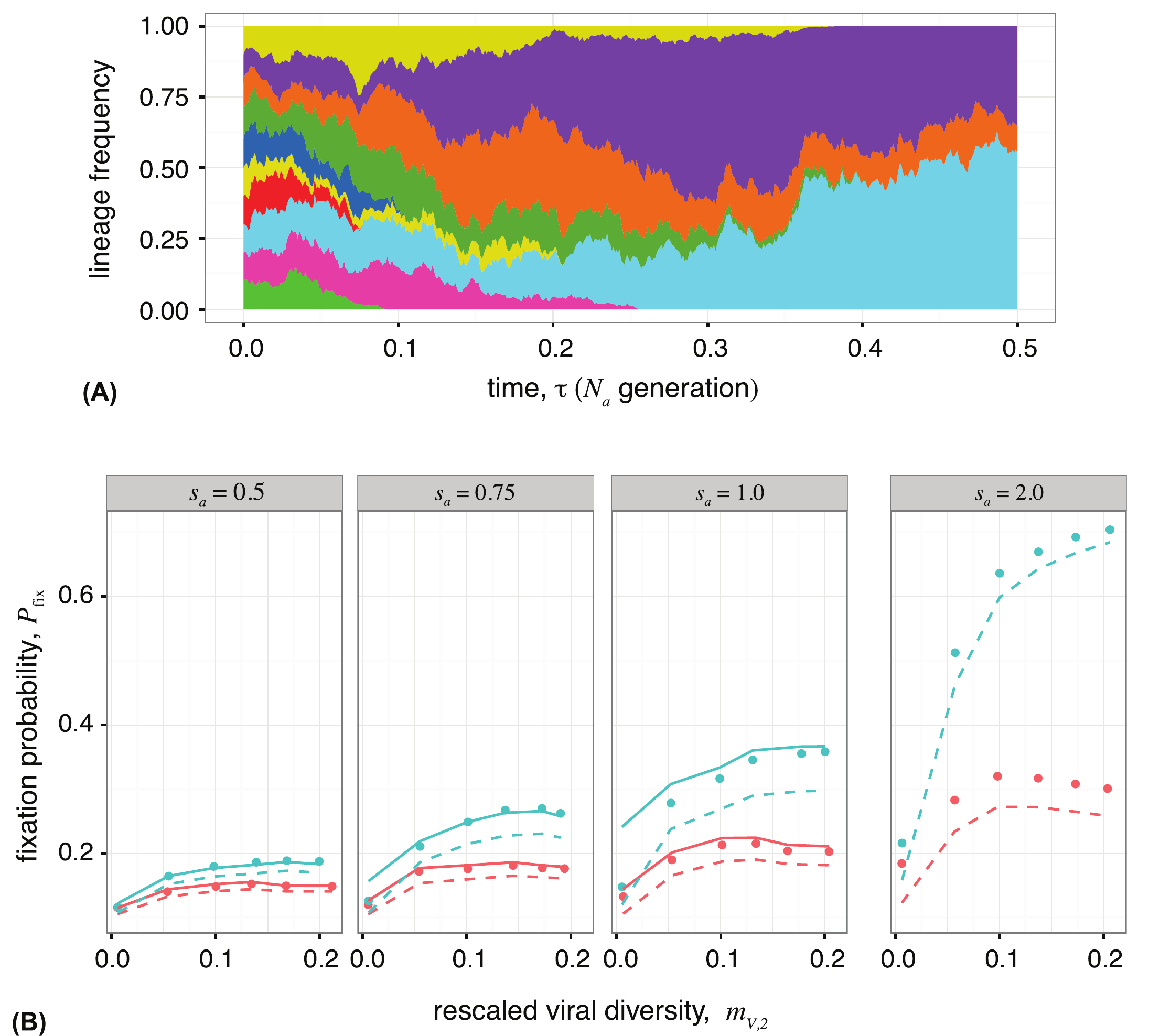}
\end{center}
\caption{{\bf Competition between antibody lineages, and fixation of broadly neutralizing antibodies.} {\bf (A)} Simulation of competition between 20 clonal antibody lineages against a viral population. Lineages with higher mean fitness, higher fitness flux, and lower transfer flux tend to dominate the antibody repertoire. Each color represents a distinct antibody lineage, however there is also diversity within each lineage from somatic hypermutations. The reduction in the number of
circulating lineages resembles the reduction in the number of active B-cell clones within the life-time of a germinal center~\cite{Victora:2012gx}. Lineages are initialized as 500 random sequences with random accessibilities $\kappa^\C$'s,
unique to each lineage, drawn from an exponential distribution with rate parameter $3$. Total population sizes are $N_a=N_v=10^4$. Other simulation parameters are as specified in the Appendix. {\bf (B)} Analytical estimates of the fixation probability $P_{\text{fix}}$ of a new antibody lineage, based on the state of the populations at the time of its introduction, compared to Wright-Fisher simulations (points) with two competing antibody lineages. A novel BnAb (blue) or non-BnAb (red) lineage is introduced at frequency 10\% into a non-BnAb resident population (simulation procedures described in the Appendix). BnAb lineages have a higher chance of fixing, compared to non-BnAb antibodies, when the viral population is diverse, whereas both types of Abs have similar chances in the limit of low viral diversity. The solid line is the analytical estimate for $P_{\text{fix}}$ given by eq.~(S140), which is valid when the rate of adaptation is similar in antibodies and viruses. The dashed line is the analytical estimate for $P_{\text{fix}}$ using the approximation in eq.~(S141), which is suitable when there is a strong imbalance between the two populations, as is the case for invasion of a BnAb with strong antibody selection $s_a>1$ or against a viral population with low diversity.  In the absence of selection (neutral regime), the fixation probability of an invading lineage is equal to its initial frequency of 10\%. Panels show different strengths of antibody selection $s_a=0.5,\,0.75,\, 1,\, 2$  against a common viral selection strength  $s_v=1$. Viral diversity is influenced mostly by the viral nucleotide diversity  $\theta_v$, which ranges from $0.002$ to $0.1$. Other simulation parameters are specified in the Appendix.}
\label{fig:pfix} 
\end{figure}

The binding preference of a clonal antibody lineage $\C$ to the viral sequence is determined by its site-specific accessibilities  {\small $\{\kappa_i^\C,\hat \kappa_i^\C\}$}, defined in Fig.~\ref{fig:schematic}B. The distribution of site-specific accessibilities over different antibody lineages {\small $ P_\C\big(\{\kappa_i^\C,\hat \kappa_i^\C\}\big) $}  characterizes the ability of an antibody repertoire to respond to a specific virus. Without continual introduction of new lineages, one lineage will ultimately dominate and the rest will go extinct within the coalescence time-scale of antibodies, $N_a$ (Fig.~\ref{fig:pfix}A). In reality, constant turn-over of lineages results in a highly diverse B-cell response, with multiple lineages acting simultaneously against an infection~\cite{Hoehn:2015kl}.

Stochastic effects are significant when the size of a lineage is small, so an important question is to find the probability that a low-frequency antibody lineage reaches an appreciable size and fixes in the population. We denote the frequency of an antibody lineage with size {\small $N_a^\C$} by  {\small $\rho^\C= N^\C_a/N_a$}. The  growth  rate of a given lineage $\C$ depends on its relative fitness {\small $F_{{}_{A^\C}}$} compared to the rest of the population, 

{ \EQ
\frac{d}{dt}  \rho^\C = (F_{{}_{A^\C}}-F_A) \rho^\C+\sqrt{\frac{\rho^\C(1- \rho^\C ) }{N_a}}\chi_{{}_\C}  \label{rhoC}
\EE}
where  {\small $F_{{}_A} =\sum_\C F_{{}_{A^\C}} \rho^\C$ }is the average fitness of the entire antibody population, and {\small $\chi_{{}_\C}$} is a standard  Gaussian noise term. For the linear fitness landscape from eq.~(\ref{ABfitness}), the mean fitness of lineage $\C$  is {\small $F_{{}_{A^\C}}=S_a(\E^\C+\hat \E^\C) $}. The probability of fixation of lineage $\C$ equals the asymptotic (i.e., long time) value of the ensemble-averaged lineage frequency, {\small $P_{\text{fix}}(\C)=\lim_{t\to\infty}   \langle \rho^\C(t)\rangle $}. 

Similar to evolution of a single lineage, the dynamics of a focal lineage are defined by an infinite hierarchy of moment equations for the fitness distribution. In the regime of substantial selection, and by neglecting terms due to mutation, a suitable truncation of the moment hierarchy allows us to estimate the long-time limit of the lineage frequency, and hence, its fixation probability  (see Section~E of the Appendix). For an arbitrary fitness function, fixation probability can be expressed in terms of the ensemble-averaged relative mean  fitness, fitness flux and transfer flux at the time of introduction of the focal lineage,

{\small \begin{align}
 \nonumber & P_{{}_\text{fix}} (\C) /P_{0_\text{fix}}\simeq 1+\left \langle N_a (F_{{}_{A^\C}}(0) -F_{{}_A (0)})  \right\rangle\\
& + \frac{N_a^2}{3} 
\left\langle \phi_{{}_{A^\C}} (0) -\phi_{{}_A}(0)\right \rangle 
-N_a N_v\left \langle \big| \T_{{}_{V\to A^\C}} (0)\big |- \big| \T_{{}_{V\to A}}(0)\big|\big] \right \rangle \label{fix_prob_general}
\end{align}}

\noindent where $P_{0_\text{fix}}$ is the fixation probability of the  lineage in neutrality, which equals its initial frequency at the time of introduction,  $P_{0_\text{fix}}=\rho^\C(0)$.  The first order term that determines the excess probability  for fixation of a  lineage is  the difference between its mean fitness and the average fitness of the whole population. Thus, a lineage with higher relative mean fitness at the time of introduction, e.g., due to its better accessibility to either the variable or conserved region, will have a higher chance of fixation.   Moreover, lineages with higher rate of adaptation, i.e., fitness flux $ \phi_{{}_{A^\C}}(t=0)$, and lower (absolute) transfer flux  from viruses $ \big| \T_{{}_{V\to A^\C}} (t=0)\big |$  tend to dominate the population. 

For evolution in the linear fitness landscape, we can calculate a more explicit expansion of the fixation probability that includes mutation effects. In this case, the fixation probability of a focal lineage can be expressed in terms of the experimentally observable  lineage-specific moments of the binding affinity distribution, instead of the moments of the fitness distribution (see Section~E of the Appendix).  

\vspace{1cm}
\begin{center}{{\bf Emergence of broadly neutralizing antibodies} }\\
\end{center}

With our multi-lineage model, we can understand the conditions for emergence of broadly neutralizing antibodies (BnAbs) in an antibody repertoire. Similar to any other lineage, the progenitor of a BnAb faces competition with other resident antibody lineages that may be dominating the population. The dominant term in the fixation probability is the relative fitness difference of the focal lineage to the total population at the time of introduction. Lineages may reach different fitnesses because they differ in their scale of interaction with the viruses, $E_0^\C$ in the variable region and $\hat E_0^\C$ in the conserved region; see Section~E of the Appendix for details. Lineages which bind primarily to the conserved region, i.e.,  $\hat E_0^\C \gg E_0^\C$,   are not vulnerable to viral escape mutations that reduce their binding affinity. Such BnAbs may be able to reach higher fitnesses compared to normal antibodies which bind to the variable region with a comparable scale of interaction. The difference in the mean fitness of the two lineages  becomes even stronger,  when viruses are more diverse (i.e., high $M_{V,2}$), so that they can strongly compromise the  affinity of the lineage that binds to  the variable region; see eq.~(\ref{fix_prob_general}). 

If the invading lineage has the same fitness as the resident lineage, then the second order terms in eq.~(\ref{fix_prob_general}) proportional to the fitness and transfer flux may be relevant.  A BnAb lineage that binds to the conserved region has a reduced transfer flux than a normal antibody lineage, all else being equal. The difference in transfer flux of the two lineages depends on the viral diversity $M_{V,2}$, and becomes more favorable for BnAbs when the viral diversity is high. Overall, a BnAb generating lineage has a higher advantage for fixation compared to normal antibodies, when the repertoire is coevolving against a highly diversified viral population, e.g., during late stages of HIV infection. 

In Fig.~\ref{fig:pfix}B we compare the fixation probability of  a BnAb lineage, that binds only to the conserved region, with a normal antibody lineage that binds only to the variable region. In both cases we assume that the  emerging lineage competes against a resident population of normal antibodies.  We compare our analytical predictions for fixation probability as a function of the initial state of the antibody and viral populations in eqs.~(S140, S141), with Wright-Fisher simulations of coevolving populations (numerical procedures detailed in Section G of the Appendix). Increasing viral diversity $M_{2,V}$ increases the fixation of BnAbs, but does not influence fixation of normal lineages. For low viral diversity,  fixation of BnAbs is similar to normal Abs, and therefore they might arise and be outcompeted by other antibody lineages.\\

\begin{center}{\Large \bf Discussion}\\
\end{center}

We have presented an analytical framework to describe coevolutionary dynamics between two antagonistic populations based on molecular interactions between them. We have focused our analysis on antibody-secreting B-cells and chronic infections, such as HIV.   We identified effective parameters for selection on B-cells during hypermutation that enhance their binding and neutralization efficacy, and conversely parameters for selection on viruses to escape antibody binding. The resulting ``red-queen" dynamics between antibodies and viruses produces a characteristic signature of coevolution in our model, i.e., viruses are resistant to antibodies from the past and are susceptible to antibodies from the future. We used our results to infer modes of immune-viral coevolution based on time-shifted neutralization measurements in two HIV-infected patients. Finally, we have shown that emergence and fixation of a given clonal antibody lineage is determined by competition between circulating antibody lineages, and that broadly neutralizing antibody lineages, in particular, are more likely to dominate in the context of a diverse viral population.

Luo and Perelson~\cite{Luo:2015gl} found that competition between lineages caused BnAbs to appear late in their simulations. In addition, they found that multiple viral founder strains dilutes the competition of BnAbs with specific antibodies, leading to higher chance of BnAb appearance. The assumptions of their simulations differ in many ways from those of our model, and yet their overall finding agrees with our analytical results: BnAbs fix more readily when there is a large diversity of viral binding. In contrast to Luo and Perelson's simulations which made assumptions about the immunogenicity of BnAbs, our analytic results show explicitly how differences in fitness of antibodies and the efficacy of viral escape affect competition between antibody lineages. 

Our model is simple enough to clarify some fundamental concepts of antibody-antigen dynamics.  However, understanding more refined aspects of B-cell-virus coevolution will require adding details specific to affinity maturation and viral reproduction, such as non-neutralizing binding between antibodies and antigens~\cite{Wyatt:1998vs,Luo:2015bi}, epitope masking by antibodies~\cite{Zhang:2013hk} and spatial structure of germinal centers~\cite{Victora:2012gx}. Importantly, viral recombination~\cite{Lemey:2006wb,Neher:2010dw,Zanini:2015vb} and latent viral reservoirs~\cite{Chun:1997iq} are also known to influence the evolution of HIV within a patient. Similarly, the repertoire of the memory B-cells and T-cells, which effectively keep a record of prior viral interactions, influence the response of the adaptive immune system against viruses with antigenic similarity.

While our analysis has focused on coevolution of chronic viruses with the immune system, our framework is general enough to apply to other systems, such as bacteria-phage coevolution. Likewise, the notions of fitness and transfer flux as measures of adaptation are independent of the underlying model. Bacteria-phage interactions have been studied by evolution experiments~\cite{Brockhurst:2013iq,Koskella:2014ds}, and by time-shifted assays of fitness~\cite{Hall:2011hh,Betts:2014ep}, but established models of coevolution typically describe only a small number of alleles with large selection effects~\cite{Agrawal:2002uj}. In contrast, our model offers a formalism for bacteria-phage coevolution where new genotypes are constantly produced by mutation, consistent with experimental observations~\cite{Hall:2011hh}. Similarly, our formalism may be applied to study the evolution of seasonal influenza virus in response to the ``global" immune challenge, imposed by a collective immune landscape of all recently infected or vaccinated individuals. Time-shifted binding assays of antibodies to influenza surface proteins are already used to gauge  the virulence and cross-reactivity of viruses~\cite{Fonville:2014fr}.  Quantifying the fitness flux and transfer flux, based on these assays, is therefore a principled way to measure rates of immunologically important adaptation in these systems.

One central challenge in HIV vaccine research is to devise a means to stimulate a lineage producing broadly neutralizing antibodies.  Common characteristics of BnAbs, such as high levels of somatic mutation or large insertions, often lead to their depletion by mechanisms of immune tolerance control~\cite{Verkoczy:2011gm}. Therefore, one strategy to elicit these antibodies is to stimulate the progenitors of their clonal lineage, which may be inferred by phylogenetic methods~\cite{Kepler:2014fs}, and to guide their affinity maturation process towards a broadly neutralizing state.  Understanding the underlying evolutionary process is necessary to make principled progress towards such strategies, and this study represents a step in that direction. For example, our results suggest that a vaccine based on a genetically diverse set of viral antigens is more likely to stimulate BnAbs.\\

\begin{center}{\large \bf Acknowledgments}\\
\end{center}

We acknowledge Francois Blanquart, Simon Frost, Mehran Kardar, Michael L\"assig and  Alan Perelson  for helpful discussions. AN thanks the Kavli Institute for Theoretical Physics (UCSB) for its hospitality, where part of this work was performed. We acknowledge funding from the US National Science Foundation grants PHY-1305525 and PHY11-25915, the James S. McDonnell Foundation 21st century science initiative-postdoctoral program in complexity science / complex systems,  the David and Lucile Packard Foundation, the U.S. Department of the Interior (D12AP00025), the U.S. Army Research Office (W911NF-12-1-0552). The funders had no role in study design, data collection and analysis, decision to publish, or preparation of the manuscript.

\begin{widetext}

\setcounter{figure}{0}
        \renewcommand{\thefigure}{S\arabic{figure}}
        \renewcommand{\thetable}{\arabic{table}}

\setcounter{equation}{0}
        \renewcommand{\theequation}{S\arabic{equation}}

\renewcommand \thesection{\Alph{section}}
\renewcommand \thesubsection{\Alph{section}.\arabic{subsection}}
\newpage{}

\noindent {\bf \Large Appendix}\\
\vspace{1cm}
\begin{center}
{\bf \Large A. Antibody-viral coevolution in genotype space}\\
\end{center}

We represent the antibody population as a set of $k$ genotypes consisting of vectors,  $\A^\alpha$  $(\alpha=1 \dots k)$, and corresponding genotype frequencies $\x$, with elements $x^\alpha$ satisfying $\sum_{\alpha=1}^k x^\alpha=1$.  Similarly, we consider a viral population  with $k'$ possible genotypes $\V^a$, and frequencies $\y$ with elements $y^\gamma$  $(\gamma=1, \dots, k')$ with $\sum_{\gamma=1}^{k'} y^\gamma=1$. Note that superscripts are indices, not exponentiation, unless next to parentheses, e.g. $(a)^b$. The frequencies change over time, although we omit explicit notation for brevity, and hence every quantity that depends on the frequencies is itself time-dependent.  In the following, we describe separately  contributions from three  evolutionary forces  (i) mutation, (ii) selection, and (iii) genetic drift, and  build a general stochastic framework for coevolution of antibodies and viruses in the space of genotypes. We assume that population sizes are large enough, and changes in frequencies are small enough to accommodate a continuous time and continuous frequency stochastic process~\cite{Gardiner:2004tx,Kimura:1964}.

\para{(i) Mutations.}
The change of the genotype frequencies due to mutations follow,
\begin{align}
\nonumber \frac{dx^\alpha}{dt} &= m_{{}_{A^\alpha}} (\x) \equiv \sum_{\beta =1}^k \mu_{{}_{A^\beta\rightarrow A^\alpha }} x^{\beta} -\left(\sum_{\beta =1}^k  \mu_{{}_{A^\alpha\rightarrow A^\beta }}\right) x^{\alpha}\\
\nonumber\frac{dy^\gamma}{dt} &= m_{{}_{V^\gamma}} (\y) \equiv \sum_{\lambda=1}^{k'} \mu_{{}_{V^\lambda\rightarrow V^\gamma}} y^\lambda -\left(\sum_{\lambda=1}^{k'}   \mu_{{}_{V^\gamma \rightarrow V^\lambda}} \right) y^\gamma\\
\label{mutation_field}
\end{align}
where we define $m_{{}_{A^\alpha}}$ and $ m_{{}_{V^\gamma}}$ as the genotype-specific components of the mutational fields in the antibodies and viruses, and $\mu_{{}_{A^\beta\rightarrow A^\alpha} }$ is the antibody mutation rate from genotype $\A^\beta$ to  $\A^\alpha$,  and similarly, $\mu_{{}_{V^\lambda \rightarrow V^\gamma}}$ is the viral mutation rate from the genotype $\V^\lambda$ to  $\V^\gamma$. We assume constant mutation rates $\mu_a$, $\mu_v$, per generation per site for antibodies and  viruses, with the exception of $\mu_v=0$ for the viral constant region, which implies that mutations in that region are lethal for the virus.  

\para{(ii) Selection and interacting fitness functions.}
The fitness of a genotype determines its growth rate at each point in time. We define fitness of genotypes in one population as a function of the genotypes in the other population. The general form of change in genotype frequencies due to selection follows, 
\EQA
\nonumber \frac{1}{x^\alpha} \frac{dx^{\alpha} }{dt} &=&  f_{{}_{A^\alpha} } (\x;\y) -\sum_{\alpha} x^{\alpha}  f_{{}_{A^\alpha} }(\x;\y)\\
\nonumber \frac{1}{y^\gamma} \frac{dy^\gamma }{dt} &=& f_{{}_{V^\gamma} } (\y;\x) -\sum_{\gamma} y^\gamma f_{{}_{V^\gamma} } (\y;\x)\\
\label{selection_fields_gen}
\EEA

The subscript for the antibody and viral fitness functions, $ f_{{}_{A^\alpha} }(\x;\y)$ and $ f_{{}_{V^\gamma} }(\y;\x)$, refer to the  genotypes in the corresponding population. The explicit  conditional dependence of the antibody fitness function $  f_{{}_{A^\alpha} } (\x;\y)$ on the viral frequency vector $\y$ emphasizes that fitness of an antibody depends on the interacting viral population $\{\V\}$. Similar notation is used for the fitness function of the viruses.  The subtraction of the population's mean fitness, $F_{{}_A} =  \sum_\alpha x^\alpha  f_{{}_{A^\alpha} } (\x;\y)$ and $F_{{}_V}=   \sum_\gamma y^\gamma f_{{}_{V^\gamma} } (\y;\x)$, ensures that the genotype frequencies remain normalized in each population.  In terms of linearly independent frequencies $\x =(x^1, \dots, x^{k-1})$ and $\y=(y^1, \dots, y^{k'-1})$, these evolution equations take the forms,
\begin{align}
\nonumber \frac{dx^\alpha}{dt} =  \sum_{\text{antibodies: } \beta} g^{\alpha\beta}\, \sigma_{{}_{A^\beta}}(\x;\y),\qquad \qquad \frac{dy^\gamma}{dt} =  \sum_{\text{viruses: } \lambda} h^{\gamma\lambda} \, \sigma_{{}_{V^\lambda}}(\y;\x)\\
\label{gen-sel}
\end{align}
where $\sigma_{{}_{A^\alpha}} (\x;\y) = f_{{}_{A^\alpha} }(\x;\y)-f_{A^k}(\x;\y) $ and $\sigma_{{}_{V^\gamma}}(\y;\x)= f_{{}_{V^\gamma} }(\y;\x)-f_{V^{k'}}(\y;\x)$ are the respective time-dependent selection coefficients of the antibody $\A^\alpha$ and the viral strain $\V^\gamma$, which depend on the state of the both populations at that moment in time. The inverse of the response matrices, $g_{\alpha\beta} =(g^{\alpha\beta})^{-1}$ and $h_{\gamma\lambda} =(h^{\gamma\lambda})^{-1}$, play the role of metric in the genotype space (see below and e.g.,~\cite{Antonelli:1977ek}). The change in the mean fitness due to selection after an infinitesimal amount of time follows, 
\begin{align}
 F_{{}_A}(\x+\delta \x;\y +\delta \y) &= \sum_\alpha\sigma_{{}_{A^\alpha}} (\x; \y)  \delta x^\alpha +  \sum_{\gamma, \alpha} x^\alpha \, \sigma_{{}_{V^\gamma \rightarrow A^\alpha}}(\x;\y)\, \delta y^\gamma\\
 F_{{}_V}(\y+\delta \y;\x+\delta \x) &= \sum_\gamma \sigma_{{}_{V^\gamma}}( \y;\x)  \delta y^\gamma + \sum_{\gamma, \alpha}  y^\gamma\,  \sigma_{{}_{A^\alpha \rightarrow V^\gamma}} (\y;\x)\, \delta x^\alpha
 \end{align}
where $\delta x^\alpha$  and $\delta y^\gamma$ are the infinitesimal  changes in the genotype frequencies, and $\sigma_{{}_{V^\gamma\rightarrow A^\alpha}}=\partial \sigma_{{}_{A^\alpha}} /\partial y^\gamma $ and, $ \sigma_{{}_{A^\alpha \rightarrow V^\gamma} } =\partial \sigma_{{}_{V^\gamma} }/\partial x^\alpha$  are respectively the change in the selection coefficient of the antibody $\A^\alpha$ and the virus $\V^\gamma$  due the evolution of opposing population. This measure of  fitness transfer is a useful concept for interacting populations. Intuitively, it can be understood as  the change of fitness in one population only due to the changes of allele or genotype frequencies in the opposing population.
 
\para{(iii) Genetic drift and stochasticity.} The stochasticity of reproduction and survival, commonly known as genetic drift, is represented as discrete random sampling of offspring genotypes from the parent's generation with the constraint that the total population size remains constant. The magnitude of this sampling noise is proportional to inverse population size. $N_a $ and $N_v$ are the effective population sizes of the antibody  and the viral populations, which represent the size of  population bottlenecks e.g.,  in a germinal center. In the continuous time, continuous frequency limit, genetic drift is represented as noise terms in a diffusion equaiton with magnitude proportional to inverse population size~\cite{Kimura:1964}. The diffusion coefficients are characteristics of the Fisher metric~\cite{Fisher:1930wf,Antonelli:1977ek},
\EQA
\nonumber g^{\alpha\beta} =  \begin{cases} 
 -x^\alpha x^{\beta} & \text{if } \alpha\neq \beta\\  
x^\alpha (1- x^\alpha) & \text{if } \alpha=\beta 
\label{gen_sampling_vir}
\end{cases}, \qquad\qquad h^{\gamma\lambda}  =
 \begin{cases} 
 -y^{\gamma} y^{\lambda} & \text{if } \gamma\neq \lambda\\  
y^{\gamma} (1- y^{\gamma}) & \text{if } \gamma=\lambda
\end{cases}\\
\label{gen_sampling}
\EEA

The generalized Kimura's diffusion equation~\cite{Kimura:1983vr} for the joint distribution of genotype frequencies $P(\x, \y,t)$ in both populations reads,
\begin{align}
\nonumber\frac{\partial}{\partial t} P(\x, \y, t) &=\nonumber \sum_{ \alpha, \beta, \gamma,\lambda} \Big[ \frac{1}{2 N_a } \frac{\partial^2}{\partial x^{\alpha} x^{\beta}} g^{\alpha\beta} (\x) +
\frac{1}{2N_v} \frac{\partial^2}{\partial y^{\gamma} \partial y^{\lambda}} h^{\gamma\lambda} (\y)\\
\nonumber&\,\,\,\,+\frac{\partial}{\partial x^{\alpha}} \big(m_{{}_A}^{\alpha} (\x)+g^{\alpha\beta}(\x)\, {\sigma_{{}_{A^\beta}}}(\x;\y)\big)+\frac{\partial}{\partial y^{\gamma}} \big(m_{{}_V}^{\gamma}(\y)+h^{\gamma\lambda} (\y){ \sigma_{{}_{V^\lambda}}}(\y;\x)\big)\Big] P(\x, \y, t)\\
\label{FP-gen}
\end{align}

This Fokker-Planck equation acts on the high-dimensional genotype space of antibodies and viruses, which are likely to be under-sampled in any biological setting. In the following, we introduce a projection from  genotype space onto a lower dimensional space of molecular traits (phenotypes) to make the problem tractable.\\

\begin{center} 
{\Large \bf B. Antibody-viral coevolution in phenotype space}\\
\end{center}

{\noindent \bf \large B.1 Molecular phenotypes for antibody-viral interaction}\\

We define the binding affinity  between an antibody and viral genotype as the molecular interaction phenotype under selection, for which we  describe the evolutionary dynamics.   Antibody and viral genotypes are represented by binary sequences of  $\pm1$. Antibody sequences are of length $\ell+\hat \ell$, while viral sequences consist of a mutable region of length $\ell$, and a conserved (i.e. sensitive) region of length $\hat \ell$, where each site is always +1, as was similarly done in \cite{Wang:2015em}. We model the binding affinity between antibody $\A^\alpha$ and virus $\V^\gamma$ as a weighted dot product over all sites,

\begin{align}
\nonumber E_{\text{tot}}(\A^\alpha, \V^\gamma) &= { \sum_{i=1}^{\ell} \kappa_{i} A^\alpha_i V^\gamma_i }+{\sum_{i=\ell+1}^{\ell+\hat \ell} \hat\kappa_{i}\,A^\alpha_i} \\
&\equiv E_{\alpha\, \gamma} +\hat E_\alpha
\end{align}
where $A^\alpha_i$, and $V^\gamma_i$  denote the $i^{th}$ site in antibody $\A^\alpha$ and virus $\V^\gamma$, respectively.  The set of coupling constants for the mutable and conserved region, $\{\kappa_{i}, \hat \kappa_{i} \geq 0\}$ represent the accessibility of a clonal antibody lineage to regions of the viral sequence. Matching bits at interacting positions enhance binding affinity between an antibody and a virus. Similar models have been used to describe  B-cell maturation in germinal centers~\cite{Wang:2015em}, and T-cell selection based on the capability to bind external antigens and avoid self proteins~\cite{Detours:1999uq,Detours:2000wz}.  In Section~\ref{sec:mult_lin}, we extend our model to multiple lineages, where each lineage has its own set of accessibilities. Antibody lineages with access to the conserved regions of the virus can potentially fix as broadly neutralizing antibodies. We denote the quantities related to the conserved sites of the virus with a hat: $\hat{\cdot}$.

We project the evolutionary forces acting on the genotype to the binding phenotype $E_\text{tot}$, and quantify the changes of the binding phenotype distribution in each population over time. For a single antibody genotype $\A^\alpha$ we characterize its interactions with the viral population by the {\em genotype-specific moments}, \\

\qquad \textbf{Statistics of the binding affinity distribution for antibody $\A^\alpha$:}
\begin{align}
\nonumber&  \text{(i) average in the variable region:}\\
 &\qquad\qquad\qquad E_{\alpha\, . } = \sum_{\gamma\in\text{ viruses}}  E_{\alpha\, \gamma} y^\gamma \label{AB_mean_mom} \\
 \nonumber& \text{(ii) average in the conserved region:}\\
 &\qquad\qquad\qquad \hat E_{\alpha\, . } = \hat E_\alpha\\
 \nonumber& \text{(iii) $ r^{th}$ central moment  in the variable region: } \\
&\qquad\qquad\qquad I_{\alpha\, . }^{(r)} = \sum_{\gamma\in\text{ viruses}}   (E_{\alpha\, \gamma} - E_{\alpha\, . } )^r y^\gamma
\label{AB_var_mom}
\end{align}

Since the viral population is monomorphic in the conserved region, the average mean binding affinity of an antibody is independent of the state of the viral population,  $\hat E_{\alpha\, . } = \hat E_\alpha$, and the higher central moments are zero, $\hat I_{\alpha\, . }^{(r)}=0$. Similarly,  we characterize the interactions of a given viral genotype $\V^\gamma$ with the antibody population,\\

 \qquad \textbf{Statistics of the binding affinity distribution for virus $\V^\gamma$:}
\begin{align}
\nonumber&  \text{(i) average in the variable region:}\\
&\qquad\qquad\qquad  E_{ .\, \gamma } = \sum_{\alpha\in \text{antibodies}} E_{\alpha\, \gamma}  x^\alpha\\
 \nonumber& \text{(ii) average in the conserved region:}\\
&\qquad\qquad\qquad   \hat E_{ .\, \gamma } = \sum_{\alpha\in \text{antibodies}} \hat E_{\alpha}  x^\alpha \equiv \hat E_{.}\\
 \nonumber& \text{(iii) $ r^{th}$ central moment  in the variable region: } \\
&\qquad\qquad\qquad  I_{ .\, \gamma }^{(r)} = \sum_{\alpha\in \text{antibodies}} (E_{\alpha\, \gamma}  - E_{ .\, \gamma } )^r x^\alpha\\
 \nonumber& \text{(iii) $ r^{th}$ central moment  in the conserved region: } \\
&\qquad\qquad\qquad \hat I_{ .\, \gamma }^{(r)} = \sum_{\alpha\in \text{antibodies}} (\hat E_{\alpha} - \hat \E  )^r x^\alpha
\end{align}

One of the most informative statistics that we characterize is the distribution of population-averaged antibody and viral binding interactions, respectively denoted by $P_A(  E_{\alpha\, . } , \hat E_\alpha)$ and $P_V(  E_{ .\, \gamma } , \hat E_.)$. The mean of these distributions are equal to each other, but the higher moments differ. We denote the {\em population-specific moments} of the average interactions by, \\

\qquad\textbf{Mean binding affinity in,}
\begin{align}
&\text{(i) the variable region:}\qquad   \E= \sum_{\alpha, \gamma} E_{\alpha\, \gamma} \, x^\alpha y^\gamma  \\
&\text{(ii) the conserved region:}\qquad \hat \E= \sum_{\alpha} \hat E_{\alpha} \, x^\alpha  \\\nonumber
\end{align}
\qquad\textbf{$r^{th}$ central moment of the average affinities in,} 
\begin{align}
\nonumber&  \qquad \text {(i) the variable region of antibody population: }\\
 &\qquad\qquad  M_{A,r} = \sum_{\alpha\in \text{antibodies}} (E_{\alpha\, . } -\E )^r x^\alpha\\
\nonumber&\qquad\text{(ii) the conserved region of antibody population:} \\
&\qquad\qquad  \hat M_{A,r} = \sum_{\alpha\in \text{antibodies}}(\hat E_{\alpha\, . } -\hat \E )^r x^\alpha\\
\nonumber&\qquad\text{(iii) the variable region of viral population:} \\
&\qquad\qquad  M_{V,r} = \sum_{\gamma\in \text{viruses}} (E_{ .\, \gamma } -\E)^r y^\gamma
\end{align}

Note that the population central moments  $M_{A,r}$ and $M_{V,r}$ are distinct from the genotype-specific moments,  $I_{\alpha\, . }^{(r)}$ and $ I_{ .\, \gamma }^{(r)}$. The  central moments of the viral population in the conserved region  are equal to zero, $\hat M_{V,r}=0$.

\para{Trait scale and dimensionless quantities.} 
It is useful to measure phenotypes in natural units, which avoids the arbitrariness of the physical units  ($\{\kappa_i,\, \hat \kappa_i\}$), and the total number of sites $\ell+\hat\ell$. As previously shown in~\cite{Nourmohammad:2013ty,NourMohammad:2013in}, there exist summary statistics of the site-specific effects, (here $\{\kappa_i,\hat\kappa_i\}$), which define a natural scale of the molecular phenotype. We  denote the moments of the site-specific effects along the genome by,

\EQ
\k_r=\frac{1}{\ell} \sum_{i=1}^{\ell}( \kappa_i)^r  , \qquad \hat \k_r= \frac{1}{\hat \ell} \sum_{i=\ell+1}^{\ell+\hat \ell}   (\hat\kappa_i)^r
\label{K_r}
\EE

We express the phenotype statistics in units of the trait scales, i.e., the squared sum of the site-specific effects, $E_0^2 = \k_2\ell$ in the variable region, and $\hat E_0^2=\hat\k_2\hat \ell$ in the conserved region. The rescaled phenotype statistics follow,

\begin{align} 
&\varepsilon  \equiv \frac{\E}{E_0}, \qquad \hat \varepsilon \equiv \frac{\E}{\hat E_0}\qquad\qquad{\text{and}}, \qquad m_{Z,r} \equiv \frac{M_{Z,r}}{E_0^r }, \qquad  \hat m_{Z,r} \equiv \frac{\hat M_{Z,r} }{\hat E_0^r}\qquad \text{(for } Z=A,V \text{)}
\label{scaling}
\end{align}

These scaled values are pure numbers (we distinguish them by use of lower case letters from the raw data). The trait scales $E_0^2$ and $\hat E_0^2$ provide natural means to standardize the relevant quantities because they are the stationary ensemble variances of the population mean binding affinity in an ensemble of genotypes undergoing neutral evolution in the weak-mutation regime (see Section B.3 for derivation of the stationary statistics),
\EQ
E_0^2= \lim_{\mu_v, \mu_a  \to 0}  \langle (\E-\langle\E\rangle)^2\rangle, \quad \quad\hat E_0^2 = \lim_{\mu_a  \to 0}  \langle (\hat \E-\langle\hat\E\rangle)^2\rangle
\EE
where $\langle \cdot\rangle$ indicates  averages over an ensemble of  independent populations.\\

\para{Binding probability.} 
The probability that an antibody is bound by an antigen determines its chance of proliferation and survival during the process of affinity maturation, and hence, defines its fitness. We describe two distinct models for antibody activation. The simplest model assumes that the binding probability of a given antibody $\A^\alpha$ is a sigmoid function of its \emph{mean binding affinity} against the viral population, 

\EQ
p_{{}_A} (\A^\alpha) = \frac{1}{1+\exp [-\beta_0 (E_{\alpha\, . }+\hat E_{\alpha\, . }-E^*)]}
\label{prob_bound_mean}
\EE
where $E^*$ is the threshold for the binding affinity and $\beta_0$ determines the amount of nonlinearity, and is related to the inverse of temperature in thermodynamics.  Following the rescaling introduced in  eq.~(\ref{scaling}), the binding threshold and the nonlinearity in eq.~(\ref{prob_bound_mean}) rescale as $e^* \equiv E^* /\sqrt{\hat E_0^2+E_0^2}$ and $\beta= \beta_0 \sqrt{\hat E_0^2+E_0^2}$.  In the following, we  will use  eq.~(\ref{prob_bound_mean}) to characterize  a biophysically grounded  fitness function for antibodies. \\

For the virus, binding to an antibody reduces the chances of its survival. Similar to eq.~(\ref{prob_bound_mean}), the probability that a given virus $\V^\gamma$ is bound by antibodies follows,

\EQ
p_{{}_V}(\V^\gamma) = \frac{1}{1+\exp [-\beta_0 (E_{ .\, \gamma }+\hat E_{ .\, \gamma }-E^*)]},
\label{prob_bound_v_mean}
\EE
where $E^*$ and $\beta_0$ are similar to eq.~(\ref{prob_bound_mean}).\\

In Section B.5, we will discuss an alternative model for activation of an antibody which is based on its  \emph{strongest binding affinity} with a subset of viruses.\\

\noindent{\bf \large B.2 Coevolutionary forces on the binding affinity}\\

Similar to genotype evolution, stochastic evolution of a molecular phenotype generates a probability distribution, $Q(\E,\hat \E,M_{A,r},\hat M_{A,r},M_{V,r} )$, which describes an ensemble of independently evolving populations, each having a phenotype distribution with  mean affinity  $\E$ and $\hat\E$ and central moments of the averaged affinity in the antibody population, $M_{A,r} $, $\hat M_{A,r}$, and in the viral population, $M_{V,r}$ (see also~\cite{Nourmohammad:2013ty}). The probability distribution $Q(\E,\hat\E ,M_{A,r}, \hat M_{A,r} ,M_{V,r})$ can be expressed in therms of the distribution for genotype frequencies, 
\begin{align}
\nonumber &Q(\E,\hat\E,M_{A,r}, \hat M_{A,r}  ,M_{V,r})=\\
\nonumber&\int  d\x d\y\,P(\x, \y,t)\,  \left[\delta (\E(\x, \y)-\E) \, \delta (\hat\E(\x)-\hat\E) \prod_r \delta(  M_{A,r}(\x, \y)-M_{A,r}) \,\delta(  \hat M_{A,r}(\x)-\hat M_{A,r}) \, \delta(  M_{V,r}(\x, \y)-M_{V,r}) \right]\\
\end{align} 
where $\delta(\cdot)$  is the Dirac delta function.  Below, we characterize the effect of mutations, selection and genetic drift on the evolution of the phenotype moments $\E$, $M_{A,r}$, $\hat M_{A,r}$ and $M_{V,r}$. 

\para{Mutation.} A mutation at site $``i"$ changes the sign of the site, and its effect on the binding affinity is proportional to $\kappa_i$ in the variable region, and $\hat \kappa_i$ in the conserved region. To compute the effect of mutations on  moments of the phenotype distribution, we classify pairs of genotypes $(\A^\alpha, \V^\gamma)$ in mutational classes, defined by the number of $+1$ positions of their product vector  ($A_1^\alpha\cdot V_1^\gamma, \dots, A_{\ell+\hat\ell}^\alpha\, \cdot V_{\ell+\hat\ell}^\gamma$), which we denote by  $n_+$ in the variable region and  by $\hat n_+$ in the conserved interaction region,
\EQ
n_+(\A^\alpha, \V^\gamma)=\sum_{i=1}^{\ell} \delta(1-  A_i^\alpha\cdot V_i^\gamma), \qquad \hat n_+(\A^\alpha, \V^\gamma)=\sum_{i=\ell+1}^{\ell+\hat\ell} \delta(1-  A_i^\alpha\cdot V_i^\gamma)
\EE
The frequency of each mutational class $\q(n_+)$ is estimated from interactions between all pairs of antibody and viral genotypes  in both variable and conserved regions of the interacting  populations,
\EQ
\q^{(1)}(n_+)=\frac{1}{N_a N_V} \sum_{\alpha, \gamma} \delta (n_+(\A^\alpha, \V^\gamma) -n_+), \qquad \q^{(2)}(n_+)=\frac{1}{N_a N_V} \sum_{\alpha, \gamma} \delta (\hat n_+(\A^\alpha, \V^\gamma) -\hat n_+)
\EE

The superscript $\lambda=1,2$  indicates the interacting region of the virus, i.e. \,$\lambda=1$ refers to the variable region of the virus with $\mu_v^{(1)}=\mu_v$ and the length $\ell^{(1)}=\ell$, and $\lambda=2$ refers to the conserved region of the viral genome with $\mu_v^{(2)}=0$ and the sequence length $\ell^{(2)}=\hat \ell$. If the mutational effects of all sites were equal to $\kappa$, phenotype moments could be simply expressed using the statistics of mutational classes: e.g., $\E= (2 [n_+]_{{}_{A,V}}-\ell)\kappa$,  where $[\cdot]_{{}_{A,V}}$ indicates averaging of a quantity in  the subscript populations, which in this case are both the  viral and the antibody populations. If the number of encoding sites of a phenotype is large, {\em annealed averages} of the  heterogeneous site-specific  contributions $\k_r,\,\hat\k_r$ can well approximate the the moments of the phenotype distribution~\cite{Higgs:1995dt,Good:2013km,Nourmohammad:2013ty}. As a result, the statistics of the variable region follow, $\E= (2 [n_+]_{{}_{A,V}}-\ell)\k_1$ for the mean binding affinity, and $M_{V,r} = 2^r \k_r  \Big [\big([n_+]_{{}_{A}}-[n_+]_{{}_{A,V}}\big)^r\Big]_{{}_{V}}$,  $M_{A,r} = 2^r \k_r  \Big [\big( [n_+]_{{}_{V}}-[n_+]_{{}_{A,V}}\big)^r\Big]_{{}_{A}}$ for the higher central moments in viruses and antibodies. Similar expressions can be derived for the statistics of the conserved region. Therefore,  evolution of the phenotype distribution can be well-approximated using projections from evolutionary dynamics of the  mutational classes. The Master equation for the evolution of the  mutational classes  under neutrality (mutation and genetic drift) follows, 

\EQA
\nonumber d\, \q^{(\lambda)}(n_+) &=&(\mu_a+\mu_v^{(\lambda)} )\Big[(\ell^{(\lambda)}-(n_+-1)) \q^{(\lambda)}(n_+-1) +(n_++1) \q^{(\lambda)}(n_++1) -\ell^{(\lambda)} \, \q^{(\lambda)}(n_+)\Big] \,dt\\
&&+ \big(\delta_{n_+',n_+} -\q^{(\lambda)}(n_+)\big) \left[ \sqrt{\frac{\q^{(\lambda)}(n_+)}{N_a }} dW_{{}_{A}}(t) +  \sqrt{\frac{\q^{(\lambda)}(n_+)}{N_v}}  dW_{{}_{V}}(t) \right]
\EEA

$W_{{}_{A}}(t)$ and $W_{{}_{V}}(t)$ are  delta-correlated Gaussian noise (Wiener process) with an ensemble mean $\langle W_i \rangle =0$  and variance, $\langle W_i (t) W_j (t')\rangle =\delta_{i,j}\, \delta (t-t')$ where $i,j\in \{A,V\}$ indicate antibodies and viruses. The stochasticity (genetic drift) is  due to finite population size of the interacting genotypes $N_a $  and $N_v$.

In neutrality, the ensemble mean for the  averaged number of positive sites $\big \langle [{n_+}^{(\lambda)}]_{{}_{A,V}}\big \rangle$ and the central moments, $\big \langle Y_{A,r}^{(\lambda)}\big \rangle \equiv \Big\langle\big[ \big( [n_+^{(\lambda)}]_{{}_{V}}-[ n_+^{(\lambda)}]_{{}_{A,V}}\big)^r\big]_{{}_{A}} \Big\rangle$ and 
$\langle Y_{V,r}^{(\lambda)}\rangle \equiv \Big\langle\big[ \big( [n_+^{(\lambda)}]_{{}_{A}}-[ n_+^{(\lambda)}]_{{}_{A,V}}\big)^r\big]_{{}_{V}} \Big\rangle$ in both variable ($\lambda=1$) and conserved ($\lambda=2$)  interaction  regions follow \cite{Higgs:1995dt,Good:2013km}, 

\EQA
\nonumber \frac{\partial \langle [{n_+}^{(\lambda)}]_{{}_{A,V}} \rangle}{\partial t} &=&\left \langle(\mu_a+  \mu_v^{(\lambda)}) \sum_{n_+} n_+  \Big[ (\ell^{(\lambda)} -n_++1) \q^{(\lambda)}(n_+-1) +(n_++1) \q^{(\lambda)} (n_++1)-\ell^{(\lambda)} \q^{(\lambda)} (n_+)\Big]\right \rangle\\
&=& \begin{cases}
-2(\mu_a+\mu_v)\Big[(\left \langle [{n_+}]_{{}_{A,V}} \right \rangle -\ell/2\Big) & \text{variable region}, \lambda=1\\
\\
-2\mu_a\Big(\left \langle [{n_+}]_{{}_{A,V}} \right \rangle -\hat \ell/2\Big)& \text{constant region}, \lambda=2\\
\end{cases}
\EEA

\EQA
\nonumber \frac{\partial  }{\partial t}\left \langle Y^{(\lambda)}_{A,r}\right \rangle &=& \mu_a \ell^{(\lambda)} \sum_{i=0}^{r-2} { r \choose i}\left \langle Y^{(\lambda)}_{A,i}\right \rangle +\frac{{r \choose 2} \left \langle Y^{(\lambda)}_{A,2} Y^{(\lambda)}_{A,r-2}\right \rangle -r \left \langle Y^{(\lambda)}_{A,r}\right \rangle }{N_a }- 2 r( \mu_a +\mu_v^{(\lambda)}) \left \langle Y^{(\lambda)}_{A,r}\right \rangle\\
&&-\mu_a  \sum_{i=0}^{r-2} {r\choose i} \left  ( \left \langle Y^{(\lambda)}_{A,i+1}\right \rangle+\left  \langle [n_+]_{{}_{A,V}} Y^{(\lambda)}_{A,i}\right \rangle \right) \, [ 1+(-1)^{r-i+1}]\label{YAR}
\EEA

\EQA
\nonumber \frac{\partial }{\partial t}\left  \langle Y^{(\lambda)}_{V,r}\right \rangle&=& \mu_v^{(\lambda)} \ell^{(\lambda)} \sum_{i=0}^{r-2} { r \choose i}\left \langle Y^{(\lambda)}_{V,i}\right \rangle +\frac{{r \choose 2}\left  \langle Y^{(\lambda)}_{V,2} Y^{(\lambda)}_{V,r-2}\right  \rangle -r \left \langle Y^{(\lambda)}_{V,r} \right\rangle }{N_v}- 2 r( \mu_a +\mu_v^{(\lambda)}) \langle Y^{(\lambda)}_{V,r}\rangle\\
&&-\mu_v^{(\lambda)}  \sum_{i=0}^{r-2} {r\choose i} \left( \left \langle Y^{(\lambda)}_{V,i+1}\right \rangle+ \left \langle [n_+]_{{}_{A,V}} Y^{(\lambda)}_{V,i}\right \rangle \right) \, [ 1+(-1)^{r-i+1}]\label{YVR}
\EEA
where $\langle \cdot\rangle$ denotes averages over independent ensembles of populations.  The second term in the right-hand side of equations (\ref{YAR}, \ref{YVR})  is a consequence of the It\^o calculus in stochastic processes~\cite{Gardiner:2004tx}. The transformations from $[n_+^{(1)}]_{{}_{A,V}}$ to $\E$ in the variable region, and from $[n_+^{(2)}]_{{}_{A,V}}$ to $\hat \E$ in the conserved region result in,

\begin{align}
& \frac{\partial \langle \E\rangle}{\partial t}= 
-2(\mu_a+\mu_v)\langle  \E \rangle
\\
& \frac{\partial \langle\hat \E \rangle}{\partial t} = -2\mu_a\langle\hat \E  \rangle\\\nonumber
\end{align}

The transformations from $Y_{A,r}^{(1)}$ to $M_{A,r}$, from $Y_{A,r}^{(2)}$ to $\hat M_{A,r}$ and from $Y_{V,r}^{(1)}$ to $M_{V,r}$ result in, 

\EQA
\nonumber \frac{\partial \langle M_{A,r}\rangle }{\partial t}&=& \mu_a \ell \sum_{i=0}^{r-2} 2^{r-i} \k_{r-i} { r \choose i}\langle M_{A,i}\rangle +\frac{{r \choose 2}  \k_2\, \k_{r-2} \langle M_{A,2} M_{A,r-2}\rangle -r \langle M_{A,r}\rangle }{N_a }- 2 r( \mu_a +\mu_v) \langle M_{A,r}\rangle\\
\nonumber &&-\mu_a  \sum_{i=0}^{r-2} 2^{r - i-1}\, {r\choose i} \Big[ {\k_{r-i-1}}   \langle M_{A,i+1}\rangle+  \frac{\k_{r-i}}{\k_1}\big \langle \E  M_{A,i}\big\rangle\Big] \, [ 1+(-1)^{r-i+1}]\\
\\
\nonumber \frac{\partial \langle \hat M_{A,r}\rangle }{\partial t}&=& \mu_a \hat \ell \sum_{i=0}^{r-2} 2^{r-i} \k_{r-i} { r \choose i}\langle \hat M_{A,i}\rangle +\frac{{r \choose 2}  \k_2\, \k_{r-2} \langle \hat M_{A,2}\hat M_{A,r-2}\rangle -r \langle\hat M_{A,r}\rangle }{N_a }- 2 r \mu_a  \langle\hat M_{A,r}\rangle\\
\nonumber &&-\mu_a  \sum_{i=0}^{r-2} 2^{r - i-1}\, {r\choose i} \Big[ {\k_{r-i-1}}   \langle\hat M_{A,i+1}\rangle+  \frac{\k_{r-i}}{\k_1}\big \langle\hat \E  \hat M_{A,i}\big\rangle\Big] \, [ 1+(-1)^{r-i+1}]\\
\\
\nonumber \frac{\partial \langle M_{V,r}\rangle }{\partial t}&=& \mu_v \ell \sum_{i=0}^{r-2} 2^{r-i} \k_{r-i} { r \choose i}\langle M_{V,i}\rangle +\frac{{r \choose 2}  \k_2\, \k_{r-2} \langle M_{V,2} M_{V,r-2}\rangle -r \langle M_{V,r}\rangle }{N_v}- 2 r( \mu_v+\mu_a) \langle M_{V,r}\rangle\\
\nonumber &&-\mu_v \sum_{i=0}^{r-2} 2^{r - i-1}\, {r\choose i} \Big[ {\k_{r-i-1}}   \langle M_{V,i+1}\rangle+  \frac{\k_{r-i}}{\k_1}\big \langle \E  M_{V,i}\big\rangle\Big] \, [ 1+(-1)^{r-i+1}]\\
\EEA

\para{Selection.}   We assume that (malthusian) fitness of an antibody is proportional to the logarithm of its activation probability given by eq.~(\ref{prob_bound_mean})  based on its average interaction strength,

\EQA
\label{non-liN_a b_av}f_{{}_{A^\alpha}}\equiv f_{{}_A}(\A^\alpha;\{ V\})&=&  c_a \log[p_{{}_A} (\A^\alpha)] = - c_a \log(1+\exp [-\beta_0 (E_{\alpha\, . }+\hat E_{\alpha\, . } -E^*)])\\
\label{liN_a b_av}&\simeq&  f_{{}_A}^*+ S_a  ( E_{\alpha\, . } +\hat E_{\alpha\, . } )
\EEA
with $f^*_{{}_A}= - c_a \log\big(1+\exp[\beta_0 E^*]\big)$ and the selection coefficient $S_a = c_a \beta_0 /(1+\exp[-\beta_0 E^*])$. The approximation in (\ref{liN_a b_av}) is by expansion of the nonlinear fitness function around the neutral binding affinity, $\E=0$. The antibody selection coefficient $S_a $ can be thought as the amount of stimulation that a bound B-cell receptor experiences, e.g. due to helper T-cells. If the chronic infection is HIV, where the virus attacks the helper T-cells,  $S_a $ may decrease as HIV progresses and the T-cell count decays. Furthermore, $f^*_{{}_A}$ affects the absolute growth rate, but does not affect the relative growth rate between genotypes. We call the fitness models based on the averaged binding affinity in eq.~(\ref{non-liN_a b_av}) as {\em nonlinear-averaged} and in eq.~(\ref{liN_a b_av}) as {\em linear-averaged}. In Section B.5 we introduce an alternative model of antibody activation,  which assumes that  proliferation of an antibody is related to its {\em  best binding affinity} against $R\leq N_v$  antigens, that are  presented to the antibody during its life time. 
The analytical results in this paper are all based on the antibody evolution in  linear-averaged fitness landscapes~(\ref{liN_a b_av}), and the other fitness models are only studied numerically. 

The viral fitness is related to the probability that it escapes the binding interactions with antibodies. We define the fitness of an antigen (virus) as the negative logarithm of its binding probability to the average antibodies that it interacts with, given by  eq.~(\ref{prob_bound_v_mean}), 

\EQA
\label{nonliN_vir_av} f_{{}_{V^\gamma}} \equiv f_{{}_V}(\V^\gamma;\{A\})&=& - c_v \log[p_{{}_V} (\V^\gamma)] =  c_v \log(1+\exp [-\beta_0 (E_{ .\, \gamma }+\hat E_{ .\, \gamma } -E^*)])\\
\label{liN_vir_av} &\simeq&  f_{{}_V}^*- S_v   (E_{ .\, \gamma } +\hat E_{ .\, \gamma } )
\EEA
with $f^*_{{}_V}=  c_v \log\big(1+\exp[\beta_0 E^*]\big)$ and the selection coefficient $S_v = c_v \beta_0 /(1+\exp[-\beta_0 E^*])$.\\

As  shown in eq.~(\ref{gen-sel}) the change in the  frequency of an antibody or a virus is proportional to its fitness, which is related to its average binding affinity.  Therefore,  the change of a given phenotype statistic  $U(\x, \y)$ due to selection follows, 
\begin{align}
&\frac{d}{dt} U(\x, \y)= \Large\sum_{\alpha, \gamma}\left[\frac{\partial U}{\partial x^\alpha} ( f_{{}_{A^\alpha}}-F_{{}_A}) \, x^\alpha +\frac{\partial U}{\partial y^\gamma} ( f_{{}_{V^\gamma}}-F_{{}_V})\, y^\gamma\right]
\end{align}
where $F_{{}_A}$ and $F_{{}_V}$ are respectively the mean fitness in the antibody and in the viral population. With this formulation we can compute the effect of selection on the statistics of the binding affinity distribution, i.e., the mean affinity $\E$, $\hat\E$, and the central moments, $M_{A,r}$, $\hat M_{A,r}$ and $M_{V,r}$, which we present in the following section. 

Similar to the rescaling procedure in  eq.~(\ref{scaling}), we  use the total  trait scales  to define the rescaled strength of selection, 
\begin{align}
&s_a =  N_a  S_a  E_0, \qquad \qquad\hat s_a  =  N_a  S_a  \hat E_0, \qquad \qquad s_v =  N_a  S_v  E_0, \qquad \qquad \hat s_v = N_v S_v \hat E_0 \label{cscaling}
\end{align}

\para{Genetic drift.} We can project the stochasticity of  the genotype space onto the phenotype  space. The projected diffusion coefficients show the correlation between the noise levels of the phenotypic statistics $A$ and $B$. 
\EQ
\G^{AB}=\frac{1}{N_a } \sum_{\alpha, \beta} \frac{\partial A}{\partial x^\alpha}\, \frac{\partial B}{\partial x^\beta} \,g^{\alpha\beta}+\frac{1}{N_v} \sum_{\gamma, \lambda}\frac{\partial A}{\partial y^\gamma}\, \frac{\partial B}{\partial y^{\lambda}} \,h^{\gamma\lambda}
\EE
and the genotypic diffusion constants $g^{\alpha\beta}$ and $h^{\gamma\lambda}$ are given by eq.~(\ref{gen_sampling}). As an example, we compute the diffusion term for the mean binding affinity in the variable region $\E$,

\EQA
\nonumber \G^{\E\E}&=&\frac{1}{N_a }\sum_{\alpha, \beta} \frac{\partial \E}{\partial x^{\alpha}}\, \frac{\partial \E}{\partial x^{\beta}} \,g^{\alpha\beta} +\frac{1}{N_v} \sum_{\gamma, \lambda}\frac{\partial \E}{\partial y^{\gamma}}\, \frac{\partial \E}{\partial y^{\lambda}} \,h^{\gamma\lambda}\\
\nonumber&=&\frac{1}{N_a } \sum_{\alpha, \beta} E_{\alpha\, . } E_{\beta\, . } \Big[-x^{\alpha} x^{\beta} (1-\delta_{\alpha\beta})+x^{\alpha} (1-x^{\alpha} ) \delta_{\alpha\beta}\Big]\\
\nonumber&&+\frac{1}{N_v} \sum_{\gamma, \lambda}E_{.\, \gamma}E_{.\, \lambda} \Big[-y^{\gamma} y^{\lambda} (1-\delta_{\gamma}^{\lambda})+y^{\gamma} (1-y^{\gamma} ) \delta_{\gamma}^{\lambda}\Big]\\
\nonumber&=&\frac{1}{N_a }\Big[ \sum_{\alpha} (E_{\alpha\, . }-\E)^2 x^{\alpha}\Big]+\frac{1}{N_v}\Big[ \sum_{\gamma} (E_{.\, \gamma}-\E)^2 y^{\gamma}\Big]\\
&=&\frac{1}{N_a }{M_{A,2}}+\frac{1}{N_v} M_{V,2} 
\EEA
where $\delta_{\alpha\beta}$ is a Kronecker delta function. A similar approach finds the diffusion terms for the second moments and the cross-correlation terms between the first and the second moments in the variable and the conserved regions (see e.g.,~\cite{Nourmohammad:2013ty} for further details),
\begin{align}
\nonumber &\G^{M_{A,2},M_{A,2}}= \frac{1}{N_a } (M_{A,4}-M_{A,2}^2), \quad \quad \G^{M_{V,2},M_{V,2}}= \frac{1}{N_v} (M_{V,4}-M_{V,2}^2),\qquad \G^{\hat M_{A,2},\hat M_{A,2}}= \frac{1}{N_a } (\hat M_{A,4}-\hat M_{A,2}^2) \\
\\
\nonumber &\G^{\E,M_{A,2}}= \frac{1}{N_a } M_{A,3}, \quad \quad \G^{\E,M_{V,2}}= \frac{1}{N_v} M_{V,3},\qquad \G^{\hat \E,\hat M_{A,2}}= \frac{1}{N_a } \hat M_{A,3}\end{align}

\vspace{1cm}

\noindent{\bf \large B.3 Stochastic evolution of molecular phenotypes ({linear-averaged} fitness)}\\

Putting all the evolutionary forces together, we can write down evolution equations for the statistics of binding affinities in a linear fitness landscape introduced in equations (\ref{liN_a b_av}, \ref{liN_vir_av}),
\EQA
  \label{d-dt-E}
  \textbf{variable region: } \qquad \frac{d}{dt}  { \E}&=& -2(\mu_v+\mu_a)  \E+S_a   M_{A,2}  -    S_v   \, M_{V,2}  +\chi_{\E} \label{deltaE}
  \\&&\nonumber \\
   \textbf{conserved region: } \qquad \frac{d}{dt}  { \hat \E}&=&S_a   \, \hat M_{A,2} -2\mu_a  \, \hat \E+\chi_{\hat\E}
\EEA
with the Gaussian correlated noise statistics due to the genetic drift, 

\begin{align}
&\langle\chi_{\E}\rangle=0, \qquad\langle\chi_{\E}(t) \chi_{\E}(t') \rangle=  \Big[\frac{ M_{A,2}}{N_a } +\frac{ M_{V,2}}{N_v}\Big]\, \delta(t-t')\\
&\langle\chi_{\hat \E}\rangle=0, \qquad\langle\chi_{\hat \E}(t) \chi_{\hat \E}(t') \rangle=  \Big[\frac{ \hat M_{A,2}}{N_a }\Big]\, \delta(t-t')
\label{E_noise} \end{align}

Similarly, we can write down the stochastic evolution equations for the second moments $M_{A,2}$, $\hat M_{A,2}$ and $M_{V,2}$,

\begin{align}
  \frac{d}{dt}   M_{A,2} &= -4\mu_a (   M_{A,2}-\ell \k_2)-4\mu_v    M_{A,2}  -\frac{   M_{A,2}}{N_a }+S_a  M_{A,3}+\chi_{ {}_{ M_{A,2}} }\\
\frac{d}{dt}  \hat M_{A,2} &= -4\mu_a (  \hat M_{A,2}  - \hat \ell \hat \k_2)  -\frac{ \hat M_{A,2} }{N_a }+S_a  \hat M_{A,3}+\chi_{ \hat  M_{A,2} }\\
 \frac{d}{dt}  M_{V,2} &=-4\mu_v(   M_{V,2}-\ell \k_2)-4\mu_a    M_{V,2} -\frac{    M_{V,2}}{N_v}-S_v   M_{V,3} +\chi_{{}_{     M_{V,2}}} 
\end{align}

with  Gaussian correlated noise statistics, 
\begin{align}
&\langle\chi_{ {}_{ M_{A,2}} }\rangle =0, \qquad\langle\chi_{ {}_{ M_{A,2}} }(t)\chi_{ {}_{ M_{A,2}} }(t') \rangle=  \left [\frac{ M_{A,4} -(M_{A,2})^2 }{N_a }\right]\, \delta(t-t')\\
& \langle\chi_{ {}_{ \hat M_{A,2}} }\rangle =0, \qquad\langle\chi_{ {}_{ \hat M_{A,2}} }(t)\chi_{ {}_{ \hat M_{A,2}} }(t') \rangle=  \left [\frac{ \hat M_{A,4} -(\hat M_{A,2})^2 }{N_a }\right]\, \delta(t-t')\\
&\langle\chi_{ {}_{ M_{V,2}} }\rangle =0, \qquad\langle\chi_{ {}_{ M_{V,2}} }(t)\chi_{ {}_{ M_{V,2}} }(t') \rangle=  \left [\frac{ M_{V,4} -(M_{V,2})^2 }{N_v}\right]\, \delta(t-t')\\
\nonumber \\
&\langle \chi_{{}_{ M_{A,2}} }(t)\chi_{\E }(t')\rangle= \frac{M_{A,3}}{N_a }\, \delta(t-t'), \quad\quad  \langle \chi_{{}_{\hat  M_{A,2}} }(t)\chi_{\hat \E }(t')\rangle= \frac{\hat M_{A,3}}{N_a }\, \delta(t-t')\\
&\langle \chi_{{}_{ M_{V,2}} }(t)\chi_{\E }(t')\rangle= \frac{\langle M_{V,3}\rangle }{N_v}\, \delta(t-t') 
\label{M_noise}\\\nonumber
 \end{align}
 
 It should be noted that we ignore the linkage correlations between the binding affinity of the variable region  $\E$ and conserved region $\hat\E$ of the virus. From the numerical analysis we see that the  covariance between the linked variable and conserved regions, $ \langle \sum_\alpha x^\alpha (\E_{\alpha\,.} - \E ) \, (\hat \E_{\alpha} -\hat  \E )  \rangle $ is  small compared to the diversity  of the average binding affinity in both regions of antibody and viral populations, $\langle M_{A,2}\rangle$, $\langle \hat M_{A,2}\rangle$ and $\langle M_{V,2}\rangle$; S2D~Fig. Lineages with access to the conserved region of the virus adapt by aligning their sites to the conserved sequence, and hence, remain relatively conserved with variations arising only from the stochastic forces of  mutation and genetic drift. In  Section B.4  we explicitly show that  the auto-correlation time  for the binding affinity in the conserved region is longer than in the variable interaction region; see equations~(\ref{autocorr_hatE_t1t2}, \ref{autocorr_E_t1t2}). Therefore, the correlation between the binding affinity of the variable and the conserved regions remains small throughout the evolutionary process.\\

\begin{figure}[t!]
\begin{center}
\includegraphics[width=  0.9\columnwidth]{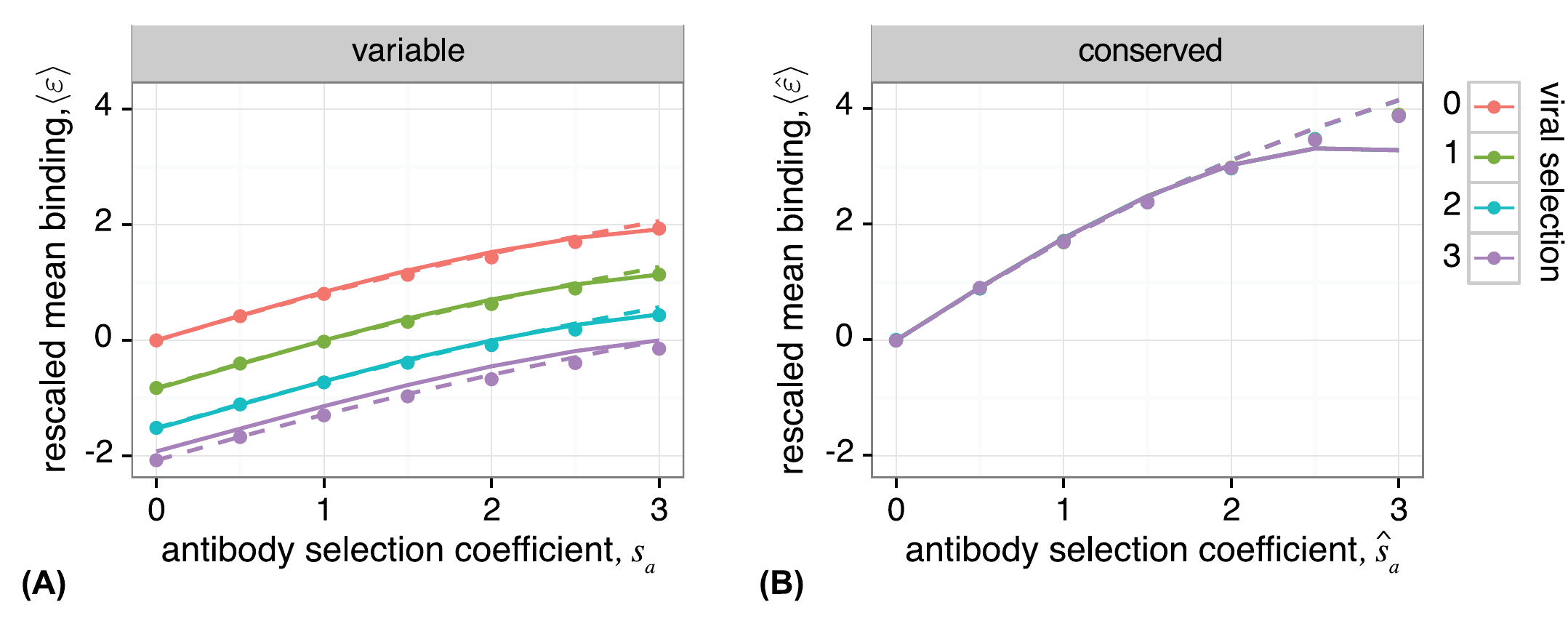}
\end{center}
\caption{{\bf Effect of selection on the mean binding affinity.} The rescaled mean binding affinity for {\bf (A)} the variable interaction region   $\varepsilon =\E/E_0$, and {\bf (B)} the conserved region $\hat \varepsilon=\hat\E/E_0$, as a function of selection coefficients. Stationary mean binding affinity is sensitive to selection on antibodies in both variable and conserved regions. The conserved region is not sensitive to viral selection strength. Points indicate simulation results,  dashed lines indicate the stationary solution  in eqs.~(S59, S61)  using estimates for the  diversity of the binding affinity from the simulations, and solid lines are the stationary solutions~(S59, S61) using the analytical estimates of the diversity from eq.~(S75). Parameters are:  $\kappa_i=\hat \kappa_i=1$ for all sites, $\ell=\hat{\ell}=50$, $N_a =N_v=1000$, $\theta_a=\theta_v=1/50$. Points are time averaged values from simulations run for $10^6 N_a$ generations, with values sampled every $N_a$ generations, and data from first $100 N_a$ generations discarded.
\label{fig:E_SI}}
\end{figure}
\vspace{1cm}
\para{Stationary solutions for trait mean and diversity.} 
From equations above we can solve for the stationary  mean binding affinity, binding diversity in both populations, and the covariance between the moments as a function of the higher moments, 

\begin{figure}[t!]
\begin{center}
\includegraphics[width=0.8\textwidth]{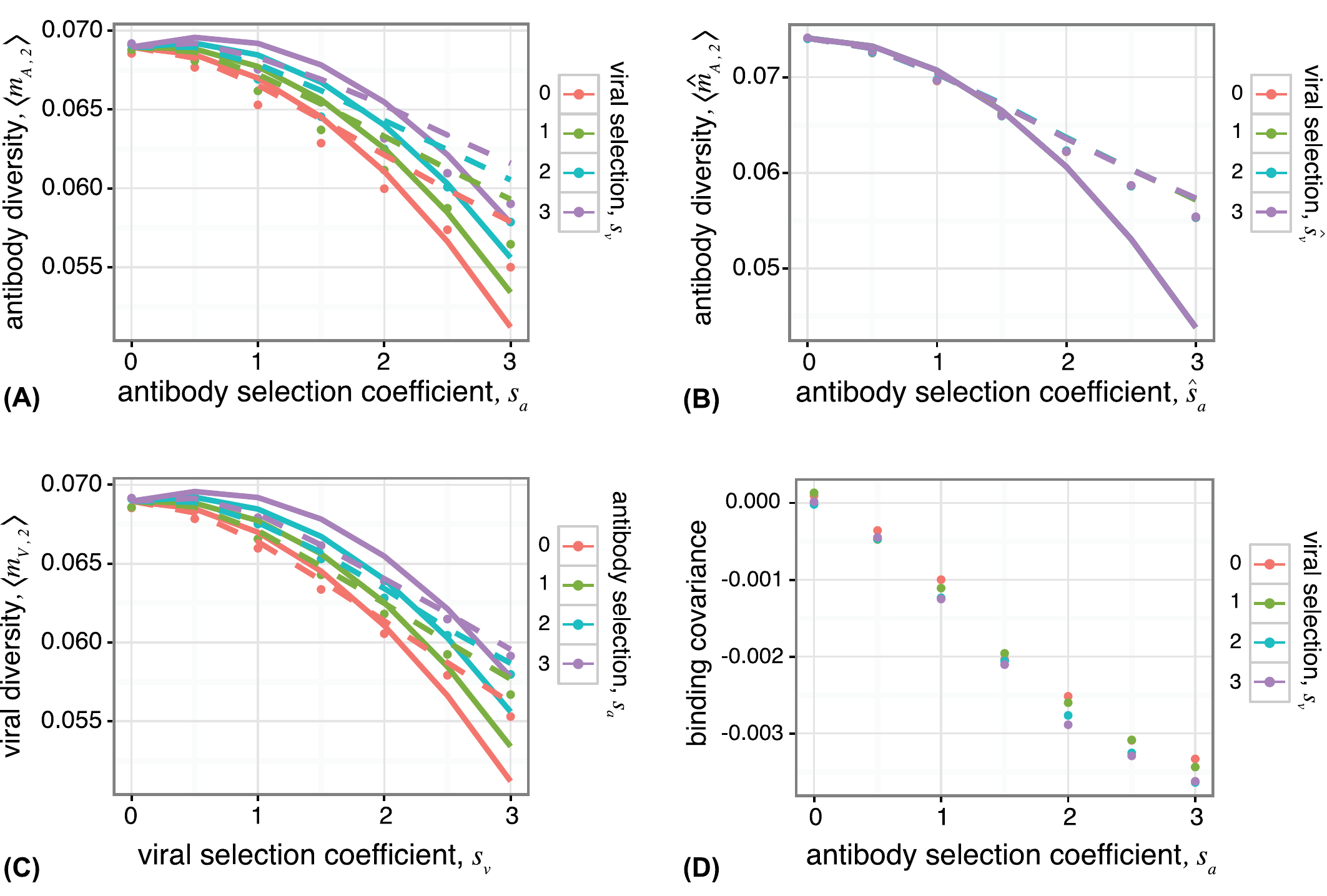}
\end{center}
\caption{{\bf Effect of selection on the diversity and covarinace of binding affinity in antibodies and viruses}. Stationary diversity of the binding affinity for {\bf (A)} the variable interaction region $m_{A,2}=M_{A,2}/E_0^2$,  {\bf (B)} the conserved interaction region $\hat m_{A,2}=\hat M_{A,2}/\hat E_0^2 $ in the antibody population, and {\bf (C)} the variable region in the viral population $m_{V,2}=M_{V,2}/E_0^2$  plotted as a function of viral and antibody selection coefficients. The diversity of binding across the antibodies in the conserved region  $\hat m_{A,2}$ in {(B)}  is not sensitive to viral selection strength. {\bf (D)} The magnitude  of  the rescaled covariance due to genetic linkage between binding of the antibody to the conserved and the variable regions, $\langle [ (E_{\alpha\,. } -\E) (\hat E_{\alpha\,. } -\hat \E) ]_{{}_A} \rangle/ E_0 \hat E_0$, is much smaller than the diversity of binding in each region, shown in {(A)} and {(B)}. Points indicate simulation results with parameters similar to Fig.~\ref{fig:E_SI}, dashed lines indicate the stationary solution using estimates for higher moments from the simulations (eqs.~(S63, S64)), and solid lines indicate the full  stationary solution given by eq.~(S75) for antibodies, and  the corresponding solution for viruses. Theory lines begin to deviate from simulation results for large selection strengths $s_a ,s_v >1$. The deviations are larger in antibodies due to neglecting the linkage correlation between the variable and the conserved regions.
}
\label{fig:M2A}
\end{figure}

\begin{align}
&\label{meanE}\langle\E \rangle = \frac{1}{2(\theta_a+ \tilde\theta_v)}{ N_a  S_a  \langle M_{A,2} \rangle}-\frac{1}{2 (\tilde\theta_a+\theta_v)} {N_v S_v  \langle M_{V,2} \rangle }\\
&\langle \E , \E \rangle = \frac{1}{4(\theta_a+\tilde\theta_v) } \big[ \langle M_{A,2}\rangle + 2N_a S_a  \langle \E,M_{A,2}\rangle \big]+\frac{1}{4(\tilde\theta_a+\theta_v)} \big[ \langle{M_{V,2}}\rangle- 2 N_v S_v  \langle \E,M_{V,2}\rangle\big]\\
&\label{meanhatE}\langle\hat\E \rangle =N_a  S_a  \langle \hat M_{A,2}  \rangle/ 2\theta_a \\
&\langle \hat\E , \hat\E \rangle =\frac{1}{ 4 \theta_a }\, \big[{ \langle \hat M_{A,2} \rangle + 2N_a  S_a    \langle \hat\E , \hat M_{A,2}  \rangle } \big] \\
\nonumber\\
\label{MA2_withMA3}&\big \langle M_{A,2}\big  \rangle =\frac{ 1}{1+4(\theta_a+\tilde\theta_v)} \Big[4\ell \k_2 \theta_a+(N_a  S_a )\,  \langle M_{A,3}\rangle  \Big]\\ 
\label{MV2_withMV3}&\big \langle M_{V,2}\big  \rangle =\frac{ 1}{1+4(\tilde\theta_a+\theta_v)} \Big[4\ell \k_2 \theta_v-(N_v S_v )\,  \langle M_{V,3}\rangle  \Big]\\ 
\nonumber\\
\label{E-MA2}& \langle \E,  M_{A,2}\rangle = 
  \frac{1}{1+ 6(\theta_a+\tilde\theta_v) }\Big[ \langle  M_{A,3} \rangle + N_a  S _a\big[\langle \E,M_{A,3}\rangle +  \langle (M_{A,2})^2\rangle\big]\Big)\\
\label{E-MV2}& \langle \E,  M_{V,2}\rangle =  \frac{1}{1+ 6(\tilde\theta_a+\theta_v) }\Big({\langle M_{V,3}\rangle }- N_v S_v  \big[\langle \E,M_{V,3}\rangle +\langle (M_{V,2})^2\rangle\big] \Big)\\
\nonumber \\
& \langle \E,  M_{A,3}\rangle =\frac{\langle M_{A,4}\rangle /3 -\langle( M_{A,2})^2\rangle }{1+8/3 (\theta_a+\tilde\theta_v)}, \qquad\qquad
 \label{cov_EMV3} \langle  \E,  M_{V,3}\rangle  =\frac{\langle M_{V,4 }\rangle /3 -\langle (M_{V,2})^2\rangle }{1+8/3(\tilde\theta_a+\theta_v)}
\end{align}
where $\tilde\theta_a= \theta_a (N_v/N_a) $ and $\tilde\theta_v= \theta_v (N_a/N_v )$. We denote the ensemble covariance  of two stochastic variables $x$ and $y$ by, 
\EQ
\big \langle x,y\big \rangle \equiv\big \langle (x-\langle x\rangle )\, (y-\langle y\rangle ) \big\rangle
\EE
and hence,  $\langle x,x \rangle$ is the ensemble variance of the variable $x$.    Similar forms  of the stationary solutions apply  to the  statistics of the binding affinity in the conserved interaction region, $\langle \hat M_{A,2}\rangle$, $\langle\hat \E, \hat M_{A,2}\rangle$, and can be found by setting the viral mutation rate $\mu_v$  and the central moments $\hat M_{V,r}$ equal to zero in equations (\ref{meanE}-\ref{cov_EMV3}).  For brevity we do not present the solutions  of the central moments in the conserved region.

In  equations~(\ref{d-dt-E}-\ref{M_noise}), the evolution of each moment depends on the higher moments in the presence of selection, which leads to an infinite moment hierarchy.   However, in the regime where rescaled coefficients satisfy $s_a\theta_a <1$ and $s_v\theta_v<1$,  we can truncate the moment hierarchy. From the comparisons of the Wright-Fisher simulations  with our theoretical results we choose  to truncate the hierarchy after the $4^{th}$ moment. Furthermore, higher central moments are fast stochastic variables (see e.g.,~\cite{Nourmohammad:2013ty} and  the discussion in Section B.4 and  Fig.~S3), and their ensemble averages can sufficiently characterize the  evolution of the mean binding affinity $\E$ and the binding diversity $M_{A,2}$,  $\hat M_{A,2}$ and $M_{V,2}$. Therefore, we will only present ensemble-averaged equations for the $3^{rd}$ and $4^{th}$ moments of the phenotype distributions. In order to clarify the truncation of the moment hierarchy, we explicitly show the evolution equations and their stationary solutions for the rescaled moments of the phenotype distribution, which are defined in eq.~(\ref{scaling}). 

\begin{align}
 &\frac{d}{d\tau_a}\langle  m_{A,3}\rangle =-6 \theta_a \langle m_{A,3}\rangle- 8 \theta_a \,  \Big( \frac{ \k_3}{E_0^2 \k_1} \, \langle \varepsilon \rangle \Big)  -6\tilde  \theta_v  \, \langle m_{A,3}\rangle-3 {\langle m_{A,3}\rangle} + s_a  \big[\langle m_{A,4}\rangle- 3\left \langle  (m_{A,2})^2 \right\rangle\big ] \label{ma3eq}\\
& \frac{d}{d\tau_v} \langle m_{V,3}\rangle 
=-6 \theta_v \langle m_{V,3}\rangle  -8\theta_v \, \Big( \frac { \k_3  }{E_0^2 \k_1}\, \langle \varepsilon \rangle \Big) - 6\tilde  \theta_a \,  \langle m_{V,3}\rangle -3 {\langle m_{V,3}\rangle } -s_v  \big[\langle m_{V,4} \rangle - 3 \left \langle (  m_{V,2} )^2\right \rangle \big]\label{mv3eq}\\
\nonumber\\
&\frac{d}{d\tau_a} \langle (m_{A,2})^2\rangle 
=-8 \theta_a \big[\langle (m_{A,2})^2\rangle  -\langle  m_{A,2}\rangle \big] -8 \tilde \theta_v \,\langle (m_{A,2})^2\rangle  + {\langle  m_{A,4}\rangle   -3\left \langle ( m_{A,2})^2\right\rangle }\\
& \frac{d}{d\tau_v} \langle (m_{V,2})^2\rangle =-8 \theta_v \big[ \langle (m_{V,2})^2\rangle  -\langle m_{V,2}\rangle\big] -8 \tilde \theta_a \,  \langle (m_{V,2})^2\rangle + {\langle m_{V,4}\rangle  -3 \left  \langle (m_{V,2})^2\right\rangle }\\
  \nonumber &\\
&     \frac{d}{d\tau_a} \langle  m_{A,4}\rangle= -8 \theta_a \Big[ \langle  m_{A,4}\rangle  -2 \frac{ \k_4}{\ell\, \k_2^2}   - (3-4/\ell  )\,   \langle m_{A,2}\rangle   \Big] -8 \tilde \theta_v \,  \langle  m_{A,4}\rangle + {6\left \langle (m_{A,2})^2\right \rangle  -4 \langle  m_{A,4}\rangle }\\
& \frac{d}{d\tau_v} \langle  m_{V,4}\rangle= -8 \theta_v\, \Big[ \langle  m_{V,4}\rangle  -2 \frac{ \k_4}{\ell\, \k_2^2} -(3-4/\ell)\,  \langle m_{V,2}\rangle \Big] - 8\tilde \theta_a\,   \langle  m_{V,4}\rangle + {6\left  \langle (m_{V,2})^2\right \rangle -4 \langle  m_{V,4}\rangle }\\\nonumber
\end{align}
with $\tilde\theta_a=\theta_a(N_v /N_a)$ and $\tilde\theta_v=\theta_v(N_a /N_v)$.  $\tau_a =t/N_a $ and $\tau_v=t/N_v$ are the evolutionary times   in natural units of  the neutral coalescence time in the antibody population $N_a $ and in the viral population $N_v$, respectively.  The term $\langle \varepsilon\rangle = 2(s_a  \theta_a -s_v  \theta_v \,(N_a /N_v)) \big/(\theta_a+\theta_v \,(N_a /N_v))$ in equations~(\ref{ma3eq}, \ref{mv3eq})  is the stationary  solution for the rescaled mean binding affinity up to orders of $\mathcal{O}(  \theta_a^2, \theta_v^2)$. The stationary solutions for the rescaled central moments of the  antibody population follow,

\begin{align}
\label{M2_mom}
 & \big \langle m_{A,2}\big  \rangle =
\frac{  4\theta_a}{1+4(\theta_a+\tilde\theta_v)} -
 \frac{8\theta_a }{3+18 (\theta_a+\tilde\theta_v)}  s_a  \Big[    \frac{ \k_3}{E_0^2 \k_1} \, \langle \varepsilon \rangle  - 4 s_a  \theta_a^2 +\mathcal{O} \big(\theta_a^3\big)  \Big] +\mathcal{O}(s_a ^2\theta_a^2) \\
 \nonumber\\
& \label{M3_sol}  \big\langle m_{A,3}\big \rangle  = -\frac{8}{3} \times \frac{\theta_a  }{1+2(\theta_a+\tilde\theta_v) } \Big (   \frac{ \k_3}{E_0^2 \k_1} \, \langle \varepsilon \rangle \Big)+\frac{32}{3}s_a  \left[ \theta_a^2 +\mathcal{O}( \theta_a^3)\right] +\mathcal{O}(s_a ^2\theta_a^3) \\
\nonumber\\
  &\big \langle  (m_{A,2})^2\big\rangle  =  \frac{8\theta_a } {3+ 28\, (\theta_a+\tilde\theta_v) }\, \Big[\frac{1}{\ell}\, \frac{ \k_4}{\k_2^2}  +  2\theta_a  (7-4 /\ell ) \Big]+\mathcal{O}( s_a  \theta_a^3)\\ 
\nonumber\\
\label{m4_stat}&   \langle m_{A,4} \rangle = \frac{ 24\theta_a}{3+ 28\, (\theta_a+\tilde\theta_v) }\Big[ \frac{1}{\ell}\, \frac{ \k_4}{\k_2^2} + 2  \theta_a (5- 4/\ell) \Big]+\mathcal{O}( s_a  \theta_a^3)\\\nonumber
\end{align}

Similar solutions can be found for the central moments of binding affinity in the viral population $m_{V,r}$, by replacing the subscripts $a$ and $v$  in the equations above.   The stationary solutions  for the central moments of the binding affinity in the conserved region of  antibody  population $\hat m_{A,r}$ can be found by setting the viral mutation rate equal to zero,  $\theta_v=0$, and by using the characteristics of the conserved region i.e., genetic length $\hat \ell$ and sites contributions $\hat \k_r$ in equations (\ref{M2_mom}-\ref{m4_stat}). Fig.~S1 shows a good agreement between the numerical results for the rescaled stationary mean binding affinity $\langle \varepsilon \rangle =\langle \E \rangle/E_0 $, $\langle \hat\varepsilon \rangle=\langle\hat \E \rangle/\hat E_0 $  from the Wright-Fisher simulations and the analytical solutions (\ref{meanE}, \ref{meanhatE}), by using the stationary ensemble averages for the  diversity of the binding affinity $\langle m_{A,2}\rangle$, $\langle\hat m_{A,2}\rangle$ and $\langle m_{V,2}\rangle$ in eq.~(\ref{M2_mom}).  Fig.~S2 compares the analytical solution for the second central moments $\langle m_{A,2}\rangle$ and $\langle m_{V,2}\rangle$ with numerical results from the Wright-Fisher simulations, by inserting the empirical estimates of the higher moments from the simulations as in equations (\ref{MA2_withMA3}) and (\ref{MV2_withMV3}), (dashed lines), and by using the analytical solutions for the higher moments to estimate the stationary value for the phenotype diversity, as given by eq.~(\ref{M2_mom}), (solid lines). \\

\vspace{1cm}

\noindent{\bf \large B.4 Time-dependent statistics and separation of time-scales}\\

\para{Statistics of the  mean phenotype.} As we show below,  the higher central  moments $M_{V,r}$ and $M_{A,r}$ for $(r>1)$ are fast stochastic variables. Therefore, it is sufficient to use their stationary ensemble averages to  compute the finite time correlation of the mean binding affinities, $\E(\tau)$ and $\hat \E(\tau)$.

\begin{figure}[t!]
\begin{center}
\includegraphics[width=0.5 \textwidth]{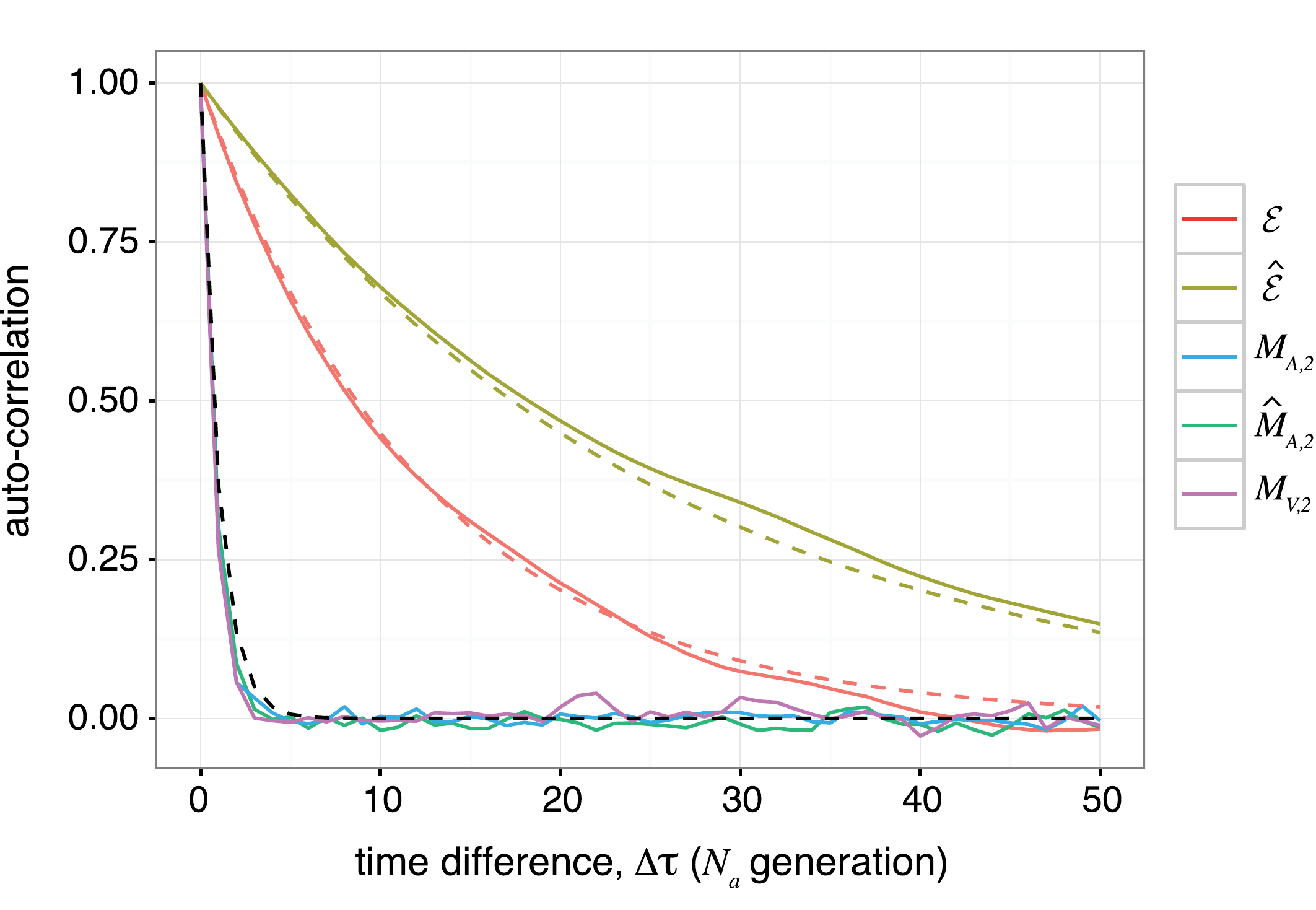}
\end{center}
\caption{{\bf Time-dependent statistics.} Auto-correlation of the stationary mean binding affinity in the variable region (red), eq.~(S81), has a  shorter decay time than in the conserved  region (yellow), eq.~(S82). The decay time for the auto-correlation of the  mean phenotype in both variable and conserved regions, which is of order of the inverse mutation rate, is much longer than the  correlation time of the second moments (green, blue, purple), which decay on a timescale of $N$ generations. Solid lines are from stationary simulations, and dashed lines are the analytical results for the auto-covariance of the moments given by, eq.~(S81) (red),  eq.~(S82) (yellow)  and eq.~(S83) (black), normalized to have magnitude 1 at separation  time $\Delta \tau=0$. Parameters are: all $\kappa_i=\hat \kappa_i=1$, $\ell=\hat{\ell}=50$, $N_a =N_v=1000$, $\theta_a=\theta_v=1/50$, $s_a=s_v=\hat s_a=1$. Simulation results are time-averaged over $10^4 N_a$ generations, with values sampled every $N_a$ generations, and first $100N_a$ generations omitted.
}
\label{fig:autocorr}
\end{figure}

The time-dependent solution for the ensemble averaged mean affinity $\langle \E(\tau)\rangle $ and $\langle \hat\E(\tau)\rangle $ at time $\tau$, and the covariance between two time-points $\tau_2\geq \tau_1$, starting from an initial condition at time $\tau_0=0$ with the ensemble averages for the mean affinities $ \langle \E(0)\rangle$, $\langle \hat \E(0)\rangle$ and the  diversities $ \langle \E(0 ), \E(0)\rangle  $,  $ \langle \hat \E(0 ), \hat \E(0)\rangle  $  follows, 

 \EQA
\langle  \E(\tau)\rangle &=&(1- \e^{-2 (\theta_a+\tilde\theta_v) \tau} )\,  \langle\E \rangle  +   \e^{-2 (\theta_a+\tilde\theta_v) \tau}\, \langle  \E(0)\rangle\\
\langle \hat \E(\tau)\rangle &=&(1- \e^{-2 \theta_a \tau} )\,  \langle\hat \E\rangle  +   \e^{-2\theta_a \tau}\, \langle \hat  \E(0)\rangle\\
\nonumber\\
\nonumber\langle \E(\tau_1), \E(\tau_2)\rangle&=&\e^{-2 (\theta_a+\tilde\theta_v) \tau_2}{ \langle \E(0), \E(0)\rangle }  
+ \left[ \frac{ \langle M_{A,2}\rangle}{N_a } +\frac{\langle M_{V,2} \rangle}{N_v}\right]  \int_0^{\tau_1} \e^{-2 (\theta_a+\tilde\theta_v)  (\tau_1-\tau')}  \e^{-2 (\theta_a+\tilde\theta_v)  (\tau_2-\tau')}  d\tau'\\
\nonumber &=&\e^{-2 (\theta_a+\tilde\theta_v) \tau_2}\,  \langle \E(0), \E(0)\rangle+ \left[ \frac{ \langle M_{A,2}\rangle}{4(\theta_a+\tilde\theta_v)} +\frac{\langle M_{V,2}\rangle}{4(\tilde\theta_a+\theta_v)}\right] \, \left [\e^{-2  (\theta_a+\tilde\theta_v)  (\tau_2-\tau_1)} -\e^{-2  (\theta_a+\tilde\theta_v)  (\tau_1+\tau_2)}\right] \\ \nonumber\\
\label{autocorr_E_t1t2}\\
\label{autocorr_hatE_t1t2} \langle \hat \E(\tau_1), \hat \E(\tau_2)\rangle&=&\e^{-2 \theta_a  \tau_2}\,  \langle \hat \E(0), \hat \E(0)\rangle+  \frac{ \langle \hat M_{A,2}\rangle}{4 \theta_a}  \, \left [\e^{-2 \theta_a  (\tau_2-\tau_1)} -\e^{-2  \theta_a  (\tau_1+\tau_2)}\right] 
\EEA
where $\langle\E\rangle$ and $\langle\hat \E\rangle$ are the  stationary values of the  mean phenotype in the variable and the conserved interaction regions, given by equations~(\ref{meanE}, \ref{meanhatE}). Time $\tau$  is measured in units of the neutral coalescence time for antibodies, $N_a $.  The characteristic time-scale for the decay of the mean binding affinity in the variable interaction region of the virus is $ 1/(2 (\theta_a+\tilde \theta_v ))$ in units of $N_a $, which is shorter than the time-scale for the conserved region,  $1/2\theta_a$. Therefore, binding affinity in the conserved region is correlated over a longer  period of time compared to the variable region (i.e., about twice as long if $\theta_a \sim \tilde \theta_v$). The difference in time-scale explains the small covariance due to the genetic linkage between the conserved and the variable region of the virus shown in Fig.~S3. 

\para{Statistics of the phenotype  diversity.} As shown in~\cite{Nourmohammad:2013ty}, the fluctuations in the phenotype diversity are scale invariant, which is a consequence of coherent, genome-wide linkage-disequilibrium fluctuations in the absence of recombination.  It is generated by sampling from a set of genotypes with  binding affinities  $E_{\alpha\,.}$ in antibodies and $E_{.\, \gamma}$ in viruses  from the underlying distributions with variance $M_{A,2}$ and $M_{V,2}$, which scale like the  genome length $\ell$. These large fluctuations result in a relatively short correlation time  for the phenotype diversity, shown in Fig.~S3. Similar to the mean binding affinity, we can estimate the typical lifetime of these  fluctuations from the stationary auto-correlation function,

\EQ
\label{autocorr-diversity} \langle M_{A,2}(\tau_a),M_{A,2}(\tau_a')\rangle \sim \e^{-(\tau_a-\tau_a')}, \qquad \langle M_{V,2}(\tau_v),M_{V,2}(\tau_v')\rangle \sim \e^{-(\tau_v-\tau_v') }
\EE
where $\tau_a$, $\tau_a'$ are measured in units of the antibody neutral coalescence time $N_a $, and $\tau_v$, $\tau_v'$ are measured in units of the viral neutral coalescence time $N_v$. Fig.~S3 shows the decay of the stationary auto-correlation for the diversity of the binding affinity $M_{A,2}$, $\hat M_{A,2}$ and  $M_{V,2}$ as a function of the evolutionary separation time $\Delta \tau=\tau-\tau'$. It is evident that the characteristic decay time for the phenotype diversity (\ref{autocorr-diversity}) is much shorter than that of the  mean phenotype, given by the auto-correlation function in eqs.~(\ref{autocorr_E_t1t2},~\ref{autocorr_hatE_t1t2}).  \\
 
 \vspace{0.5cm}
 
\noindent{\bf \large B.5 Alternative fitness models}\\

\para{Nonlinear activation probability based on average binding (nonlinear-averaged).} 
 We assume that the growth rate (fitness) of an antibody is proportional to the logarithm of its activation probability given by eq.~(\ref{non-liN_a b_av}), which may be  approximated by a linear function if the nonlinearity is small (\ref{liN_a b_av}). Here, we numerically study the effect of nonlinear sigmoidal fitness functions by  comparing the evolutionary dynamics of populations in fitness landscapes with different values  of nonlinearity $\beta = \beta_0\sqrt{ E_0^2+\hat E_0^2}$ and binding threshold $e^* =E^*/\sqrt{E_0^2+\hat E_0^2}$, while keeping the overall strength of (rescaled) selection, $s_a = c_a \beta/(1+\exp[-\beta e^*])$ constant. The strength of selection corresponds to the slope of the approximate {linear-averaged}  fitness function in eq.~(\ref{liN_a b_av}).  
 
As the rescaled nonlinearity $\beta=\beta_0 E_0$ of the fitness  function~(\ref{non-liN_a b_av}) increases, the mean binding affinity $\E$ becomes closer to the neutral value; see Fig.~S4A. This is due to the sigmoid form of the fitness function, which reduces fitness differences between genotypes at extreme values of binding affinity. Since mutations push the mean binding affinity towards zero, the reduced advantage of binding at the extremes moves the stationary binding affinity towards zero.

Similar arguments suggest that the rate of adaptation in the antibody population should decrease as the  fitness landscapes become more non-linear. The rate of adaptation is determined by  fitness flux~\cite{Mustonen:2009vu,Mustonen:2010iga}, and is approximately  equal to the variance of fitness in the population~\cite{Fisher:1930wy}; see Section~C for detailed discussion. Due to the sigmoidal shape of the fitness function, fitness differences become small at large values of binding affinity (i.e., the functional antibodies), resulting in a  reduction of the  fitness variance in the population, and hence, a lower rate of adaptation. However, this effect is less pronounced when the threshold for specific interaction is very large, $e^*\gg 1/\beta$. In this case,  the fitness function is nearly linear for most antibodies, since their binding affinity fall below the binding threshold $e^*$.  In this regime, the fitness variance and the rate of adaptation are only sensitive to the selection strength $s_a$  (i.e., slope of fitness at $e=0$), and not the nonlinearity of the fitness  landscape. Evidently, the fitness variance (Fig.~S4B) is less sensitive to the non-linearity, than the mean binding affinity (Fig.~S4A). \\

\para{Nonlinear activation probability based on the strongest binding (nonlinear-EVD).} We  study a model for activation of antibodies which is based on their \emph{strongest binding affinity} with a subset of viruses. The basic assumption is that an antibody attempts to bind to a set of viruses (which may be smaller than the viral population size), and once a high affinity binding occurs, it begins to proliferate. Similar treatments have been introduced in the context of T-cell activation~\cite{Kosmrlj:2008kk,Kosmrlj:2009wm}.  The probability distribution function, $\Pi(E_{\alpha\, .}^*)$ of the strongest of $R$ independent binding interactions between the antibody $\A^\alpha$ and the viral population  $\{\V\}$ can be obtained using extreme value statistics. According to extreme value theory, if the distribution of binding affinities for a given antibody has an exponential tail,  the corresponding distribution for its strongest  binding affinity belongs to the class of Gumbel distributions~\cite{deHaan:2006tc}. In the evolutionary regime that we study here, the amount of genetic polymorphism in the population of antibodies results in a Gaussian-like distribution for the  binding affinities, with mean $E_{\alpha\, . }+\hat E_{\alpha\, . }$, and variance $I_{\alpha \, .}^{(2)}$ given by eq.~(\ref{AB_var_mom}). Therefore, the corresponding probability distribution for the strongest binding affinity out of $R$ independent trials, is a Gumbel distribution~\cite{deHaan:2006tc} with  a peak at,
\begin{figure}
\begin{center}
\includegraphics[width=.95\textwidth]{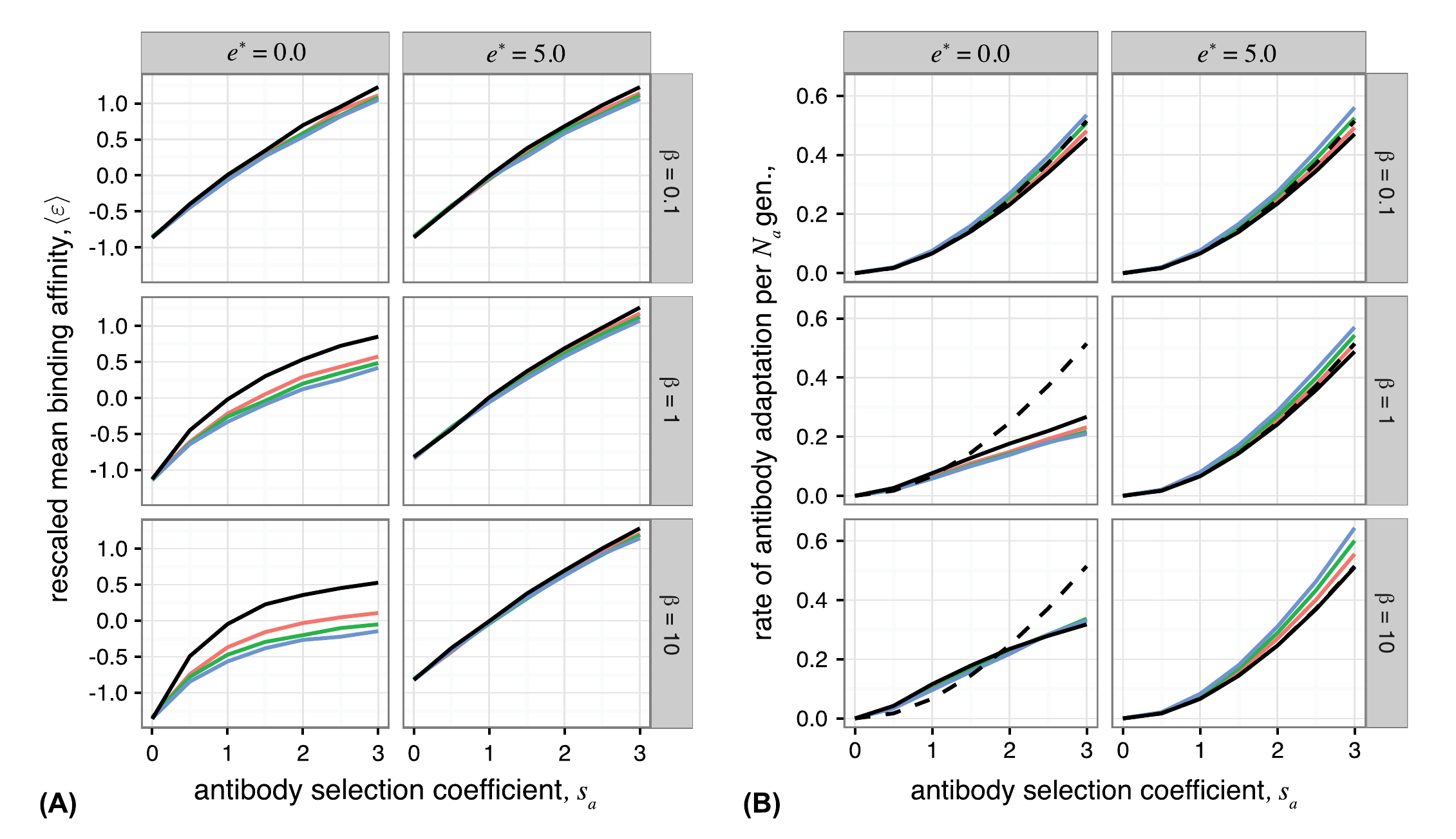}
\end{center}
\caption{ {\bf Alternative fitness models.} {\bf (A)} Stationary mean binding affinity and {\bf (B)} rate of antibody adaptation (fitness flux) due to selection, estimated by population fitness variance, for the nonlinear-averaged fitness model (black) and the nonlinear-EVD fitness model with the number of interactions, $R=10$ (red), $R=100$ (green), and $R=1000$ (blue). The mean binding affinity is sensitive to the degree of non-linearity $\beta$, and binding threshold $e^*$, but it is not very sensitive to the number of interactions $R$. The selection coefficient $s_a$ is defined as in eq.~(S39). Dashed line in {(B)} indicates the expected fitness variance for a linear-averaged fitness model, $ \langle \phi_{{}_A}\rangle \simeq s_a ^2 \langle m_{A,2}\rangle $, which is the selection component of the fitness flux in eq.~(S91). Parameters are:  $\kappa_i=\hat \kappa_i=1$ for all sites, $\ell=50$, $\hat \ell = 0$, $N_a =N_v=1000$, $\theta_a=\theta_v=1/50$.   Points are time averaged values from simulations run for $10^5 N_a$ generations, with values sampled every $N_a$ generations, and data from first $100N_a$ generations discarded.} 
\label{fig:altfit}
\end{figure}
\EQA
E_{\text{max}}^\alpha = E_{\alpha\, . }+\hat E_{\alpha\, . } +\sqrt{ 2 I^{(2)}_{\alpha\, . } \ln R}
\label{Emax}
\EEA
and a width  $\Sigma^\alpha =\sqrt {\pi  I^{(2)}_{\alpha\, . } /(12  \ln R)}$. If we assume that $\ln R\gg 1$, the distribution is sharply peaked, and $E_{\text{max}}^\alpha$ is sufficient to describe it. In addition, we assume the activation probability is a sigmoid function of $E_{\text{max}}^\alpha$,

\EQ
p_{{}_{A, \text{max}}} (\A^\alpha) = \frac{1}{1+\exp [-\beta_0 (E^\alpha_{\text{max}}-E^*)]}.
\label{prob_bound_max}
\EE

The fitness function $f_{A, \text{max}}(\A^\alpha;\{ V\})$ for the {\em nonlinear-EVD} model is related to the logarithm of the activation probability,

\EQA
\label{non-liN_a b_mx}f_{A, \text{max}}(\A^\alpha;\{ V\})&=&  c_a \log[p_\text{max} (\A^\alpha)] = - c_a \log(1+\exp [-\beta_0 (E_{\text{max}}^\alpha  -E^*)])
\EEA
where the coefficients are similarly defined as in eq.~(\ref{non-liN_a b_av}). Fig.~S4A shows the stationary mean binding affinity  for  {\em nonlinear-EVD} fitness model. While the mean binding affinity is sensitive to the nonlinearity parameter $\beta$, it is relatively insensitive to the number of interactions $R$, and behaves similarly to the {nonlinear-averaged} model. This is not surprising given the logarithmic dependence of binding affinity on the number of interactions $R$ in eq.~(\ref{Emax}).  \\

\vspace{1cm}
\begin{center}
 {\Large \bf C. Fitness flux and coevolutionary transfer flux}\\
\end{center}
\vspace{0.5cm}
The fitness flux $\phi (t)$ characterizes the adaptive response of a population by genotypic or phenotypic changes in a population~\cite{Lassig:2007vb,Mustonen:2009vu,Mustonen:2010iga,Held:2014di,Nourmohammad:2015vc}. The cumulative fitness flux, $\Phi(\tau) = \int_t^{t+\tau} N \phi (t') dt' $, measures the total amount of adaptation over an evolutionary period $\tau$~\cite{Lassig:2007vb,Mustonen:2010iga}. The evolutionary statistics of this quantity is specified by the fitness flux theorem~\cite{Mustonen:2010iga}. In our model, the fitness flux for the antibodies $\phi_{{}_A}(t)$ and the viruses $\phi_{{}_V}(t)$ follow,

\EQA
\label{ABflux}\phi_{{}_A}  (t) & = &\sum_{\alpha\in \text{antibodies}}\, \frac{\partial F_{{}_A} (t)}{\partial x^\alpha}\times  \, \frac{d x^\alpha(t)}{d t} \\
\label{Virflix}\phi_{{}_V}  (t) & = &\sum_{\gamma\in \text{viruses}}\, \frac{\partial F_{{}_V} (t)}{\partial y^\gamma} \times \, \frac{dy^\gamma(t)}{dt} 
\EEA
where, $F_{{}_A}(t)$ and $F_{{}_V}(t)$ are  the  mean fitness of the antibody and the viral populations at time $t$, and time is measured in units of generations. {It should be noted that the cumulative fitness flux for evolution in a constant environment (equilibrium) is equal to the difference of the mean fitness between the final and the initial time points. However, the cumulative fitness flux in time-dependent environments (non-equilibrium) depends on the whole evolutionary history of the population, and captures its incremental adaptive response to the underlying environmental fluctuations.}

We introduce a  new measure  of interaction between coevolving populations ``{\em transfer flux}", which is the change in the mean fitness of a  population due to the evolution of the opposing population. The transfer flux from antibodies to viruses  $\T_{A\rightarrow V}$  and from viruses to antibodies $\T_{V\rightarrow A}$ follow, 

\EQA
\label{A-to-V}\T_{A\to V}  (t) & = &\sum_{\alpha\in \text{antibodies} }\, \frac{\partial F_{{}_V} (t)}{\partial x^\alpha}\times  \, \frac{d x^\alpha(t)}{dt} \\
\label{V-to-A}\T_{V\to A}  (t) & = &\sum_{\gamma\in \text{viruses}}\, \frac{\partial F_{{}_A} (t)}{\partial y^\gamma} \times \, \frac{dy^\gamma(t)}{dt} 
\EEA

In the regime of substantial selection  $s_a,s_v\gtrsim1$, the transfer flux in  antagonistically interacting populations of antibodies and viruses is always negative, implying that adaptation of one population reduces the fitness of the opposing population. 

The fitness flux and transfer flux are rates of adaptation and interaction that are time-independent only in the  stationary state. The total amount of adaptation for non-stationary evolution, where the  fluxes change in time, can be generally measured by the cumulative  fitness and transfer flux.  {For coevolution in the  {linear-averaged} fitness landscape of equations~(\ref{liN_a b_av}, \ref{liN_vir_av}) the  cumulative fitness flux over an evolutionary period $[t_0:t]$ for antibodies and viruses follow from a simple genotype-to-phenotype projection,

\begin{align}
\nonumber \left \langle   \Phi_A(t_0:t) \right \rangle&=\left\langle  N_a\int_{t'=t_0}^{t} \phi_{{}_A} (t')dt' \right\rangle \\
\nonumber&=   \left\langle   N_a\Large\int_{t'=t_0}^{t} dt'\, \left(\frac{\partial F_{{}_A}(t')} {\partial \E(t')}\, \frac{\partial \E(t')}{\partial t'}  \Big |_{\{\V\}}  +  \frac{\partial F_{{}_A}(t')} {\partial \hat \E(t')}\, \frac{\partial \hat \E(t')}{\partial t'} \Big |_{\{\V\}} \right) \right \rangle \\
\label{cum_fit_flux_A_general}&= \left\langle   \int_{t'=t_0/N_a}^{t/N_a} dt' \left [-2\theta_a \left ( s_a \varepsilon(t') +\hat s_a \hat \varepsilon(t') \right)+ \left(s_a^2 m_{A,2}(t') +\hat s_a^2 \hat m_{A,2}(t') \right)\right]\right\rangle\\
\nonumber\\
\nonumber\left \langle  \Phi_V(t_0:t)\right \rangle &=  \left\langle   N_v \int_{t'=t_00}^{t} \phi_{{}_V}(t') dt'  \right\rangle  \\
\label{cum_fit_flux_V_general}&=\left\langle \int_{t'=t_0/N_v}^{t/N_v} dt'\,\big[2\theta_v s_v  \varepsilon (t') +s_v ^2\,  m_{V,2}(t') \big] \right\rangle  
\end{align}}
The first terms (proportional to $\theta$) in the integrants of  eqs.~(\ref{cum_fit_flux_A_general},~\ref{cum_fit_flux_V_general})  are the fitness changes due to mutations and the second terms are due to selection;  the changes due to genetic drift are zero for the ensemble-averaged fitness flux  of the linear-averaged fitness landscapes in eqs.~(\ref{liN_a b_av},~\ref{liN_vir_av}). In the regime of substantial selection $s_a ,s_v  \gtrsim 1$, the fitness flux in a polymorphic population asymptotically converges to the variance of the stationary fitness distribution in the population (e.g., $s_a^2\, m_{A,2}+\hat s_a^2\, \hat m_{A,2}$ for antibodies)~\cite{Mustonen:2010iga}, which is in accordance with the rate of adaptation given by Fisher's fundamental theorem and Price's equation~\cite{Fisher:1930wy,Price:1970vw}.\\ 

Similarly, the cumulative transfer fluxes over an evolutionary period $[t_0:t]$ read,
\begin{align}
\nonumber \left \langle   {\bf T}_{A\rightarrow  V}(t_0:t)\right \rangle&= \left\langle  N_v \int_{t'=t_0}^{t} \T_{A\to V} (t') dt'  \right\rangle \\
\nonumber &=\left \langle  N_v \int_{t'=t_0}^{t} \left( \frac{\partial F_{{}_V}(t')} {\partial \E(t')}\, \frac{\partial \E(t')}{\partial t'} \Big |_{\{\V\}}+  \frac{\partial F_{{}_V}(t')}{\partial \hat \E(t')}\, \frac{\partial \hat \E(t')}{\partial t'} \Big |_{\{\V\}} \right) \right \rangle\\
&=(N_v/N_a) \left \langle \int_{t'=t_0/N_v}^{t/N_v} dt'\, \left [ 2\theta_a s_v \big(  \varepsilon (t')+ \hat \varepsilon (t')\big) - s_v   \, \big( s_a  m_{A,2} (t') +\hat s_a  \hat m_{A,2} (t')\big)\right]\right\rangle \label{TransferAV-cum_gen}\\
\nonumber\\
\nonumber \left\langle  {\bf T}_{V\rightarrow  A}(t_0:t) \right\rangle&= \left\langle  N_a\int_{t'=t_0}^{t} \T_{V\to A} (t') dt'  \right\rangle\\
&=(N_a/N_v) \left \langle \int_{t'=t_0/N_a}^{t/N_a} dt'\,\left[ -2\theta_v s_a  \varepsilon(t') -s_a  s_v  \,  m_{V,2} (t')\right] \right\rangle \label{TransferVA-cum_gen}
\end{align}
The first terms in equations~(\ref{TransferAV-cum_gen}, \ref{TransferVA-cum_gen})  are the fitness changes due to mutation, the second terms are due to selection.  \\

In the stationary state, the  cumulative flux values grow linearly with the evolutionary time, and simplify to,

\begin{align}
\left \langle   \Phi_A(\tau_{a}) \right \rangle_{\text{st.}}=-\left \langle   {\bf T}_{V\rightarrow  A}(\tau_a)\right  \rangle_{\text{st.}} &=\frac{s_a }{\tilde \theta_a+ \theta_v} \left( s_a \langle m_{A,2} \rangle \theta_v + s_v\langle m_{V,2}\rangle\theta_a\right) \,\tau_a \\
\left \langle \Phi_V (\tau_v)\right\rangle_{\text{st.}} =-\left \langle   {\bf T}_{A\rightarrow  V}(\tau_v)\right \rangle_{\text{st.}} &=\frac{s_v}{\tilde\theta_v+ \theta_a } \left( s_a \langle m_{A,2}\rangle\theta_v +s_v \langle m_{V,2}\rangle \theta_a \right)\,\tau_v
\end{align}
where we have substituted the expected values for the  ensemble averaged binding affinities in the stationary state, given by  eqs.~(\ref{meanE}, \ref{meanhatE}). $\tau_a =(t-t_0)/N_a$ and $\tau_v =(t-t_0)/N_v$ are the evolutionary times respectively  in natural units of  the neutral coalescence time in the antibody population $N_a $ and in the viral population $N_v$.   In the stationary state, the fitness flux in each population and the transfer flux from the opposing population  sum up to 0, keeping the mean fitness of both populations constant. Non-stationary states occur during transient evolutionary dynamics of the whole population, or when considering a subset of the population, such as a clonal lineage, whose size fluctuates to fixation or extinction. In particular, the imbalance between the fitness flux and the transfer flux may determine the evolutionary fate of a clonal lineage, which we discuss in  Section~E. A convenient way to measure  fitness  and transfer flux is from time-shifted fitness measurements, for the  stationary (Fig.~4 and Fig.~S5) and non-stationary  (Fig.~S6) conditions.\\

\vspace{0.9cm}
\begin{center}
{\Large \bf D. Signature of coevolution from time-shifted fitness measurements}\\
\end{center}

Measuring interactions between antibody and viral populations sampled at different time points provides means to quantify the amount of host-pathogen co-adaptation.  We introduce the time-shifted binding affinity between viruses at time $t$ and antibodies at time $t +\tau$ in the variable and in the conserved regions, 
\EQA
\E_\tau(t) &=& \sum_{\alpha,\gamma} E_{\alpha\,\gamma} x^\alpha(t+\tau)\, y^\gamma(t)\\
\hat \E_\tau(t) &=& \sum_{\alpha, \gamma} \hat E_\alpha x^\alpha(t+\tau)y^\gamma(t) = \hat \E(t+\tau)
\EEA
and the corresponding rescaled quantities, $\varepsilon_\tau(t) =\E_\tau(t) /E_0$ and $\hat \varepsilon_\tau(t) =\hat \E_\tau(t) /\hat E_0$. Since the virus cannot evolve in the conserved  region, the time-shifted binding affinity in this region $\hat \E_\tau(t)$  is identical to the non-shifted affinity $\hat \E(t+\tau)$ at time $t+\tau$. The  time-shifted fitness  for antibodies and viruses at time $t$ in interaction with the opposing population sampled at time $t+\tau$ follow,
\begin{itemize}
\item time-shifted viral fitness:
\EQ
\label{Eqtimeshift_vir} N_v F_{V;\tau}(t)=-s_v\varepsilon_\tau(t) 
 \EE
\item time-shifted antibody fitness:
\EQ
N_a F_{A;\tau}(t)= s_a \varepsilon_{-\tau}(t+\tau) + \hat s_a \hat\varepsilon(t) 
\EE
\end{itemize}

As shown in Fig.~4, the behavior of the time-shifted binding affinity (or fitness) is primarily determined by the strength of selection on the phenotype at short values of time-lag $\tau$,  and  is characterized by randomizing mutations  at large separation times. Here, we analytically characterize  the stationary state behavior of the time-shifted binding affinity as a function of the separation time $\tau$. The change in time-shifted binding affinity due to the affinity maturation of antibodies (adaptation) to neutralize the  focal viral population  (i.e., for positive separation times $\tau>0$)  follows, 

\begin{figure}
\begin{center}
\includegraphics[width=0.7\textwidth]{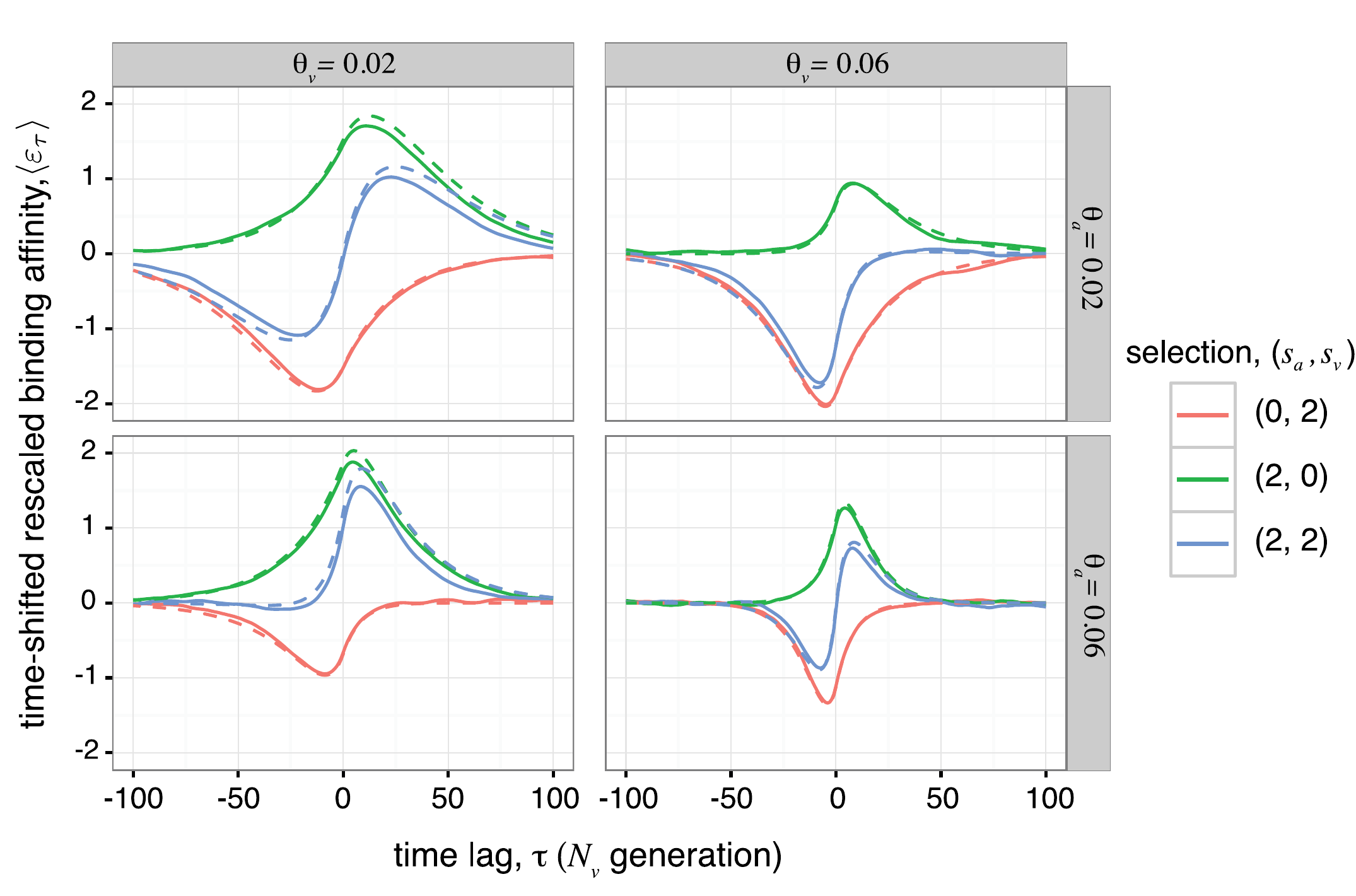}
\end{center}
\caption{\label{fig:S-curve_theory_SI}
{{\bf Stationary time-shifted binding affinity between antigens and antibodies.} Analytical estimates (dashed lines, eqs.~(S102, S103)) for the ensemble-averaged  time-shifted binding affinity $\langle \varepsilon_\tau\rangle$ between the viral population sampled at time $t$, and the antibody population at time $t+\tau$ averaged over all $t$ in the stationary state, show good agreements with the numerical estimates of Wright-Fisher simulations (full lines), over a range of evolutionary parameters. Parameters are $N_a=N_v=1000$ and $\kappa_i=1$ for all sites, and the selection coefficients and the nucleotide diversity as indicated by the legend. Results are time-averaged over $10^4N_a$ generations, with first $100N_a$ generations omitted.
}}
\end{figure}
\EQA
\nonumber \frac{d}{d\tau} \langle \varepsilon_\tau(t) \rangle &=&\left\langle \frac{1}{E_0}\, \sum_{\alpha,\gamma} E_{\alpha\,\gamma} y^\gamma(t)\, \frac{d}{d\tau} x^\alpha(t+\tau)\right\rangle \\
\nonumber &=&  -2 \tilde \theta_a \langle \varepsilon_\tau(t) \rangle+\left\langle \frac{s_a}{E_0^2} \sum_\alpha E_{\alpha\, .}(t) \left(E_{\alpha\, .}(t+\tau) -\E(t+\tau) \right)  x^\alpha(t+\tau)\right\rangle\\
&\simeq&-2 \tilde \theta_a \langle \varepsilon_\tau(t) \rangle+ s_a \langle m_{A,2}\rangle  e^{-2\theta_v \tau}  \label{fw_dyn_vir_timeshift}
\EEA
where time  is measured in units of the viral coalescence time, $N_v$.  We used a mean-field approach  in the stationary state  to approximate the finite-time divergence of the averaged binding affinity for a given antibody in a time-varying environment of evolving viruses, i.e., $\left\langle \sum_\alpha x^\alpha(t+\tau) (E_{\alpha\,.}(t+\tau )- \E(t+\tau)  ) \,  (E_{\alpha\, .}(t )- \E(t)) \right \rangle \simeq e^{-2\theta_v\tau} \left \langle  \sum_\alpha x^\alpha(t+\tau) (E_{\alpha\, .}(t+\tau )- \E(t+\tau)  )^2\right \rangle= e^{-2 \theta_v \tau} \langle M_{A,2}\rangle$. The behavior of the time-shifted binding  affinity at negative separation times $\tau<0$  is mainly determined by the adaptation (escape) of the viruses to the antibodies in the past. In the stationary state, the backward dynamics of the time-shifted binding affinity with respect to the focal viral population is equivalent to the  forward dynamics  with respect to the focal  antibody population, which can be evaluated similarly to eq.~(\ref{fw_dyn_vir_timeshift}). Combining the forward and the backward dynamics results in the following functional form for the rescaled time-shifted binding affinity,
\begin{figure}
\begin{center}
\includegraphics[width=0.4\textwidth]{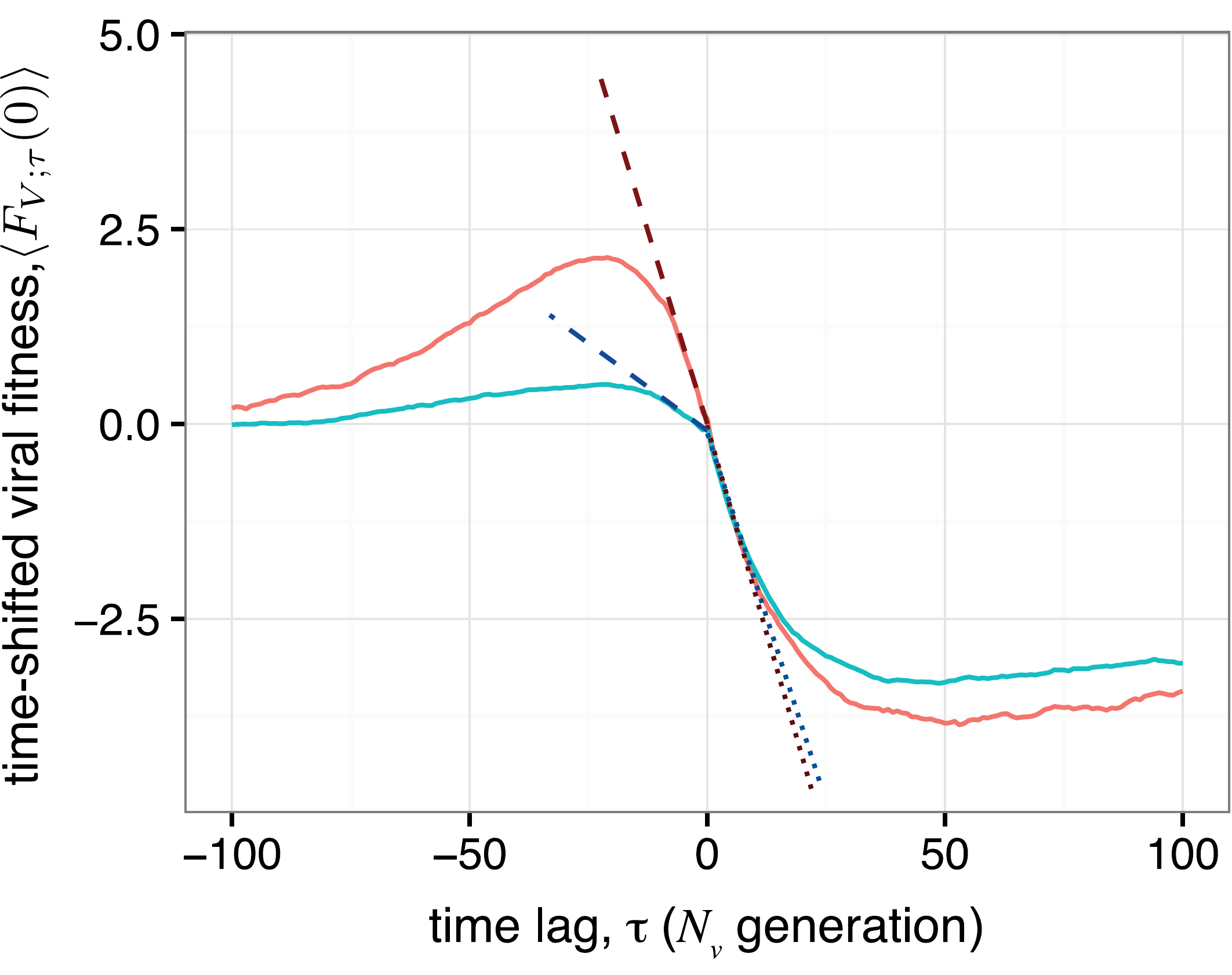}
\end{center}
\caption{\label{fig:delay_nonstat}
{{\bf Non-stationary signature of coevolution from time-shifted fitness.}
Transient (non-stationary) coevolution is quantified by the ensemble-averaged time-shifted mean fitness of the viral population sampled at a reference time point of $N_v$ generations after the beginning of the simulation, that is before the system reaches a stationary state; see eqs.~(S81, S82). For $\tau>0$, the time-shifted fitness $\langle F_{V;\tau}(0)\rangle$ (S99)  measures the fitness of the focal viral population at reference time 0 against the antibodies sampled at time $+\tau$. For $\tau<0$,  we  show  $\langle F_{V;-\tau}(\tau)\rangle$, i.e., the time-shifted fitness with antibodies from $t=0$ and viruses from time $+\tau$. The fitness function is shown for two evolutionary regimes, (i) stronger viral selection, $s_v=2$, $s_a=1$ (red) and  (ii) weaker viral selection, $s_v=1$, $s_a=2$ (blue). The slope of time-shifted fitness at $\tau=0$ measures the population's fitness flux (dashed lines) and the transfer flux from the opposing population (dotted lines), estimated based on the phenotype statistics measured in the simulations (S92, S93). Fitness flux and transfer flux do not have equal values in a non-stationary state, leading to the discontinuity in the slope of the time-shifted fitness function at $\tau=0$. Parameters are $\ell=\hat \ell=50$, $N_a=N_v=1000$, $\theta_a=\theta_v=1/50$. Populations are evolved for $N_v$ generations to reach the reference time $\tau=0$, then data is collected over $100N_v$ generations. Results are ensemble-averaged over $10^3 $ initializations.}}
\end{figure}

\begin{itemize}
\item for antibody affinity maturation, $\tau\geq0$
\EQA
\langle \varepsilon_\tau(t)\rangle =\begin{cases}
\frac{ s_a m_{A,2}}{2 (\theta_a -\tilde \theta_v)}\,e^{-2\theta_v \tau} -   \left(\frac{s_v m_{V,2} }{2(\tilde\theta_a+\theta_v)}   + \frac{\theta_v s_a m_{A,2}}{\theta_a\tilde\theta_a -\theta_v\tilde\theta_v}\right) e^{-2 \tilde \theta_a \tau} 
 & \theta_a \neq\theta_v\\\\
 \left( \frac{s_a m_{A,2}(N_v/N_a)-s_v m_{V,2}}{4\theta}+ s_a m_{A,2} {\small (N_v/N_a) }\tau\right) e^{-2 \theta\tau} &\tilde \theta_a= \theta_v=\theta
  \end{cases}\label{time_shift_sol_fwd}
\EEA
\item for viral escape, $\tau<0$
\EQA
\langle \varepsilon_\tau(t)\rangle =\begin{cases}
\frac{ s_v m_{V,2}}{2 (\tilde \theta_a -\theta_v)}\, e^{-2\tilde \theta_a |\tau|} +     \left(\frac{s_a m_{A,2} }{2(\theta_a+\tilde \theta_v)}  -\frac{\theta_a s_v m_{V,2}}{\theta_a\tilde\theta_a-\theta_v\tilde\theta_v}\right)e^{-2\theta_v |\tau|}
 & \theta_a \neq\theta_v\\\\
 \left( \frac{s_a m_{A,2} (N_v/N_a)-s_v m_{V,2}}{4\theta}-s_v m_{V,2}| \tau|\right)  e^{-2 \theta|\tau|}&\tilde \theta_a=\theta_v=\theta
  \end{cases}\label{time_shift_sol_bwd}
\EEA
\end{itemize}
with $\tilde \theta_a= \theta_a (N_v/N_a)$ and $\tilde\theta_v=\theta_v(N_a/N_v)$. Fig.~4 and Fig.~S5 show  good agreements between the numerical results for the time-shifted  fitness $N_v\langle F_{V;\tau}\rangle = -s_v \langle\varepsilon_\tau\rangle$   from the Wright-Fisher simulations and the analytical solutions~(\ref{time_shift_sol_fwd}, \ref{time_shift_sol_bwd}),  in the stationary state. The slope of time-shifted fitness at time-lag $\tau=0$ is a measure of the antibody population's fitness flux (towards the past) and the transfer flux from the opposing population (towards the future), which are equal in stationary state as depicted in Fig.~4 and Fig.~S5.  In the non-stationary state, the time-shifted fitness $\langle F_{V;\tau}(t)\rangle$ may have a discontinuous derivative at $\tau=0$, due to an imbalance  between  fitness flux and transfer flux (Fig.~S6).\\

\begin{center} 
{\Large \bf E. Evolution of multiple antibody lineages}\\
\end{center}
\vspace{0.5cm}
\label{sec:mult_lin}

\para{Fixation probability in a general fitness landscape.}  We extend our results to multiple clonal antibody lineages evolving with a viral population.  We denote the frequency of an antibody lineage with size $N_a^\C$ by  $\rho^\C= N^\C_a/N_a$.  Assuming that mutations cannot change the identity of one lineage to another, the growth of a given lineage $\C$  depends on the relative mean fitness of the lineage $F_{{}_{A^\C}}$  to the mean fitness of the whole population $F_{{}_A}(t)=\sum_\C F_{{}_{A^\C}}(t) \rho_{{}_\C}(t)$, and on the strength of stochasticity due to genetic drift,

 \begin{align}
\label{rho-change} \frac{d}{dt} \rho_{{}_\C} (t)&=   \sum_\alpha \big( f_{{}_{\C^\alpha}}(t)-F_{{}_A}(t) \big)\, x_{{}_\C}^\alpha(t) + \sqrt{\frac{ \rho_{{}_\C} (1- \rho_{{}_\C} )}{N_a}}
\end{align}
where $f_{{}_{\C^\alpha}}(t)$ is the fitness of the genotype $\A^\alpha$ in the lineage $\C$, and $x_{{}_\C}^\alpha\equiv x_\C(\A^\alpha) $ is the frequency of the genotype $\A^\alpha$ from lineage $\C$ in the total population. Similar to the evolution of a single lineage, the growth of multiple lineages  follows an infinite hierarchy of moment equations for the fitness distribution. Here, we truncate these equations at the second central moment of fitness, which relates to the  lineage-specific fitness flux $ \phi_{{}_{A^\C}} $ and the transfer flux $\T_{V\to A^\C}$. The changes of the ensemble-averaged  mean fitness of a lineage $F_{{}_{A^\C}}(t)$ and the mean fitness of the whole population $F_{{}_A}(t)$,  weighted by the lineage frequency  $ \rho_{{}_\C}(t)$  follow,

 \begin{align}
\nonumber \left\langle \frac{d}{dt} \sum_\alpha f_{{}_{\C^\alpha}}(t)\, x_{{}_\C}^\alpha(t) \right\rangle &= \left \langle  \rho_{{}_\C}(t)\sum_{\alpha\in \C} \frac{\partial F_{{}_{A^\C}} }{\partial x^\alpha_{{}_\C}}  \, \times \frac{d x^\alpha_{{}_\C}}{dt} \right\rangle + \left  \langle  \rho_{{}_\C}(t) \sum_\gamma \frac{\partial  F_{{}_{A^\C}} }{\partial y^\gamma}  \, \times\frac{d  y^\gamma}{dt}\right \rangle-\frac{1}{N_a}  \left \langle F_{{}_{A^\C}}(t) \, \rho_{{}_\C}(t) \right \rangle \\
 &\equiv \big \langle  \rho_{{}_\C} (t)\,   \phi_{{}_{A^\C}} (t) \big \rangle +\big \langle  \rho_{{}_\C}(t) \,  \T_{V\to A^\C} (t) \big\rangle -\frac{1}{N_a} \left \langle F_{{}_{A^\C}}(t) \, \rho_{{}_\C}(t) \right \rangle\\\nonumber\\
  \left \langle \frac{d}{dt}  \sum_\alpha F_{{}_A} (t)\, x_{{}_\C}^\alpha (t)\right \rangle
   &= \big \langle  \rho_{{}_\C}(t)\, \phi_{{}_A} (t) \big \rangle +\big \langle  \rho_{{}_\C}(t)\, \T_{V\to A} (t)\big\rangle
  -\frac{1 }{N_a}\Big \langle F_{{}_A} (t) \, \rho_{{}_\C}(t) \Big \rangle
 \end{align}

Here, we assume that the mean fitness of a lineage only depends on the genotypes within the lineage, as is the case for the fitness functions given by eqs.~(\ref{liN_a b_av}, \ref{liN_vir_av}). The ensemble-averaged changes of the fitness flux and the transfer flux due to selection  depend on higher central moments of the fitness distribution,  which we neglect in our analysis.  The effects of mutation and genetic drift (using It\^o calculus) on the flux quantities follow, 

\begin{align}
\label{drho}&\frac{d}{dt} \Big \langle  \rho_{{}_\C}(t)\, \phi_{{}_{A^\C}} (t)\Big\rangle \simeq \Big  \langle  \rho_{{}_\C}(t)\, m_{{}_{A^\alpha}}  \,\frac{\partial }{\partial x^\alpha} \phi_{{}_{A^\C}}  (t) \Big\rangle +\frac{1}{N_a} \, \Big[\big \langle  \rho_{{}_\C}(t)\, \phi_{{}_{A}} (t)\big\rangle -2  \big \langle  \rho_{{}_\C}(t)\, \phi_{{}_{A^\C}} (t)\big\rangle\Big]  \\
 &\frac{d}{dt} \Big \langle  \rho_{{}_\C}(t)\, \phi_{{}_{A}} (t)\Big\rangle \simeq \Big  \langle  \rho_{{}_\C}(t)\, m_{{}_{A^\alpha}} \, \frac{\partial }{\partial x^\alpha} \phi_{{}_{A}}  (t) \Big\rangle +\frac{1}{N_a} \, \Big[\big \langle  \rho_{{}_\C}(t)\, \phi_{{}_{A^\C}} (t)\big\rangle -2  \big \langle  \rho_{{}_\C}(t)\, \phi_{{}_{A}} (t)\big\rangle\Big]  \\
& \frac{d}{dt} \Big \langle  \rho_{{}_\C}(t)\, \T_{V\to A^\C} (t)\Big\rangle \simeq\Big  \langle  \rho_{{}_\C}(t)\, \Big[ m_{{}_{A^\alpha}}  \, \frac{\partial }{ \partial x^\alpha} \T_{V\to A^\C} (t)  +m_{{}_{V^\gamma}} \,  \frac{\partial }{ \partial y^\gamma} \T_{V\to A^\C} (t)  \Big] \Big \rangle -\frac{1}{N_v}  \Big \langle  \rho_{{}_\C}(t)\, \T_{V\to A^\C} (t)\Big\rangle  \\
\label{rho-TVA-change}& \frac{d}{dt} \Big \langle  \rho_{{}_\C}(t)\, \T_{V\to A} (t)\Big\rangle \simeq\Big  \langle  \rho_{{}_\C}(t)\, \Big[ m_{{}_{A^\alpha}}  \, \frac{\partial }{\partial x^\alpha} \T_{V\to } (t)  +m_{{}_{V^\gamma}} \,  \frac{\partial }{ \partial y^\gamma} \T_{V\to A} (t)  \Big] \Big \rangle -\frac{1}{N_v}  \Big \langle  \rho_{{}_\C}(t)\, \T_{V\to A} (t)\Big\rangle \\\nonumber
\end{align}
where $m_{{}_{A^\alpha}}$ and $m_{{}_{V^\gamma}}$  are the mutational fields associated with  the changes in genotype frequencies due to mutations in  antibodies and  viruses, as defined by eq.~(\ref{mutation_field}). 
  
In order to compute the fixation probability $P_{\text{fix}}= \lim_{t\to \infty} \langle \rho_{{}_\C}(t)\rangle $, it is convenient to use the Laplace transform of the lineage  frequency,  and compute its asymptotic behavior at large time (see e.g., \cite{Desai:2007wv}). The Laplace transform of a given function $A(t)$ can be computed as, $\mathcal{A} (z) = \sum_t A(t) \exp[- z t] $ with the inverse transform: $ A(t)=\underset{T\rightarrow \infty}{\lim} \frac{1}{2 \pi i}\int_{ \gamma-iT}^{\gamma+ i T} \exp[ z t] \mathcal A(z)$. Following this procedure for the hierarchy of equations~(\ref{rho-change}-\ref{rho-TVA-change}) entails a general form  for the fixation probability of a lineage, depending on the initial states of the antibody and the viral populations,   

\begin{align}
\nonumber P_{\text{fix}} (\C)&= \lim_{t\rightarrow\infty}   \langle \rho_{{}_\C}(t)\rangle \\
\nonumber&=  \big \langle \rho_{{}_\C}(0) \big \rangle + \Big  \langle N_a \big( F_{{}_{A^\C}}(0) - F(0) \big) \, \rho_{{}_\C}(0)\Big \rangle+\frac{1}{3} \Big  \langle N_a^2 \big( \phi_{{}_{A^\C}}(0) - \phi_{{}_A}(0) \big) \, \rho_{{}_\C}(0)\Big \rangle\\
&\quad- \Big  \langle N_aN_v\Big(\big| \T_{V\rightarrow A^\C}(0)\big| - \big| \T_{V\rightarrow A} (0)\big|\Big) \, \rho_{{}_\C}(0) \Big \rangle+\mathcal{O}(\theta \langle (N\delta f)^2\rangle , \langle (N\delta f)^3\rangle) \label{pfix_general}
\end{align}
where $\langle (\delta f )^r\rangle$ denotes the $r^{th}$ central moment of the fitness distribution. Here, we have neglected the change in fitness and transfer flux due to mutations, which is of the order  of  $\mathcal{O}(\theta \langle (N\delta f)^2\rangle)$. Below, we will explicitly study the mutational terms for the specific case of the linear fitness model in eqs.~(\ref{liN_a b_av}, \ref{liN_vir_av}).
 The first term in  eq.~(\ref{pfix_general}) is the  ensemble-averaged  initial frequency of the lineage at time $t=0$, and equals its fixation probability  in neutrality.  In the presence of selection, lineages of antibodies with higher relative mean fitness, $ F_{{}_{A^\C}}(0) - F(0) $,  higher rate of adaptation, $\phi_{{}_{A^\C}}(0) - \phi _{{}_A}(0)$, and lower (absolute)  transfer flux from viruses, $\big| \T_{V\rightarrow A^\C}(0)\big| - \big| \T_{V\rightarrow A} (0)\big|   $,  tend to dominate the population. \\

\para{Fixation probability in the linear fitness landscape.} In the linear-averaged fitness model~(\ref{liN_a b_av}, \ref{liN_vir_av}), the growth of a lineage depends on its relative binding affinity compared to  the rest of the population. In order to quantify the competition between the lineages, we define the following lineage-specific moments,

\begin{align}
&L^{{}^\C}_{A_m} =  \big\langle \sum _{\alpha} ( E_{\alpha \, .} -  \E)^m\,  x_{{}_\C}^\alpha \big\rangle ,\quad \quad \hat L^{{}^\C}_{A_m} =  \big\langle \sum _{\alpha} ( \hat E_{\alpha \, .}  - \hat \E)^m \,  x_{{}_\C}^\alpha \big\rangle\\
&L^{{}^\C}_{A_{(m;n)}}  = \Big\langle \sum_{\alpha} (E_{\alpha\,.}-\E)^{m} \,  x_{{}_\C}^\alpha \, \sum_{\beta, \C'} (E_{\beta\,.}-\E)^{n} \,  x_{{}_{\C'}}^\beta \Big\rangle\\
&\hat L^{{}^\C}_{A_{(m;n)}}  = \Big\langle \sum_{\alpha} (\hat E_{\alpha\,.}-\hat \E)^{m} \,  x_{{}_\C}^\alpha \, \sum_{\beta, {\C'}} (\hat E_{\beta\,.}-\hat \E)^{n} \,  x_{{}_{\C'}}^\beta 
\Big\rangle\\
&L^{{}^\C}_{A_m,V_k} =  \Big\langle \sum_{\gamma} (E_{.\, \gamma}-\E)^k y^\gamma \sum_{\alpha} (E_{\alpha\, \gamma}-\E)^{m} \,  x_{{}_\C}^\alpha \Big\rangle\\\nonumber
\end{align}

In this notation the zeroth order lineage-specific moment is equal to the ensemble-averaged frequency of  the focal lineage $L_{A_0}^{{}^\C}\equiv \langle \rho_{{}_\C}\rangle $.  As given by eq.~(\ref{drho}), the change in the frequency of the lineage $\C$ follows from the evolution equation,
\begin{align}
\frac{d}{dt} L_{A_0}^{{}^\C}= S_a   (L_{A_1}^{{}^\C}+\hat L_{A_1}^{{}^\C})
\end{align}

The evolutionary dynamics of multiple lineages  follows from an infinite hierarchy of moment equations. In order to estimate the fixation probability of a lineage up to the order of $\mathcal{O}((NS)^2)$, it is sufficient to truncate the hierarchy at the second moment. These hierarchy of evolution  equations for the lineage-specific moments $L_{A_m}^{{}^\C}$ and the cross-statistics $L_{A_m,V_k}^{{}^\C}$ follow,  

\begin{align}
& \nonumber {\textbf{variable region:}}\\
 \label{var_Lin_mom}&\frac{d}{dt} L_{A_1}^{{}^\C}= S_a  \left( L_{A_2}^{{}^\C} -  L_{A_{(0;2)}}^{{}^\C}\right) - S_v   \left( L_{A_1,V_1}^{{}^\C}  -  L_{A_0,V_2}^{{}^\C} \right)-2 (\mu_a +\mu_v) L_{A_1}^{{}^\C} - \frac{L_{A_1}^{{}^\C}}{N_a }\\
&\frac{d}{dt}L_{A_2}^{{}^\C}=-4 \mu_a \left(L_{A_2}^{{}^\C} -   \ell\, Q_{2}^\C\right) -4\mu_v  L_{A_2}^{{}^\C}  + \frac{L_{A_{(0;2)}}^{{}^\C} - 2 L_{A_2}^{{}^\C}}{N_a } +\mathcal{O}(S_a  )\\
&\frac{d}{dt}L_{A_{(0;2)}}^{{}^\C}=-4 \mu_a\left (L_{A_{(0;2)}}^{{}^\C} - \ell \, Q_{(0;2)}^\C\right) -4\mu_v  L_{A_{(0;2)}}^{{}^\C}   +\frac{L_{A_2}^{{}^\C} -2L_{A_{(0;2)}}^{{}^\C}}{N_a }+\mathcal{O}(S_a )\\
&\frac{d}{dt}L_{A_1,V_1}^{{}^\C}=-4 \mu_a L_{A_1,V_1}^{{}^\C} -4\mu_v\left ( L_{A_1,V_1}^{{}^\C} - \ell \, \sqrt{Q_{2}^\C\, Q_{(0;2)}^\C} \,\right)  -\frac{{L_{A_1,V_1}^{{}^\C}}}{N_v} +\mathcal{O}(S_a,S_v )\\
&\frac{d}{dt}L_{A_0,V_2}^{{}^\C}=-4 \mu_a L_{A_0,V_2}^{{}^\C} -4\mu_v \left(L_{A_0,V_2}^{{}^\C}  -  \ell\, Q_{(0;2)}^\C\right)-\frac{L_{A_0,V_2}^{{}^\C}}{N_v} +\mathcal{O}(S_a,S_v )
\\\nonumber\\
& \nonumber {\textbf{conserved region:}}\\
\label{cons_Lin_mom}&\frac{d}{dt} \hat L_{A_1}^{{}^\C}= S_a \left ( \hat L_{A_2}^{{}^\C} - \hat L_{A_{(0;2)}}^{{}^\C}\right)-2 \mu_a  \hat L_{A_1}^{{}^\C} - \frac{\hat L_{A_1}^{{}^\C}}{N_a }\\
&\frac{d}{dt}\hat L_{A_2}^{{}^\C}=-4 \mu_a\left (\hat L_{A_2}^{{}^\C} - \hat \ell\, \hat Q_{2}^\C\right)   +\frac{\hat L_{A_{(0;2)}}^{{}^\C} - 2 \hat L_{A_2}^{{}^\C}}{N_a }+\mathcal{O}(S_a)\\
&\frac{d}{dt}\hat L_{A_{(0;2)}}^{{}^\C}=-4 \mu_a\left (\hat L_{A_{(0;2)}}^{{}^\C} - \hat\ell\,  \hat Q_{(0;2)}^\C\right)    +\frac{\hat L_{A_2}^{{}^\C} -2\hat L_{A_{(0;2)}}^{{}^\C}}{N_a }+\mathcal{O}(S_a)\label{cons_Lin_mom}\\\nonumber
\end{align}
with the  lineage-specific statistics of the trait scale, 
\begin{align}
&Q_{2}^\C= \left\langle \rho_{{}_\C}\, \k_{2}^{\C}\right\rangle,\qquad \qquad Q_{(0;2)}^\C=\left\langle\rho_{{}_\C} \sum_{\text{lineages }\C'}\, \k_{2}^{\C'} \rho_{{}_{\C'}}\right\rangle  \\
&\hat Q_{2}^\C= \left\langle \rho_{{}_\C} \, \hat \k_{2}^{\C} \right\rangle,\qquad \qquad\hat Q_{(0;2)}^\C=\left\langle\rho_{{}_\C} \sum_{\text{lineages }\C'} \,\hat \k_{2}^{\C'} \rho_{{}_{\C'}}\right\rangle 
 \end{align}
 $\k_{2}^{\C}= \sum_{i=1}^{\ell}( \kappa_i^{\tiny \C})^2\big/ {\ell}$ and $\hat \k_{2}^{\C}=\sum_{i=1+\ell}^{\ell+\hat\ell }(\hat \kappa_i^{\tiny \C})^2\big / {\hat \ell} $ are the averaged accessibilities for a given lineage $\C$,  similar to the definition in eq.~(\ref{K_r}). As indicated by eqs.~(\ref{var_Lin_mom}-\ref{cons_Lin_mom}), the composite lineage-specific trait statistics $Q_2^\C-Q_{(0;2)}^\C$ and $\hat Q_2^\C-\hat Q_{(0;2)}^\C$ influence the evolution of  the lineage frequency. These quantities vary over time due to  changes in the lineage composition of the population,
   \begin{align}
&  \frac{d}{dt} \left(Q_{2}^\C -Q_{(0;2)}^\C\right) = \frac{1}{N_a} \left (Q_2^\C-Q_{(0;2)}^\C \right) +\mathcal{O}(S_a) \\
 &\frac{d}{dt} \left(  \hat Q_{2}^\C - \hat Q_{(0;2) }^\C \right )=\frac{1}{N_a} \left (\hat Q_2^\C-\hat Q_{(0;2)}^\C \right)+ \mathcal{O}(S_a)
 \end{align}
 
In order to compute the fixation probability, we use the Laplace transform of the lineage-specific moments $\L_{A_m,V_k}^\C(z)$ and the lineage-specific statistics of the trait scale $\q_2^\C(z)-\q_{(0;2)}^\C(z)$, and compute the asymptotic behavior of  the $0^{th}$ moment $L_0^{{}^\C}$, after the inverse transform (see e.g., \cite{Desai:2007wv,Good:2013km}).  The Laplace transform of  the moment hierarchy~(\ref{var_Lin_mom}-\ref{cons_Lin_mom}) up to order of $\mathcal{O} ((NS)^2)$ in $\L_{A_0}^{{}^\C}$ follows,\\

 \begin{align}
& z \L_{A_0}^{{}^\C} (z)-L_{A,0}^{{}^\C}(0) = S_ a (\L_{A_1}^{{}^\C}(z) +\hat \L_{1}^{{}^\C}(z))\\\nonumber\\
& \textbf{variable region:}  \nonumber \\
& z \L_{A_1}^{{}^\C}(z)-L_{A_1}^{{}^\C}(0) = S_a  (\L_{A_2}^{{}^\C} (z)- \L_{A_{(0;2)}}^{{}^\C}(z))-S_v   ( \L_{A_1,V_1}^{{}^\C}  -  \L_{A_0,V_2}^{{}^\C} )-2 (\mu_a+\mu_v) \L_{A_1}^{{}^\C}(z) - \frac{\L_{A_1}^{{}^\C}(z)}{N_a }\\
& z \L_{A_2}^{{}^\C}(z) -L_{A_2}^{{}^\C}(0) =-4 \mu_a\left (\L_{A_2}^{{}^\C}(z) - \ell\, \q_{2}^\C (z)\right) -4\mu_v  \L_{A_2}^{{}^\C}(z) +\frac{\L_{A_{(0;2})}^{{}^\C}(z) - 2 \L_{A_2}^{{}^\C}(z)}{N_a }\\
& z \L_{A_{(0;2)}}^{{}^\C} - L_{A_{(0;2})}^{{}^\C}(0) = -4 \mu_a \left (\L_{A_{(0;2})}^{{}^\C} - \ell\, \q_{(0;2)}^\C(z)\right  ) -4\mu_v  \L_{A_{(0;2})}^{{}^\C}   +\frac{\L_{A_2}^{{}^\C} -2\L_{A_{(0;2})}^{{}^\C}}{N_a }  \\
 &z \L_{A_1,V_1}^{{}^\C}-L_{A_1,V_1}^{{}^\C}(0)=-4 \mu_a \L_{A_1,V_1}^{{}^\C} -4\mu_v\left ( \L_{A_1,V_1}^{{}^\C} - \ell \, \sqrt{  \q_{2}^\C(z)  \q_{(0;2)}^\C(z)  }\,\,\,\right)   -\frac{\L_{A_1,V_1}^{{}^\C}}{N_v}\\
 &z \L_{A_0,V_2}^{{}^\C}- L_{A_0,V_2}^{{}^\C}(0)=-4 \mu_a \L_{A_0,V_2}^{{}^\C} -4\mu_v \left (\L_{A_0,V_2}^{{}^\C}  - \ell \q_{(0;2)}^\C(z)\right)  -\frac{{\L_{A_0,V_2}^{{}^\C}}}{N_v} \\
 & z\left( \q_2^\C(z)-\q_{(0;2)}^\C(z) \right)-\left(Q_2^\C(0)-Q_{(0;2)}^\C(0)\right)= \frac{1}{N_a} \left( \q_2^\C(z)-\q_{(0;2)}^\C(z) \right)
\\\nonumber\\\nonumber
 &\textbf{conserved region:}  \nonumber \\
& z \hat \L_{A_1}^{{}^\C}(z)-\hat L_{A_1}^{{}^\C}(0)  = S_a  (\hat \L_{A_2}^{{}^\C} (z)- \hat \L_{A_{(0;2})}^{{}^\C}(z))-2 \mu_a \hat \L_{A_1}^{{}^\C}(z) - \frac{\hat \L_{A_1}^{{}^\C}(z)}{N_a }\\
& z \hat \L_{A_2}^{{}^\C}(z) -\hat L_{A_2}^{{}^\C}(0) = -4 \mu_a\left (\hat \L_{A_2}^{{}^\C}(z) - \hat \ell \,\hat \q_2^\C(z)\right) +\frac{\hat \L_{A_{(0;2})}^{{}^\C}(z) - 2 \hat \L_{A_2}^{{}^\C}(z)}{N_a } \\
& z \hat \L_{A_{(0;2})}^{{}^\C} - \hat L_{A_{(0;2})}^{{}^\C}(0) = -4 \mu_a \left(\hat \L_{A_{(0;2})}^{{}^\C} - \hat \ell\,\hat\q_{(0;2)}^\C(z)\right)  +\frac{\hat \L_{A_2}^{{}^\C} -2\hat \L_{A_{(0;2})}^{{}^\C}}{N_a }\\
& z\left( \hat \q_2^\C(z)-\hat \q_{(0;2)}^\C(z) \right)-\left(\hat Q_2^\C(0)-\hat Q_{(0;2)}^\C(0)\right)= \frac{1}{N_a} \left( \hat \q_2^\C(z)-\hat \q_{(0;2)}^\C(z) \right)
 \end{align}
 
The inverse transform of $\L_{A_0}^{{}^\C}(z)$ in the  limit of $z\rightarrow 0$  results in the asymptotic behavior of the ensemble-averaged frequency of the lineage $\C$, $\underset{t\rightarrow\infty}{\lim} \,L_{A_0}^{{}^\C}$, which corresponds to the fixation probability $P_{{}_\text{fix}}(\C)$ of the lineage,

\begin{align}
\nonumber  P_{{}_\text{fix}} (\C)=& \underset{t\rightarrow\infty}{\lim} {L_{A_0}^{{}^\C}(t) } \\
\nonumber=&L_{A_0}^{{}^\C}(0) +\frac{ N_a \,S_a }{1+2 (\theta_a+\tilde\theta_v)} L_{A_1}^{{}^\C}(0) +  \frac{ N_a \,S_a }{1+2  \theta_a}\hat L_{A_1}^{{}^\C}(0) \\
\nonumber&+\frac{  (N_a \,S_a)^2  }{(1+2(\theta_a+\tilde\theta_v)) }\times \frac{1}{3+4(\theta_a+\tilde \theta_v)}\left[ L_{A_2}^{{}^\C} (0)-L_{A_{(0;2)}}^{{}^\C}(0) +4\theta_a\ell\left (Q_2(0)-Q_{(0;2)}(0)\right) \right]\\
\nonumber& +\frac{  (N_a \,S_a)^2  }{(1+2 \theta_a ) }\times \frac{1}{3+4\theta_a} \left[ \hat L_{A_2}^{{}^\C}(0)-\hat L_{A_{(0;2)}}^{{}^\C}(0)+4\theta_a \hat \ell \left (\hat Q_2(0)-\hat Q_{(0;2)}(0)\right)  \right]\\
\nonumber& -\frac{(N_v \,S_v)(N_a S_a)}{1+2(\theta_a+\tilde\theta_v)}\times \frac{1}{1+4(\theta_a+\tilde\theta_v)} \left[ L_{A_1,V_1}^{{}^\C}(0) - L_{A_0,V_2}^{{}^\C}(0) +4\theta_v\ell\left( \sqrt{Q_{(0;2)}(0)Q_2(0)} -Q_{(0;2)}(0) \right)\right]\\
\label{pfix}
\end{align}

The fixation probability of a lineage can be characterized by the state of the antibody  and the viral population upon its introduction. The first term in  eq.~(\ref{pfix}) is the frequency of the antibody lineage at the time of introduction, and is equal to the neutral fixation probability. The terms proportional to  the antibody selection coefficient $(N_a S_a)$ measure the  relative fitness  of the lineage $\C$ to the mean fitness of the population. The terms proportional to the $(N_a  S_a )^2$ measure the relative fitness flux of the lineage  $\C$ to the fitness flux of the whole  population. The terms proportional to $(N_a S_a) \times (N_v S_v)$ measure the  transfer flux from the viral population to the antibody lineage $\C$ relative to the total transfer flux from viruses to the  antibody population. $L^{{}^\C}_{A_{(0;2)}}  $ and $L_{A_0,V_2} = \langle \rho^{\C} M_{V,2}\rangle $ are respectively the total diversity of binding in the antibody and in the viral population, scaled by the frequency of the lineage $\C$, and determine the fitness flux  and transfer flux associated with  the whole antibody population. The diversity of binding affinity in viruses is a population observable which affects the lineage fixation probability, as shown in Fig.~5.

The higher viral diversity favors the fixation of  broadly neutralizing antibodies for two reasons. First, the larger viral diversity compromises the mean fitness  of the resident non-broad antibody population, and makes it easier for the potential BnAb lineage to take over the existing antibody lineages. This effect is captured by terms proportional to $N_a S_a$ in eq.~(\ref{pfix}). Second,  the transfer flux from  the viral population to the lineage with access to the conserved interaction regions (i.e, a lineage with $\hat E_0^2/E_0^2 \gg 1$) is  small. Therefore,  the viral escape from binding to a  potential BnAb lineage is less efficient than from the resident non-broad antibody population, which increases the chance of fixation for a potential BnAb lineage.  This effect is captured by terms proportional to $(N_a S_a) \times (N_v S_v)$ in eq.~(\ref{pfix}). 

The approximation used to estimate the fixation probability in eq.~(\ref{pfix}) is valid when the effective selection pressure on the lineage (rescaled by the nucleotide diversity) is comparable to the effective pressure on viruses, i.e., $(s_a^\C- \sum_{\C'}s_a^{\C'}\rho_{{}_{\C'}}+\hat s_a^\C -\sum_{\C'}\hat s_a^{\C'} \rho_{{}_{\C'}}) \theta_a  \sim s_v\theta_v$, where $s_a^\C=N_a S_a (\ell \,\k_2^\C)^{1/2}$ and $\hat s_a^\C= N_a S_a (\hat\ell \,\hat\k_2^\C)^{1/2}$ are the rescaled selection coefficients of the focal lineage $\C$ in the variable and the conserved regions. Fig.~5   shows deviations between  analytical expectations from eq.~(\ref{pfix}) and the outcome of the Wright-Fisher simulations beyond this approximation regime. Specifically, the analytical predictions become less reliable for the case of an emerging BnAb lineage  on the background of  a  neutralizing resident population, which causes a strong selection imbalance between the two populations. Including the higher order terms of the lineage-specific moments would improve the analytical predictions. However, in the regime of very strong selection, the higher order terms of the  series expansion in eq.~(\ref{pfix}) become very large (and of alternating sign),  so that  the fixation probability remains bounded ($0\leq P_{\text{fix}}\leq 1$). In this regime, we show that substituting the second order lineage-specific moments in eqs.~(\ref{var_Lin_mom}, \ref{cons_Lin_mom}) by their ensemble-averaged expectation in neutrality,

\begin{align}
L_{A_2}^\C-L_{A_{(0;2)}}^\C \simeq  4\theta_a \ell  \left (Q_2^{{}^\C}(0)-Q_{(0;2)}^{{}^\C}(0)\right ), \qquad\qquad \hat L_{A_2}^\C-\hat L_{A_{(0;2)}}^\C \simeq  4\theta_a \hat  \ell \left (\hat Q_2^{{}^\C}(0)-\hat Q_{(0;2)}^{{}^\C}(0)\right ) 
\end{align}
could provide a more reliable approximation to the fixation probability as opposed to a higher order yet incomplete expansion; see Fig.~5 (dashed lines).\\
\vspace{1cm}
\begin{center}
{\Large \bf F. Analysis of time-shifted neutralization data}\\
\end{center}

The empirical study by  Richman {\em et~al.}~\cite{Richman:2003dc} provides time-shifted measurements of viral neutralization by a patient's circulating antibodies, as the percent inhibition of viral replication at various levels of antibody dilution compared to an antibody-negative control. 
The inhibition of the virus for a given  concentration of antibodies in the serum $[AB]$ is,

\EQ
{I} = \frac{ [AB]}{ [AB]+K }
\EE
where  $K$ is a  constant that equals the antibody concentration which inhibits $50\%$ of viruses. The inhibition can be written in terms of the plasma dilution $d_{AB}$,

\EQ
I (\V(t_1),\A(t_2))= \frac{ d_{AB} (t_2)}{ d_{AB}(t_2) +1/\text{titer}(V_{t_1},A_{t_2})}
\EE
where titer  is the reciprocal of antibody dilution where inhibition reaches $50\%$ ($\text{IC}_{50}$). Inhibition by antibodies reduces the replication rate of viruses from the maximum value in the absence of antibodies $r_{\text{max}}$ by a factor $1-I$, and results in population growth, $N_v(t+1)= r_{\text{max}} (1-I) N_v(t)$, with a malthusian mean fitness for the viral population $F_v= \frac{1}{t-t_0} \log\left( N_v(t)/N_v(t_0)\right)= \log\left( r_{\text{max}} (1-I) \right)$. In the patient, the plasma is  not diluted i.e., $d_{AB}\sim1$. Therefore, the viral fitness during infection can be approximately expressed in terms of the neutralization titer $F_v\sim-\log (\text{titer})$. A similar relation between neutralization titers and viral fitness has been previously suggested by Blanquart \& Gandon~\cite{Blanquart:2013fq}.

Additional control experiments  show inhibition of a neutralization-sensitive virus (NL43)~\cite{Richman:2003dc}, which we denote by $I^*$. In the stationary state, we  expect that the titers (and fitness) associated with the neutralization-sensitive virus to be comparable across serums of various time-points. However, due to a low antibody response at the initial stages of the infection, the neutralization titers for both autologous viruses and the control  NL43 virus grow as the infection progresses; see Fig.~S7. In order to account for this non-stationary antibody response, we evaluate the fitness as the relative titers of the autologous viruses and the neutralization-sensitive virus (NL43) at each time-point. We define the relative  time-shifted mean fitness of the viral population at time $t$ against the antibody serum sampled at time $t+\tau$  as,
\EQ
F_{V;\tau}(t) = c_0- \log\left(\text{titer}(V_t,A_{t+\tau}) \Big / \text{titer}^*(A_{t+\tau})\right)
\label{titerfitness}
\EE 
where $\text{titer}^*(A_{t+\tau})$ is the neutralization titer for NL43 virus against the serum sampled at time $t+\tau$, and $c_0$ is a constant that relates the relative neutralization titers to the viral fitness. 
\begin{figure}
\begin{center}
\includegraphics[width=  0.8\columnwidth]{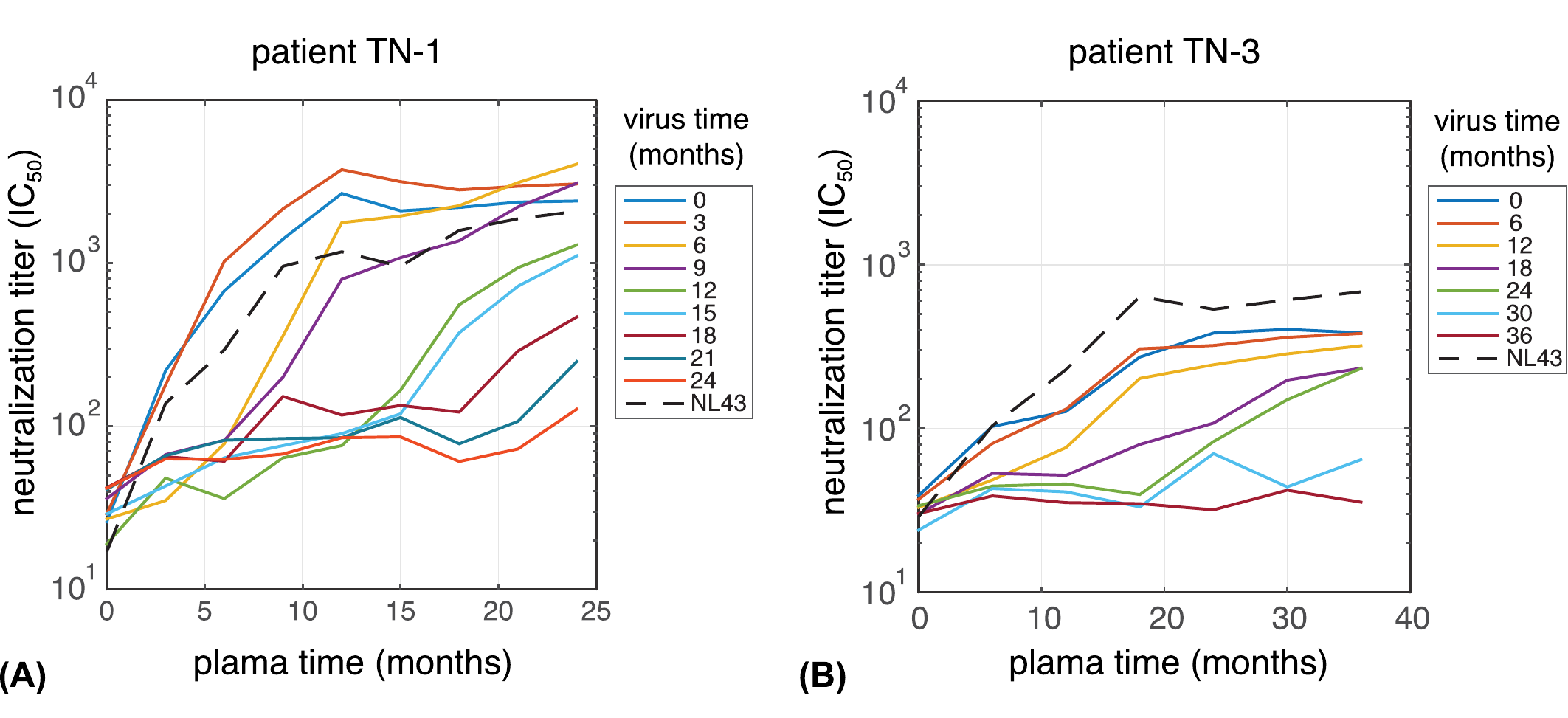}
\end{center}
\caption{ {\bf Neutralization titers of HIV against  patients plasma.}  Neutralization activity (titer) of plasma against autologous viruses collected at various time-points (colors) from two HIV patients, {\bf(A)} TN-1 and {\bf (B)} TN-3 as reported by~\cite{Richman:2003dc}. Neutralization titers are defined as the reciprocal of antibody dilution at the level that  inhibition reaches 50\% ($\text{IC}_{50}$). In addition, plasma activity against a neutralization-sensitive virus (NL43) is taken as a control measurement (dashed line), which indicates an increasing antibody response over time. 
\label{fig:HIV_data}}
\end{figure}
Fig.~4C  shows the time-shifted relative mean fitness $F_{V;\tau}(t)$ averaged over all time-points $t$, evaluated for two  patients (TN-1 \& TN-3) from the data provided by Richman {\em et~al.}~\cite{Richman:2003dc}.  Before averaging, we linearly interpolate the raw data to produce equal time shifts (3 months for TN-1 and 6 months for TN-3). Due to  the functional form of time-shifted fitness in eqs.~(\ref{time_shift_sol_fwd}-\ref{time_shift_sol_bwd}), which involves sums of two exponentials, brute force parameter  scanning is necessary for a convergent solution. Our results indicate comparable values of nucleotide diversity  in antibodies and viruses $\tilde \theta_a$ and $\theta_v$. Therefore, we report fits to the simpler analytical forms of time-shifted fitness with common $\theta$'s  given by  eqs.~(\ref{time_shift_sol_fwd}-\ref{time_shift_sol_bwd}), that use a single exponential function to both sides of the data. Fits are found by scanning parameters and calculating the mean squared errors with appropriate weights due to averaging over equal time-shifts. Each fit contains 4  composite variables which are  functions of the underlying evolutionary parameters: (i) nucleotide diversity $\theta$, (ii) selection component of the  fitness flux in the viral population $S_v^2 M_{V,2}$, (iii)  selection component of the transfer flux  from antibodies to viruses, $-S_a S_v M_{A,2} (N_v/N_a)$, and (iv) the constant $c_0$ in eq.~(\ref{titerfitness}). Assuming that the derivative of the time-shifted fitness function is continuous at the separation time $\tau=0$, the mean fitness of viruses interacting with their co-residing antibody population can be evaluated dependent on the other fitted parameters, $ F_{V;0}= (S_v^2 M_{V,2}-S_a S_v M_{a,2})/ 4\theta$. The fitted variables are listed below for  both patients, 

{\small
\EQA
\nonumber\begin{array}{ll ccccccccc}
  && \mbox{diversity / month,} &&\mbox{sel. part of $\phi_V\,/\,\text{month}$,} && \mbox{sel. part of $\T_{{}_{ A\to V}}\, / \,\text{month}$,}  && \mbox{offset,} \\
&&\mbox{ $\theta\,\cdot\,(\text{month} /N_v)$} && \mbox{$S_v^2 M_{V,2} $}  &&\mbox{ $-S_a S_v  M_{A,2}$} &&\mbox{$c_0$} \\\\
\mbox{patient TN-1} &&0.07&& 0.69 && -0.24 &&  -0.24
\\
\mbox{patient TN-3} &&0.05 && 0.20 &&  0 && 0.52
\end{array}\\
\label{scoretable}
\EEA
}

The time-shifted fitness measurements match well with the analytical fits and indicate two distinct regimes of coevolutionary dynamics in the two patients.  In patient TN-1,  viruses and antibodies experience a comparable adaptive pressure, as indicated by the ``S-curve" in Fig.~4C (blue line), with $s_v m_{V,2}/(s_a m_{A,2}) =2.9$.  In patient TN-3,  adaptation in viruses is much stronger   than in antibodies, resulting in an imbalanced shape of the time-shifted fitness curve in Fig.~4C (red line). The lower overall neutralization titers in patient TN-3 (Fig.~S7) is indicative of  such imbalance between the immune response and HIV escape in the patient.   It is likely that a longer monitoring of patient TN-3 would capture a stronger antibody response in later stages of infection.

Note that  in these studies time is measured in units of months rather than coalescence time of the populations. Estimating the coalescence time-scale in units of months would require analysis of genealogical relations between sequences  of antibodies and viruses extracted from each patient over the course of infection, which is not available for this study. \\

\begin{center}
{ \bf  \Large G. Simulations}\\
\end{center}

Simulations of the full genotype model (Wright-Fisher dynamics) were implemented as follows. Viral and antibody populations consist of genotypes as strings of $\pm1$ with length $\ell + \hat \ell$. Binding interactions are calculated between all pairs of antibodies and viruses as in eq.~(\ref{binding}), which define the fitness as in eqs.~(\ref{ABfitness}, \ref{Virfitness}). Genotypes within an antibody lineage share the same accessibilities, $\{\kappa_i,\,\hat\kappa_i\}$. For each generation, a poisson distributed number of mutations occur, with each mutation flipping the sign of a site. Each generation is replaced by their offspring which inherit their parents' genotype. Each parent generates a binomially distributed number of offspring, with probability proportional to the exponential of its fitness, with the constraint that the total number of individuals remains constant $N_a$ in antibodies and  $N_v$ in viruses, which is equivalent to multinomial sampling. Note that we define fitness as ``malthusian", which means that fitness is the relative growth rate of genotypes, and the expected number of offspring is proportional to the exponential of fitness.
 
Simulation parameters for all figures are $N_a=N_v=10^3$, $\ell=\hat \ell = 50$, $\theta_a=\theta_v=1/50$, and all $\kappa_i=\hat\kappa_i=1$, unless otherwise stated.  Populations are initialized with all individuals having the same randomly generated genotype. To measure quantities in the stationary state (Figs~\ref{fig:E}, \ref{fig:delay}) simulations are run for $10^4 N_a$ generations, and quantities are averaged from samples every $N_a$ generations. Data from the beginning of the simulations are omitted from the calculations, where the cutoff is $\tau =2\mu_a^{-1}$, the correlation time for the mean binding (Fig.~S3 and Section~B.4 of the Appendix). To produce the simulations shown in Fig.~\ref{fig:pfix}B, the newly emerging antibody lineages compete with the resident  population  as follows. First, the resident lineage is evolved with the virus for $50 N_a$ generations to build up diversity. Simultaneously, the invading lineage is evolved with the virus, except that the viral fitness is determined only by the resident lineage. This  ensures that invading lineages can marginally bind to the viral population, and are functional lineage progenitors; a process that happens prior to affinity maturation in germinal centers. The pre-adaptation of the invading lineage can also be interpreted as initial rounds of affinity maturation in germinal centers isolated from competition with  adapted antibody lineages.  Then the two antibody lineages are combined with resident at 90\% and  invader at 10\%, with a total size of $10^3$, and the state of the system is recorded. The two lineages are evolved until one is extinct, repeated over 100 replicates to estimate the fixation probability. The whole procedure is repeated $10^3$ times for ensemble averaging.  The invader, is either a normal lineage with all $\kappa_i=1$ and $\hat \kappa_i = 0$ or a BnAb that binds only to the conserved region, $\kappa_i=0$ and $\hat \kappa_i = 1$. 

Simulations are written in julia and code is available at https://github.com/jotwin/coevolution.

{\small
\bibliography{bib_armita_abbrv}

}

\vfill{}
	\end{widetext}	
\end{document}